\documentclass[12pt,preprint,longnamesfirst]{aastex}

\newcommand{\kms}{\ifmmode \mathrm{km~s^{-1}}\else km~s$^{-1}$\fi}
\newcommand{\smpy}{\ifmmode M_\sun~\mathrm{yr}^{-1}\else M$_\sun$~yr$^{-1}$\fi}
\newcommand{\lir}{\ifmmode L_\mathrm{IR}\else $L_\mathrm{IR}$\fi}
\newcommand{\lsun}{\ifmmode L_\sun\else $L_\sun$\fi}
\newcommand{\msun}{\ifmmode M_\sun\else $M_\sun$\fi}
\newcommand{\nags}{\ion{Na}{1}}
\newcommand{\nad}{\ion{Na}{1}~D}

\newcommand{\ot}{[\ion{O}{3}]}
\newcommand{\otl}{[\ion{O}{3}] $\lambda5007$}
\newcommand{\nt}{[\ion{N}{2}]}
\newcommand{\ntl}{[\ion{N}{2}] $\lambda6583$}
\newcommand{\cf}{\ifmmode C_f\else $C_f$\fi}
\newcommand{\co}{\ifmmode C_\Omega\else $C_\Omega$\fi}
\newcommand{\dvmax}{\ifmmode \Delta v_{max}\else $\Delta v_{max}$\fi}
\newcommand{\dvtau}{\ifmmode \Delta v_{maxN}\else $\Delta v_{maxN}$\fi}

\shorttitle{Starburst Outflows. II. Analysis and Discussion}
\shortauthors{Rupke, Veilleux, \& Sanders}

\begin{document}

\title{Outflows in Infrared-Luminous Starbursts at $z < 0.5$.  II.  Analysis and Discussion\footnotemark[1] \footnotemark[2] \footnotemark[3]}

\author{David S. Rupke, Sylvain Veilleux}
\affil{Department of Astronomy, University of Maryland, College Park, MD 20742}
\email{drupke@astro.umd.edu, veilleux@astro.umd.edu}
\and
\author{D.~B. Sanders}
\affil{Institute for Astronomy, University of Hawaii, 2680 Woodlawn Drive, Honolulu, HI 96822}
\email{sanders@ifa.hawaii.edu}
\footnotetext[1]{Some of the observations reported here were obtained at the W. M. Keck Observatory, which is operated as a scientific partnership among the California Institute of Technology, the University of California, and the National Aeronautics and Space Administration. The Observatory was made possible by the generous financial support of the W. M. Keck Foundation.}
\footnotetext[2]{Some of the observations reported here were obtained at the MMT Observatory, which is a joint facility of the Smithsonian Institution and the University of Arizona.}
\footnotetext[3]{Some of the observations reported here were obtained at the Kitt Peak National Observatory, National Optical Astronomy Observatory, which is operated by the Association of Universities for Research in Astronomy, Inc. (AURA) under cooperative agreement with the National Science Foundation.}

\begin{abstract}
We have performed an absorption-line survey of outflowing gas in 78 starburst-dominated, infrared-luminous galaxies.  This is the largest study of superwinds at $z \la 3$.  Superwinds are found in almost all infrared-luminous galaxies, and changes in detection rate with SFR--winds are found twice as often in ultraluminous infrared galaxies (ULIRGs) as in less-luminous galaxies--reflect different wind geometries.  The maximum velocities we measure are 600~\kms, though most of the outflowing gas has lower velocities ($100-200$~\kms).  (One galaxy has velocities exceeding 1000~\kms.)  Velocities in LINERs are higher than in \ion{H}{2} galaxies, and outflowing ionized gas often has higher velocities than the neutral gas.  Wind properties (velocity, mass, momentum, and energy) scale with galaxy properties (SFR, luminosity, and galaxy mass), consistent with ram-pressure driving of the wind.  Wind properties increase strongly with increasing galactic mass, contrary to expectation.  These correlations flatten at high SFR ($\ga$10$-$100~\smpy), luminosities, and masses.  This saturation is due to a lack of  gas remaining in the wind's path, a common neutral gas terminal velocity, and/or a decrease in the efficiency of thermalization of the supernovae energy.  It means that mass entrainment efficiency, rather than remaining constant, declines in galaxies with SFR $> 10$~\smpy\ and $M_K < -24$.  Half of our sample consists of ULIRGs, which host as much as half of the star formation in the universe at $z \ga 1$.  The powerful, ubiquitous winds we observe in these galaxies imply that superwinds in massive galaxies at redshifts above unity play an important role in the evolution of galaxies and the intergalactic medium.
\end{abstract}


\section{INTRODUCTION} \label{intro}

The number of theoretical papers discussing or incorporating galactic winds has grown sharply in recent years.  Superwinds have been invoked to explain a myriad of issues relating to contemporary cosmology \citep{vr04,vcb05}.  Numerical simulations and semi-analytic models typically incorporate simplified prescriptions for superwinds, rather than detailed microphysics.  This is appropriate not only due to the current level of computing power, but also because hydrodynamic simulations of superwinds cannot yet accurately predict many large-scale properties of the outflows.  Thus, theorists depend on observers to demonstrate the macroscopic behavior of real outflows over a range of galaxy types and environments.

Previous large surveys of local starburst galaxies have shown that superwinds are common.  \citet{lh95,lh96} undertook an imaging and spectroscopic survey of $\sim$50 edge-on galaxies with infrared luminosities $\lir < 10^{12}$~\lsun\ and redshifts $cz < 15000$~\kms.  Evidence for superwinds in these galaxies includes: extended line emission, shock-like line ratios, broad emission lines, velocity shear, and blue emission line asymmetries.  Several of these properties also show a positive correlation with infrared activity ($=\lir/\mathrm{L_{opt}}$).  \citet{lh96} find that the average (deprojected) outflow speed of the ionized gas in these galaxies is $170^{+150}_{-80}$~\kms.

Using the complementary technique of optical absorption-line spectroscopy, \citet{hlsa00} studied a sample of 32 infrared-luminous galaxies with $z \la 0.1$.  Their survey includes galaxies whose luminosities are dominated by a starburst as well as some powered by an active galactic nucleus (AGN).  They include galaxies with a range of luminosities, including five ultraluminous infrared galaxies (ULIRGs; \lir~$\geq 10^{12}$~\lsun).  \citet{hlsa00} also find that the detection rate of winds is high.  They show that the outflows are dusty, and measure outflow velocities up to $400-600$~\kms; however, they do not measure an individual mass outflow rate for each galaxy and fit only a single velocity component to each absorption feature.

Most recently, detailed absorption-line studies of $10-20$ ULIRGs \citep{rvs02,m05} have found that the detection rate of winds in ULIRGs is higher than that in luminous infrared galaxies (LIRGs; \lir~$\geq 10^{11}$~\lsun).  These studies show that ULIRGs have high mass outflow rates, of tens to hundreds of \smpy.  \citet{m05} also argues that wind velocity scales with star formation rate and galactic mass.

At higher redshifts, directly studying winds is possible only using deep observations of bright or gravitationally-lensed galaxies.  One of the best-studied high-redshift galaxy populations is the Lyman-break galaxies (LBGs).  Extensive spectroscopic observations of LBGs have shown that most of them host high-velocity outflows \citep{p_ea01,ss_ea03,ad_ea03}.  Lyman-$\alpha$ emission is redshifted (perhaps due to resonant scattering on the receding half of the outflow), while UV absorption lines are blueshifted due to interstellar absorption in the outflow.  The average offset of the two is 614~\kms\ \citep{ad_ea03}, implying that an average (projected) outflow velocity is approximately 300~\kms.  Early evidence suggested that these outflows are quite large ($r \la 500$~kpc) and metal-enriched \citep{ad_ea03}, but more recent work has revised the probable sizes of these outflows to more modest radii ($r \la 100$~kpc; \citealt{ad_ea05}).

Though the observational data set on outflows is steadily increasing, both in quality and quantity, it is in many ways still quite limited.  Over most of cosmic history, the frequency of occurrence and impact of superwinds remains unquantified.  We also know very little about how the properties of superwinds depend on the properties of their host galaxies.  Observations have typically focused on dwarf galaxies or edge-on disk galaxies with moderate star formation rates (as a well as a set of high-redshift galaxies with as yet uncertain properties).  ULIRGs are of special interest, as they may host over 50\% of global star formation at $z\sim2$ \citep{p_ea05}.  By studying ULIRGs in the local universe, we can learn how winds are affecting the cosmos at high redshift, when many of the present-day stars were formed and much of the metal enrichment of the IGM may have occurred.

These observations are needed in part to inform theoretical prescriptions of winds in numerical simulations.  Many theorists assume that the mass outflow rate in the wind and star formation rate in the corresponding host galaxy are comparable and linearly proportional to each other \citep[e.g.,][]{kc98,a_ea01a,a_ea01b,s03}.  In other words, the ratio of the mass outflow rate to the star formation rate in a galaxy is generically unity.  This is presumably based on the claim of some observers that this is so, for luminous infrared galaxies \citep{hlsa00} and dwarfs or spirals \citep{m99}.  However, preliminary data from our pilot study suggests that this ratio is an order of magnitude lower in ULIRGs \citep{rvs02}, at least for the neutral gas phase.

The role of this work is to fill in the gaps mentioned above and produce a systematic study of large-scale outflows.  We have made the largest survey to date of massive outflows in infrared-luminous galaxies.  Our survey includes over 100 galaxies and extends over a broad redshift range ($z = 0.0-0.5$).  We have included 78 starburst galaxies with a wide range of properties and, for the first time, searched for superwinds in a large number (43) of starburst-dominated ULIRGs.  (In a forthcoming paper, we discuss observations of 26 ULIRGs which have Seyfert nuclei and compare them to the starburst ULIRGs; \citealt{rvs05b}.)

The existence of outflows can be inferred from the presence of absorption lines that are blueshifted with respect to the systemic velocity of the host galaxy.  We apply this technique using moderately high resolution spectroscopy (FWHM~$= 65-85$~\kms) of the \nad\ $\lambda\lambda5890,~5896$ doublet.  Our observations were obtained with echelle and long-slit spectrographs on the Keck II, MMT, and KPNO 4m telescopes.  We perform detailed profile fitting using multiple velocity components and assuming a Gaussian in optical depth; from these fits we measure velocity, velocity width, optical depth, covering fraction, and column density for each component.  The details of the sample selection, observations, and data reduction are discussed in a companion paper \citep[][hereafter Paper I]{rvs05a}.  In Paper I, we also present the absorption- and emission-line spectra and a discussion of the absorption-line fitting.

The size and makeup of our sample allows us to study the properties of these winds as a function of host galaxy properties over a large dynamic range.  Wind properties that we can measure include: velocity; mass, momentum, energy, and their outflow rates; and mass entrainment efficiency, which is a measure of the relative efficiency with which these winds entrain interstellar gas clouds.  The relevant galaxy properties are discussed in Paper I, and include star formation rate, optical and near-infrared luminosity, circular velocity, spectral type, and redshift.

In \S\S\ref{dv}$-$\ref{cf}, we describe in detail the distributions of velocity, optical depth, column density, and covering fraction in our sample.  \S\S\ref{detrate}, \ref{mme}, and \ref{eta} cover our measurements of outflow detection rate; mass, momentum, energy, and their outflow rates; and mass entrainment efficiency.  \S\ref{spatialinfo} discusses the spatial distribution of absorbing gas.  In \S\ref{ofprop_v}, we look at the dependence of outflow detection rate and properties on host galaxy properties.  We also compare absorption and emission lines in \S\ref{eml}.  \S\ref{discuss} covers, in this order: our discussion of alternatives to the superwind scenario, the global covering factor and escape fraction of outflows, comparison to theory, and discussion of redshift evolution.  We summarize in \S\ref{concl}.

For all calculations, we assume present-day values for the cosmological parameters of $H_0 = 75$~\kms~Mpc$^{-1}$ and the standard $\Omega_m = 0.3$, $\Omega_{\Lambda} = 0.7$ cosmology.  All wavelengths quoted are vacuum wavelengths (except those used as labels for spectral lines) and are generally taken from the NIST Atomic Spectra Database\footnote{\texttt{http://physics.nist.gov/cgi-bin/AtData/main\_asd}}.  (The vacuum wavelengths of \nad\ are 5891.58 and 5897.55~\AA.)

\section{OUTFLOW PROPERTIES} \label{ofprop}

In this section, we discuss our measurements of outflow properties.  Table \ref{avgprop} lists the average outflow properties for our three subsamples.  Table \ref{objprop} lists the outflow properties of each galaxy.  As discussed in Paper I, the full sample of 78 galaxies is divided into three subsamples based on infrared luminosity and redshift: (a)~the IRGs (InfraRed Galaxies), with \lir~$< 10^{12}$~\lsun; (b)~the low-$z$ ULIRGs, with \lir~$\geq 10^{12}$~\lsun\ and $z < 0.25$; and (c)~the high-$z$ ULIRGs, with \lir~$\geq 10^{12}$~\lsun\ and $0.25 < z < 0.50$.

These outflow quantities are calculated from the fitted parameters of the \nad\ feature in each galaxy, as listed in the tables of Paper I.  Paper I describes the details of the \nad\ fitting procedure.  In short, we fit multiple velocity components to the \nad\ feature in each object.  The large velocity dispersions in the galaxies we observe cause the components of the \nad\ doublet, as well as velocity components if there are more than one, to blend with one another.  Our fitting procedure addresses this problem.  We assume Gaussians in optical depth, which translates into observable non-Gaussian intensity profiles for optical depths greater than unity.  We also fit a constant covering fraction for each velocity component.  The measured parameters resulting from our fits are: outflow velocity, $\Delta v$; Doppler width, $b$; central optical depth of the \nad$_1$ line, $\tau_{1,c}$; and the covering fraction, \cf.  The errors in these parameters are determined using Monte Carlo simulations.

In Paper I, we also dicuss the method of computation of the \nags\ and H column densities from these fitted parameters.  The values of $N$(\nags) and $N$(H) for each galaxy are listed in Table 4 of Paper I.

\subsection{Rate of Detection} \label{detrate}

The current survey contains over $50\%$ more starburst galaxies than previous superwind surveys of infrared sources \citep{lh95,lh96,hlsa00,rvs02,m05}.  It also has a much higher number of ULIRGs (43 vs. $<$20).  We are thus in a better position to quantify the frequency of occurrence of winds in ULIRGs, especially as a function of galaxy properties.

We use a velocity cutoff of $-50$~\kms\ to delineate galaxies with superwinds, a slightly more liberal cutoff than in our pilot study (in which we used $-70$~\kms).  Velocity components blueshifted by more than 50~\kms\ are assumed to be outflowing.  This cutoff is chosen to avoid contamination due to systematic errors and measurement errors in wavelength calibration ($\la$10~\kms), line fitting (10~\kms\ on average), and redshift determination (15~\kms).  The vast majority of redshifted components have $|\Delta v| < 50$~\kms, suggesting that this cutoff is a good one.

Table \ref{avgprop} lists the detection rate of massive outflows in each subsample.  We observe that outflows are common in each subsample, are detected in a strong majority of ULIRGs, and are detected more frequently in ULIRGs than in less-luminous galaxies (see \S\S\ref{detrate_v} and \ref{gcf}).  The high-$z$ ULIRGs also show a lower detection rate than the low-$z$ ULIRGs; we discuss this further in \S\ref{zevol}.

Applying this same cutoff to the galaxies in \citet{hlsa00} that are optically classified as \ion{H}{2} galaxies or LINERs and have \lir~$< 10^{12}$~\lsun, we compute a wind detection rate of ($32\pm12$)\% ($7 / 22$ galaxies, with a median log[\lir/\lsun] $=11.0$).  This is consistent with the measurement from our IRG subsample.  Our measurement of ($42\pm8$)\% is slightly higher than that of \citet{hlsa00}, but so is the median infrared luminosity of our subsample.  Our detection rate in ULIRGs, $(70\pm7)$\% overall, is consistent with those from our preliminary report, $(75\pm15)$\% \citep{rvs02}, and another recent survey, $(85\pm8)$\% \citep{m05}.  High-$z$ ($z \sim 3$) galaxies of comparable SFR to the IRGs, the LBGs (with SFR $\sim$ 10$-$100~\smpy; \citealt{p_ea98}), have a much higher detection rate than in our IRG subsample; almost 100\% of LBGs have detected outflows \citep{ad_ea03}.

In \S\ref{gcf}, we consider the detection rates in light of the global covering factor of the gas.  In short, these detection rates are lower limits to the actual fraction of these galaxies that host outflows, which may be close to unity.

\subsection{Velocity Distribution} \label{dv}

Previous surveys of winds in starbursting galaxies that measured the distribution of outflow velocities in galaxies include \citet{hlsa00} and \citet{ad_ea03}.  The latter show that the distribution of the velocity offset between the Ly$\alpha$ line and UV absorption lines is approximately a Gaussian of mean 614~\kms\ and width $\sigma = 316$~\kms, including a slight upturn at $\Delta v = 0$.  This implies that every $z \sim 3$ Lyman-break galaxy has outflowing gas, with an average projected outflow velocity of $\sim$300~\kms.

The distribution of gas velocities in local galaxies of comparable SFR is different from that of LBGs, however.  The strong-stellar-contamination subsample of \citet{hlsa00} shows a roughly symmetric \nad\ velocity distribution around 0~\kms.  The distribution of the interstellar-dominated subsample is asymmetric; it lies completely in the blue for $|\Delta v| > 100$~\kms.

\begin{figure}[t]
\plotone{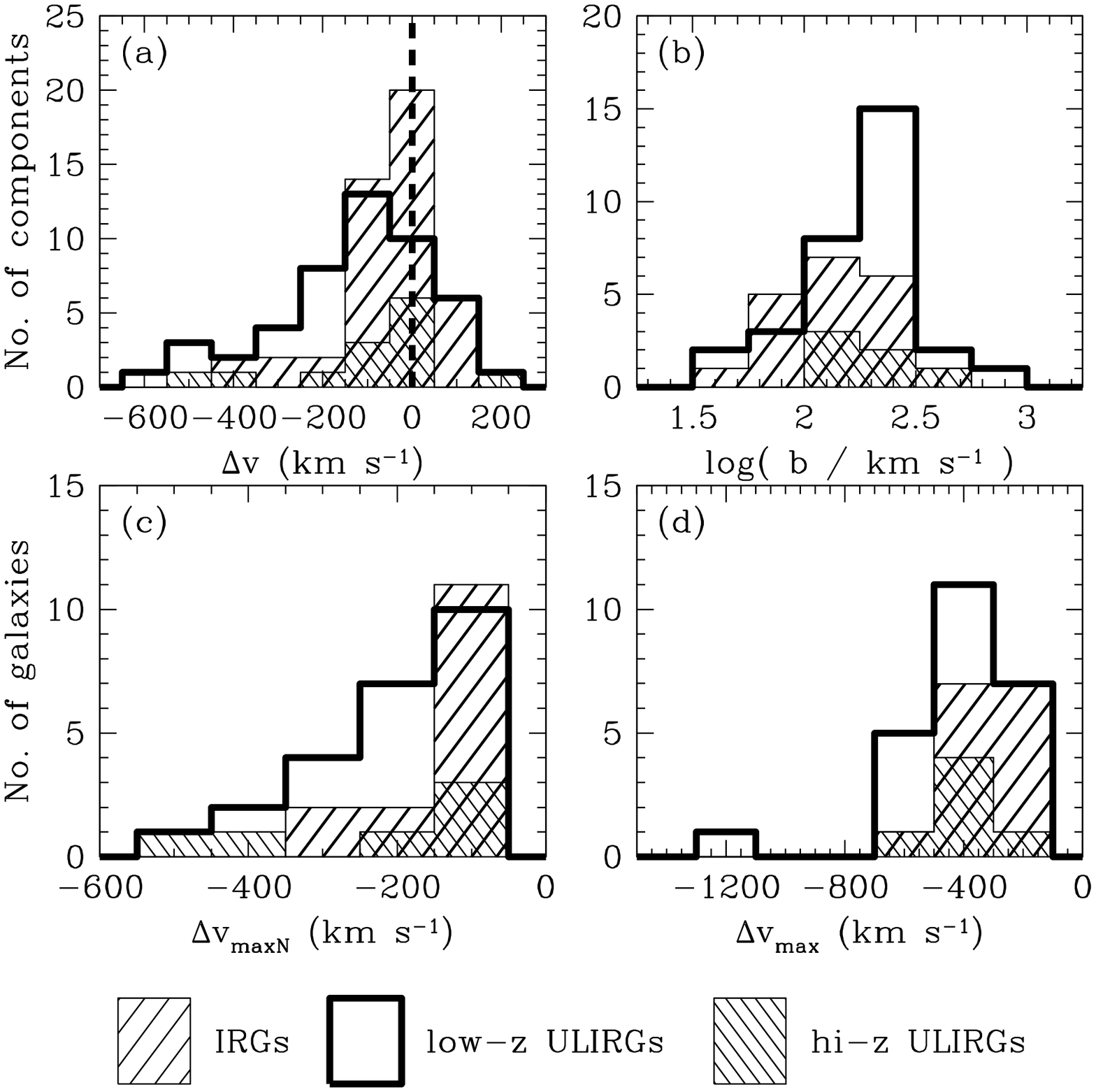}
\caption{Velocity distributions.  ($a$)~Central velocity of each component.  ($b$)~Logarithmic distributions of the Doppler parameter for outflowing components.  ($c$)~Distribution of the velocity of the highest column density gas in each galaxy.  ($d$)~Distribution of `maximum' velocity ($\dvmax\equiv\Delta v - \mathrm{FWHM}/2$, computed for the most blueshifted component) in each galaxy.  The maximum velocities for all galaxies are $\sim$600~\kms\ or below except for F10378$+$1108.  K-S and Kuiper tests show that the distributions of \dvmax\ are consistent with having the same parent distribution, but that the distributions of \dvtau\ for the IRGs and low-$z$ ULIRGs are different at $>$90\% confidence.  The lines in ULIRGs are slightly broader on average than in the IRGs, though K-S amd Kuiper tests do not show significant differences between the distributions.  See \S\ref{dv} for further discussion.}
\label{histdv}
\end{figure}

In Figure \ref{histdv}$a$, we show the distributions of the central velocity of each component in our sample using 100~\kms\ bins.  The distribution of velocities for IRGs is fairly symmetric around $\Delta v = 0$ but has a clear tail toward blue velocities.  The blue asymmetry in the distribution of velocities for ULIRGs above $|\Delta v| = 100$~\kms\ is even more pronounced than in IRGs.  Note also that there are several components with redshifted velocities greater than 100~\kms\ in the ULIRG distributions (though not in the IRG distribution).

The distributions of the velocity of the highest column density outflowing gas in each galaxy, \dvtau, are shown in Figure \ref{histdv}$c$.  (Essentially, these distributions show the $\Delta v$'s which are less than -50~\kms, but we pick only one component from each galaxy.)  The median value for each subsample is listed in Table \ref{avgprop}.  The median value for our full sample is -140~\kms, with a dispersion of 120~\kms\ and a median error of 7~\kms.  The range of $|\dvtau|$ is $50-520$~\kms.  Kolmogorov-Smirnov (K-S) and Kuiper tests show that the distributions of \dvtau\ for the IRGs and low-$z$ ULIRGs are different at $90\%$ confidence; ULIRGs have higher \dvtau\ on average.

The median \dvtau\ of the IRGs, 100~\kms, is smaller than the deprojected velocities of ionized gas in the halos of edge-on galaxies of comparable SFR ($170^{+150}_{-80}$~\kms; \citealt{lh96}) and the mean projected velocity of ionized gas in high-$z$ LBGs (300~\kms; \citealt{ad_ea03}).  We address the relative velocities of ionized and neutral gas in \S\ref{eml}.

The distributions of `maximum' velocity in each galaxy ($\dvmax \equiv \Delta v - \mathrm{FWHM}/2$, computed for the most blueshifted component) are shown in Figure \ref{histdv}$d$.  ($\ga$90\% of the detectable gas in the wind has a velocity less than or equal to \dvmax.)  The median value for each subsample is listed in Table \ref{avgprop}.  The median value for our full sample is $-350$~\kms, with a dispersion of 170~\kms\ and median error of 20~\kms.  The range for $|\dvmax|$ is  $130-600$~\kms, though there is a single galaxy (F10378$+$1108) which has $\dvmax = -1100$~\kms.  K-S and Kuiper tests show that the distributions of each subsample are consistent with being drawn from the same parent distribution.

The Doppler widths ($b=\mathrm{FWHM}/[2\sqrt{\mathrm{ln}~2}]$) of each velocity component are shown as a histogram in Figure \ref{histdv}$b$.  These values have been corrected downward (in quadrature) for the instrumental resolution.  The range of values is 50~\kms\ (the resolution limit) up to almost 700~\kms, with a median error of 25~\kms.  The median value for the IRGs is 150~\kms, while the average value in the low-$z$ ULIRGs is higher, at 200~\kms.  K-S and Kuiper tests show that the overall distributions of the different subsamples are not significantly different, however.

The large values that we measure for $b$ imply that the broadening is due to the sum of large-scale motions in the wind.  These values are far higher than those expected from thermal broadening of warm Na gas ($b_{th}=2.7$~\kms\ at $T=10^4$~K).  However, our spectral resolution is not high enough to resolve clouds with widths $b\la50$~\kms; the wind may be a superposition of such clouds, each at a different velocity.  \citet{sm04} find that the narrowest \nad\ component in a sample of dwarf starbursts is 15~\kms.  In a future paper (Rupke \&\ Veilleux 2005, in prep.), we will present high-resolution observations designed to search for such components in ULIRGs.

\begin{figure}[t]
\plotone{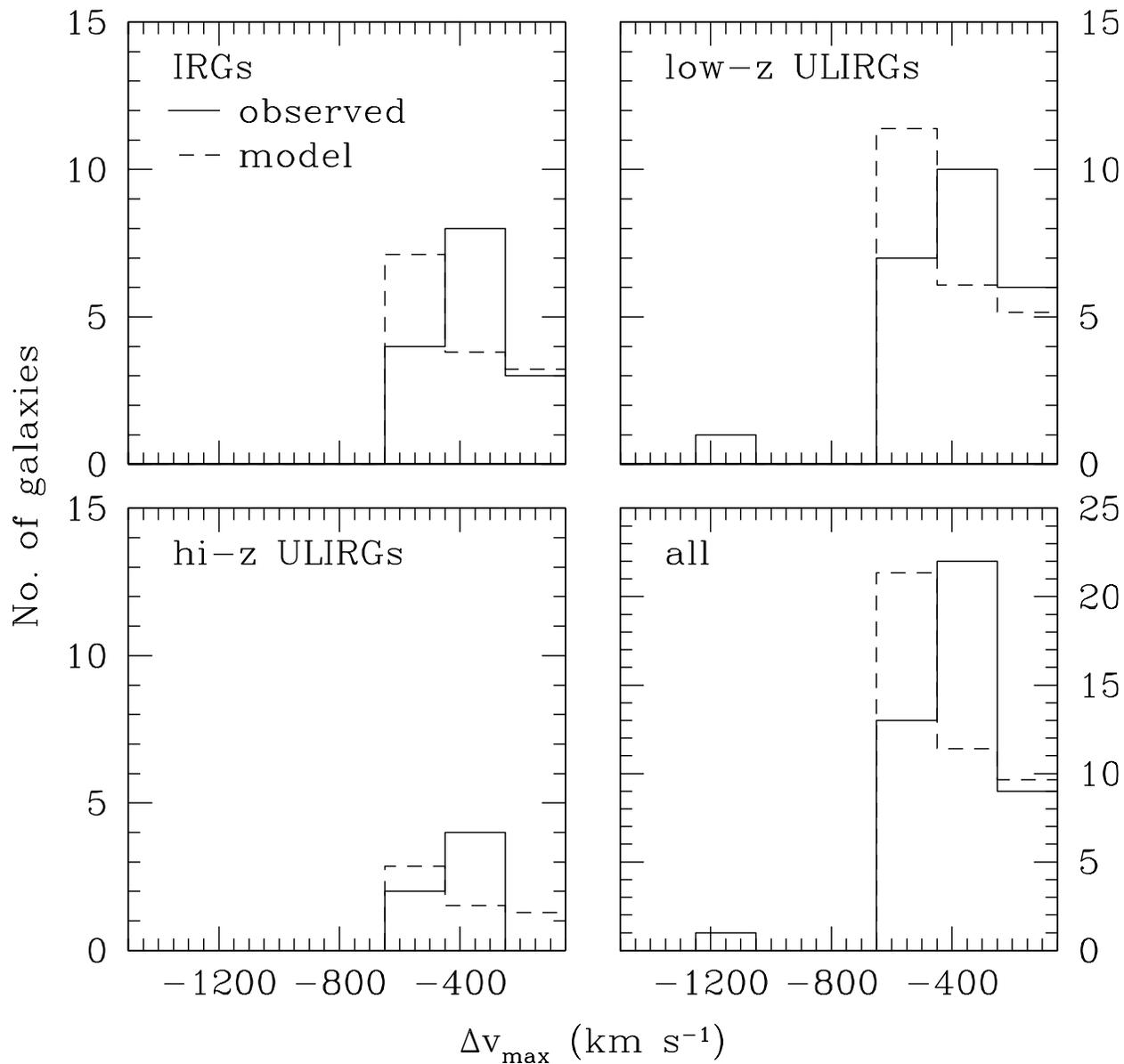}
\caption{Distribution of maximum velocities for each subsample compared to the predicted distribution from simple orientation effects (assuming a constant-velocity, 600~\kms\ wind emerging perpendicular to the galactic disk).  The observed distributions all differ from the predicted one at $>$99\% confidence.  See \S\ref{dv} for further discussion.}
\label{histdvpred}
\end{figure}

What about projection effects?  If the wind is not spherically symmetric and the wind velocities are not purely radial, the measured velocities will be smaller than the actual wind velocities due to the inclination of the wind with respect to the line of sight.  Consider a simple model of a constant-velocity wind emerging perpendicular to the disk of the galaxy, where the highest velocity we observe ($\sim$600~\kms) is the wind velocity.  Lower velocities could then be due to the inclination of the wind.  Assuming a uniform distribution of inclinations, we can compute the expected distribution of velocities for such a model.  In Figure \ref{histdvpred}, we compare the observed velocities with those predicted from this simple model.  The observed and predicted distributions differ at $>$99\% confidence in each case.  Obviously projection effects are more complex than this simple model.  If there really is a uniform distribution of inclinations, the real maximum velocities must be typically smaller than 600~\kms.

\subsection{Equivalent Width, Optical Depth, and Column Density} \label{nh}

\begin{figure}[t]
\plotone{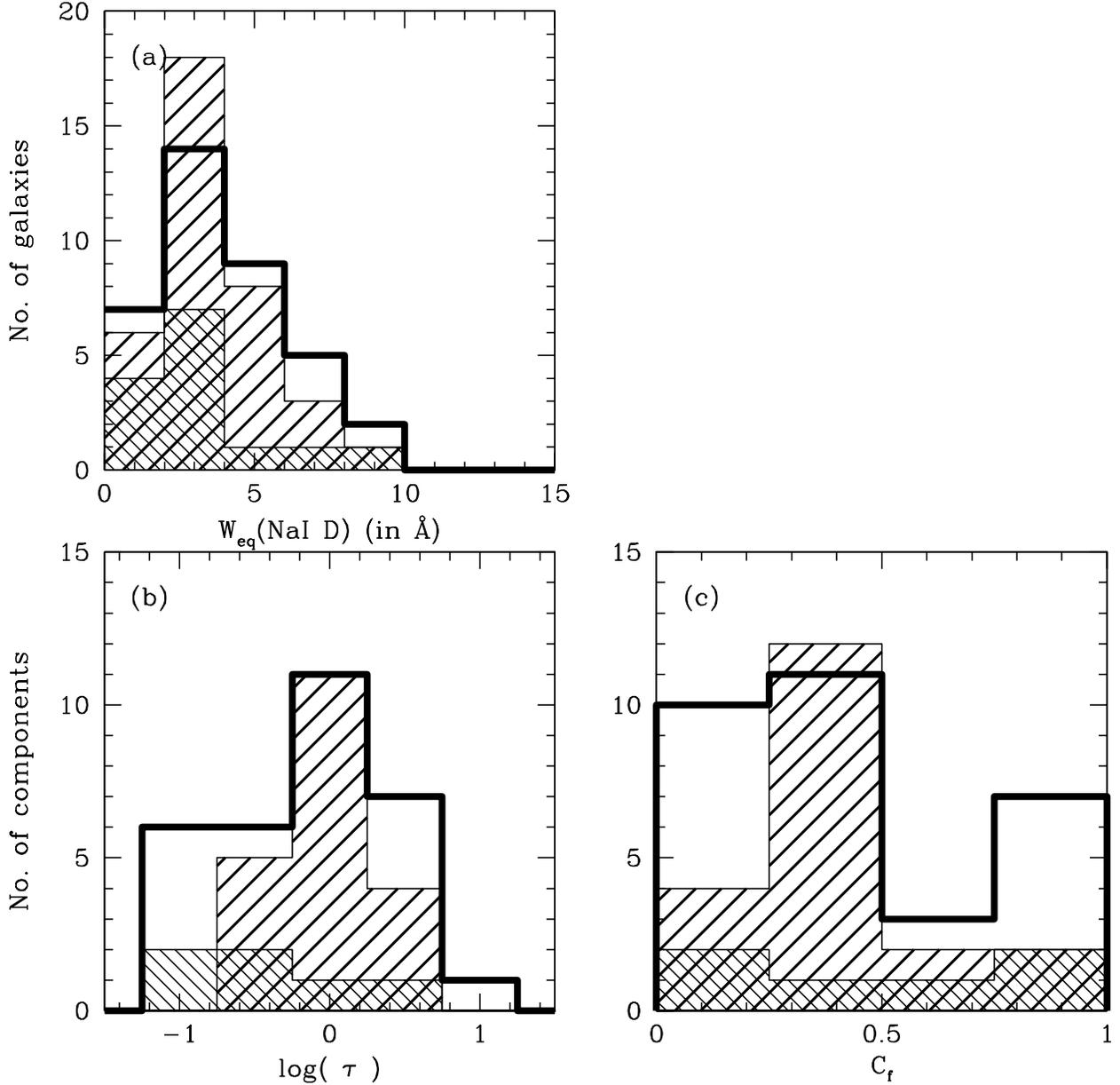}
\caption{Distributions of \nad\ parameters.  ($a$)~Distribution of rest-frame equivalent widths for all spectra.  ($b$)~Logarithmic distributions of the central optical depth in the D$_1$ line for outflowing components.  Notice the peak at $\tau \sim 1$; galaxies with lower limits on $\tau$ are typically assigned $\tau = 5$.  ($c$)~Logarithmic distributions of the covering fraction for outflowing components.  Note the single peak at \cf\ $\sim 0.25-0.50$ for the IRGs and the double peak at $\cf \sim 0.00-0.50$ and $0.75-1.00$ for the ULIRGs.  K-S tests show that the distributions of \cf\ and $\tau$ for different subsamples are not significantly different from one another, though Kuiper tests reveal differences in \cf\ at 95\%\ confidence.  The shading follows the pattern of Figure \ref{histdv}.  See \S\S\ref{nh} and \ref{cf} for more details.}
\label{histline}
\end{figure}

Figure \ref{histline}$a$ plots the distribution of the total rest-frame equivalent width of the \nad\ feature for every galaxy nucleus in our sample.  These values are based on our fits to the data.  Table 2 of Paper I lists these values explicitly.  We compute a median of 3.3 \AA, a dispersion of 2.1 \AA, and a maximum of 9.1 \AA.

Figure \ref{histline}$b$ plots the distribution of the central optical depth of the \nad$_1$ line (the D$_2$ line has twice the optical depth of the D$_1$ line).  The distributions for IRGs and ULIRGs are quite similar, and K-S and Kuiper tests show that they are consistent with being drawn from the same parent distribution.  We see that the peaks of the distributions occur at $\tau \sim 1$; in other words, most of the gas is moderately optically thick (if our assumptions about a constant covering fraction and a Maxwellian velocity distribution are correct; see Paper I).  The median error in an individual measurement of $\tau$ is $\sim$20\%.

\begin{figure}[t]
\plotone{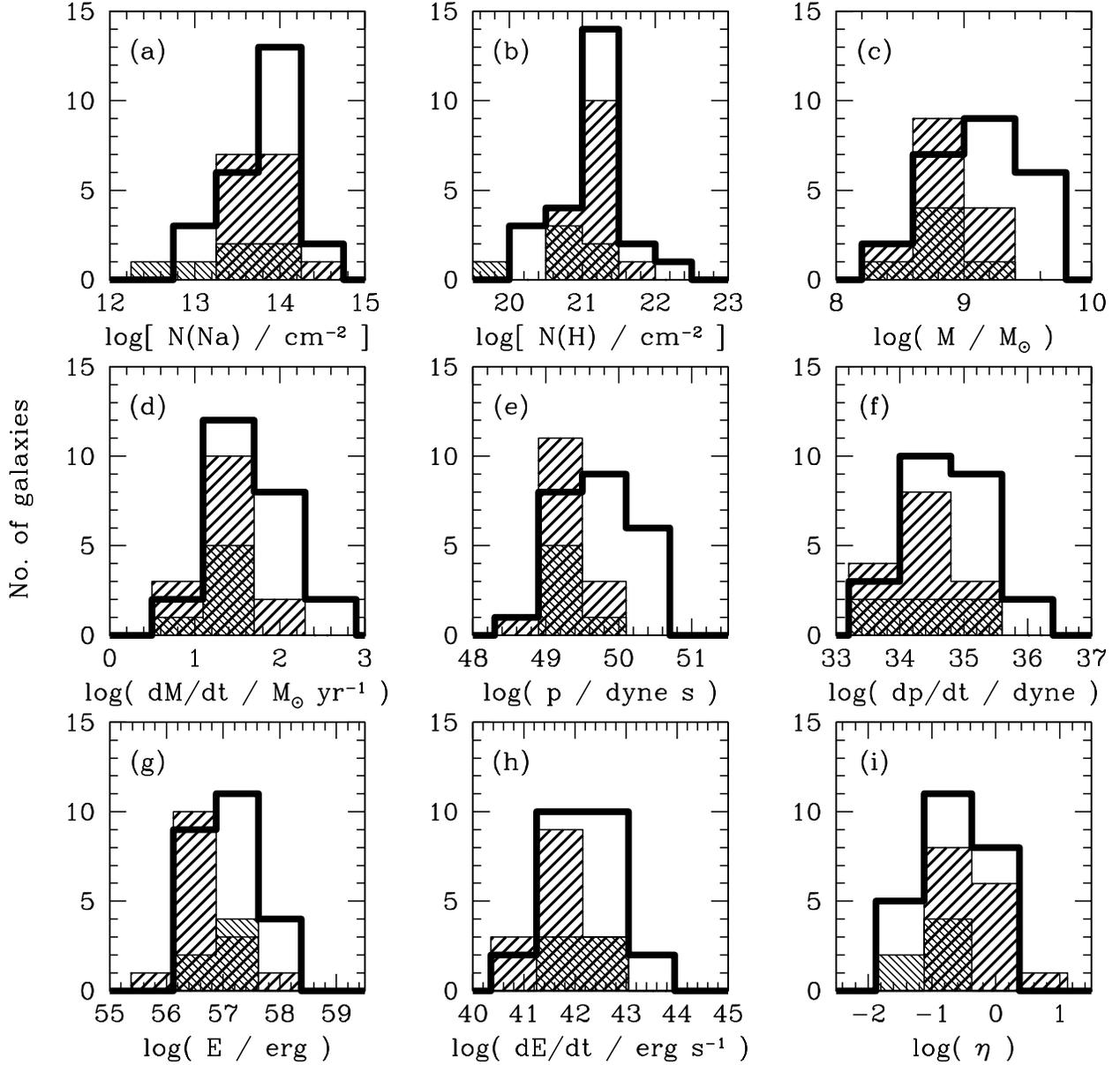}
\caption{Distributions of properties of the outflows in our sample.  The quantities are ($a$)~column density of \nags, ($b$)~column density of hydrogen, ($c$)~mass, ($d$)~mass outflow rate, ($e$)~momentum, ($f$)~momentum outflow rate, ($g$)~total kinetic energy, ($h$)~total kinetic energy outflow rate, and ($i$)~mass entrainment efficiency.  K-S and Kuiper tests show significant differences between the subsamples only for $\eta$ (panel $i$).  The shading is as in Figure \ref{histdv}.  See \S\S\ref{nh} and \ref{mme} for more details.}
\label{histof}
\end{figure}

The distributions of \nags\ and H column densities are shown in Figure \ref{histof}$a$ and $b$.  The distributions peak at log[$N$(\nags)]~$\sim 13.5-14$ and log[$N$(H)]~$\sim21$.  As a check on our results, we have one galaxy (NGC 1614) in common with \citet{sm04}.  Our \nags\ column density measurement matches theirs within 0.25 dex (despite their much higher spectral resolution).  Our measurement of the difference in velocity between the red and blue components in this galaxy ($\Delta v = 200$~\kms) also matches theirs.

\subsection{Covering Fraction} \label{cf}

In Figure \ref{histline}$c$, we show the distributions of covering fraction in each subsample.  Recall that \cf\ is assumed to be constant within each velocity component.  The mean value of \cf\ is $0.37-0.45$ for all three subsamples, and the error in an individual measurement is $0.05-0.1$.  For the IRGs, the \cf\ distribution peaks in the $0.25-0.5$ bin.  The distribution of \cf\ in ULIRGs has a broader peak in the range $0.0-0.5$, and a possible second peak in the $0.75-1.00$ bin.  K-S tests show that the differences between these distributions are not significant, though these tests are not as sensitive to data far from the mean (such as the hypothesized second peak at $\cf = 0.75-1.00$ for ULIRGs).  The Kuiper test, which is more evenly weighted, shows a significant difference between the IRGs and low-$z$ ULIRGs, at 95\%\ confidence.

The physical model behind \cf\ is described in Paper I.  In the superwind context, \cf\ may reflect (a)~the clumpiness of the wind and/or (b)~the global solid angle subtended by the wind with respect to the galaxy center.  A covering fraction less than unity may also be caused by light scattered into the line-of-sight, or by the wind having a small size as projected against the background stellar continuum.  If the latter is the case, we would expect \cf to vary significantly with redshift.  However, we observe only a negligible dependence on $z$ (Table \ref{avgprop}).  See \S\ref{gcf} for further discussion of \cf\ in the context of the global covering factor of the outflow.

\subsection{Spatial Distribution of Absorbing Gas} \label{spatialinfo}

We know from observations of local galaxies that superwind gas extends over a range of scales from sub-kiloparsec to 10$+$~kpc \citep[e.g.,][]{v_ea03}.  The hot wind outer radius in starburst galaxies apparently increases as galaxy size and mass increase \citep{s_ea04}.  In the large, massive galaxies we are studying, we expect winds to extend to radii of several kpc or greater.  For instance, recent integral field spectroscopy and deep {\it Chandra} imaging and spectroscopy of the nearest ULIRG, Arp 220, show that it contains a superwind with a velocity of 200~\kms\ at radii of a few kpc \citep{acc01,m_ea03}.  NGC 6240, a nearby high-luminosity LIRG, contains a superbubble with radius 5~kpc and a bow shock at 20~kpc \citep{v_ea03}.

As we discuss below (\S\ref{mme}), our mass, momentum, energy, and mass entrainment efficiency measurements are sensitive to the radial distribution of the gas.  The inner radius of the wind is constrained by the size of the starburst region, and this radius cannot be smaller than a few hundred pc \citep{ds98}.  Are there any constraints on either the inner or outer radius of the wind from our spectra, obtained by measuring the size of the absorbing region across the spatially extended background continuum light?

The continua in the spectra of most ULIRGs in our sample have angular sizes comparable to the seeing limit.  There are a few nearby exceptions.  F17207$-$0014 ($z = 0.043$) possesses blueshifted absorption across the entirety of the spatial profile, implying that this absorption occurs at projected radii of $\la$5~kpc.  {\it IRAS} 20046$-$0623 ($z = 0.084$) shows a velocity gradient in the \nad\ profile across $6\arcsec$, or $\sim$9~kpc.  The gradient in \nad\ matches the emission-line profile, which has a blueshifted tail westward of the western nucleus \citep[see also][]{m_ea01}.  The velocity of this tail matches that of the blueshifted absorption in the western nucleus, which extends over $\ga$2~kpc.  As we discuss below, F10190$+$1322:E ($z = 0.076$) has {\it redshifted} absorption that is extended on scales of $5-6$~kpc (\S \ref{overlap}) and arises in the disk of the western nucleus.  Finally, we observe blueshifted absorption in F10565$+$2448 ($z = 0.043$) that indicates projected radii of $\la$2~kpc.

Many more of the IRGs than the ULIRGs have spatially extended continua due to their lower redshifts.  Several show clear evidence for stellar rotation in their \nad\ profiles; typically these rotation components are narrow.  A few IRGs also show more complex structures.  The blueshifted \nad\ in F16504$+$0228 (NGC 6240; $z = 0.024$) is constant in velocity across most of the continuum, but it becomes more blueshifted in concert with the line emission to the east of the galaxy.  This suggests absorbing radii of up to 8~kpc in projection.  In F08354$+$2555 (NGC 2623; $z = 0.018$), the blueshifted absorption extends across 22$\arcsec$, which corresponds to 4~kpc in projected radius (though as we argue below, this is probably not outflowing gas; \S \ref{overlap}).

These values are consistent with our assumption of a `thin-shell' outflow located at radii of a few kpc (\S\ref{mme}).  However, they do not rule out a thicker shell extended over several kpc in radius.  Furthermore, the absorbing shell cannot be {\it too} thin, due to the broad profiles we observe (\S\ref{dv}).  The large global covering factors we measure (\S\ref{gcf}) also argue against narrow radial filaments, which would tend to have small global covering factors.

\subsection{Mass, Momentum, and Energy} \label{mme}

\subsubsection{Formulas}

We would like to measure the mass, momentum, and energy (and their outflow rates) in these winds for the phase of the ISM that we probe with our observations (roughly, the neutral gas phase).  We choose a simple model for the wind that depends on the physical parameters output from our fitting.  An alternative is to produce a database of synthetic profiles based on a more complex model; comparing these synthetic profiles directly to our spectra would then yield properties of the wind such as mass, density, etc.  Unfortunately, a complex model of this sort is under-constrained by our data.

As in our preliminary report \citep{rvs02}, we assume a spherically symmetric mass-conserving free wind, with an instantaneous mass outflow rate and velocity that are independent of radius within the wind and zero outside.  The mass outflow rate at a radius $r$ within the wind is given by
\begin{equation} \label{mdot(r)}
dM/dt(r) = \Omega \mu m_p n(r) v r^2,
\end{equation}
where $\Omega$ is the solid angle subtended by the wind as seen from its origin (i.e., the wind's global covering factor), $\mu m_p$ is the mass per particle ($m_p$ being the proton mass and $\mu$ a correction for relative He abundance), $n(r)$ is the wind number density, and $v$ is the wind velocity.  We wish to re-write this equation in terms of the (measurable) column density $N$ rather than the space density $n$.  Rearranging to solve for $n(r)$, we require a density profile $n(r) \propto r^{-2}$ (e.g., an isothermal sphere).  

For a wind of finite thickness (inner and outer radii $r_1$ and $r_2$), the total observed column density in the outflow is the integral of $n(r)$ along our line-of-sight:
\begin{equation} \label{Nint}
N = \int_{r_1}^{r_2} n(r) dr.
\end{equation}
Solving eq. (\ref{mdot(r)}) for $n(r)$, substituting this into eq. (\ref{Nint}), integrating, and rearranging leaves us with the instantaneous mass outflow rate across any radius $r$ within the wind,
\begin{equation} \label{dmdtinstthick}
dM/dt_{thick}^{inst} = \sum_i \Omega_i \mu m_p N_i v_i \frac{r_1 r_2}{r_2 - r_1}.
\end{equation}
The sum is taken over the number of outflowing components in the galaxy: $\Omega_i$ is the solid angle subtended by the component as seen from the wind's origin, $N_i$ is the column density of each component, and $v_i$ is the central wind velocity in each component.  The corresponding mass of the wind, obtained by simple volume integration of the density profile, is
\begin{equation} \label{mthick}
M_{thick} = \sum_i \Omega_i \mu m_p N_i r_1 r_2.
\end{equation}
To average the mass outflow rate across a radius $r < r_1$ over the wind lifetime, we divide $M_{thick}$ by $t_{wind} = r_2/v_i$ to give
\begin{equation} \label{dmdtavgthick}
dM/dt_{thick}^{avg} = \sum_i \Omega_i \mu m_p N_i v_i r_1.
\end{equation}

Suppose we assume instead a thin shell, with $r_1 \sim r_2 = r$ and $\Delta r = r_2 - r_1$.  Equations (\ref{dmdtinstthick})$-$(\ref{dmdtavgthick}) then become
\begin{eqnarray}
             & dM/dt_{thin}^{inst} & = \sum_i \Omega_i \mu m_p N_i v_i \frac{r^2}{\Delta r}, \\
             & dM/dt_{thin}^{avg}  & = \sum_i \Omega_i \mu m_p N_i v_i r, \\
\mathrm{and} & M_{thin}            & = \sum_i \Omega_i \mu m_p N_i r^2.
\end{eqnarray}

The energies and momenta and their outflow rates are computed using similar sums:
\begin{eqnarray}
             & dp/dt  & = \sum_i (dM/dt)_i v_i, \\
             & p      & = \sum_i M_i v_i, \\
             & dE/dt  & = \sum_i (dM/dt)_i \times (v_i^2 / 2 + 3 \sigma_i^2 / 2),\\
\mathrm{and} & E      & = \sum_i M_i (v_i^2 / 2 + 3 \sigma_i^2 / 2),
\end{eqnarray}
where the energy includes both the `bulk' kinetic energy due to the outflowing gas and 'turbulent' kinetic energy (where we assume the same $\sigma$ in each dimension).

\subsubsection{Wind Geometry, Cloud Lifetimes, and Time-Averaging} \label{windgeom}

The masses, momenta, and energies we compute are sensitive to the assumed geometry.  In \citet{rvs02}, we assumed a thick wind that extends from an inner radius $r_\star$ to infinity.  Here, we will revert to a thin shell of uniform radius 5~kpc.  This radius is motivated both by observations of local starbursts \citep[e.g.,][]{s_ea04,vcb05} and by our own data (\S\ref{spatialinfo}).

The `thin shell' geometry is, however, uncertain.  Theory and simulations predict a thin shell of warm gas behind the forward shock of the wind \citep[e.g.,][]{cmw75,s_ea94,ss00}.  This gas may or may not be accompanied by neutral material (see the discussion in \S7 of Paper I), but the shell is a natural interpretation of our simple model.  The shell may be partly or completely broken up due to Rayleigh-Taylor instabilities \citep{s_ea94}.  Observations of local starbursts also find dusty filaments (which presumably contain neutral gas) entrained on the edges of superwind bicones \citep[see][and references therein]{vcb05}.  Our model does not account for these filaments, which would have a broader radial range and subtend a much smaller angle.

Clouds may also exist in the wind interior.  These clouds can be ablated and destroyed by thermal evaporation (conduction) or hydrodynamic ablation.  Simple formulas for thermal evaporation in the absence of magnetic fields \citep{cmo81} and cloud destruction by hydrodynamic ablation \citep{kmc94,pfb02} show that small clouds of size 1~pc and density 100~cm$^{-2}$ can survive for a time of order $10^6$~yr in a wind fluid of density 10$^{-3}$~cm$^{-2}$ and temperature $3\times10^7$~K.  This is not much smaller than the wind lifetime ($5\times10^6-5\times10^7$~yr) of a 5~kpc shell with a constant velocity of $100-1000$~\kms.  Various effects can increase cloud lifetimes, including: (1) the presence of azimuthal magnetic fields, which make cloud evaporation less likely, as they decrease conduction relative to the classical Spitzer value (as observed in 30 Dor; \citealt{sw04}); (2) conversely, increased heat conduction, which may inhibit hydrodynamic instabilities \citep{m_ea05}; (3) clouds stop evaporating when they are crushed to high optical depth \citep{fs93}; and (4) Kelvin-Helmholtz instabilities are suppressed by ablation of small bumps in the cloud \citep{sck95}.  Cloud lifetimes can be decreased, however, by interactions among nearby clouds \citep{p_ea04}.

In short, the survival times of clouds in the wind are not yet certain, but it is entirely possible that they may survive for a substantial fraction of the wind lifetime and be observable.  If they do exist throughout the wind's interior, then a thick wind model is more applicable.  Assuming a thick wind instead of a thin wind will decrease our measured masses, momenta, energies, and outflow rates, since the radial factor in the above equations is the inner radius in the thick wind case, rather than the outer radius.  For instance, a thick wind extending from $1-5$~kpc has a mass 5 times lower than the $r=5$~kpc thin shell that we assume.  However, there are limits to the inner radius of the wind based on the size of the starburst \citep[e.g.,][]{ds98}, such that our masses, etc., will not decrease by more than a factor of $\sim$10.  More complicated geometries (e.g., irregular, filamentary structures) will alter our results in ways that we cannot easily predict, though we do not expect dramatic departures from our predictions.

A second geometric consideration is the solid angle subtended by the wind, as seen from the wind's origin (i.e., its global covering factor).  Local winds emerging from disk galaxies are typically biconical along the galaxy's minor axis.  We account for this by letting the solid angle subtended by the wind $\Omega$ be less than $4\pi$.  Furthermore, the wind may be clumpy, rather than a smooth shell.  In our model, we divide $\Omega$ into two parts, the large-scale covering factor (related to the wind's opening angle), given by \co, and the local covering factor (related to the wind's clumpiness), given by \cf.  Thus, $\Omega/4\pi = \co~\cf$.  For the IRGs, we assume $\co = 0.4$, where 0.4 is based on the average opening angle of winds in local starbursts ($\sim$65$\degr$; e.g., \citealt{vcb05}).  For ULIRGs, we use a larger value of $\co = 0.8$, which is motivated by our detection rate (\S\ref{gcf}).  We assume that the measured covering fraction \cf\ describes the wind's local clumpiness.

Furthermore, for the outflow rates, we must specify whether we wish to compute the instantaneous values in the wind or the values averaged over the lifetime of the wind.  In \citet{rvs02}, we quoted instantaneous mass outflow rates.  In this paper, we will quote time-averaged outflow rates, which are a more useful quantity when comparing to the star formation rate.  We address the relative wind and starburst lifetimes in \S\ref{eta}.

\subsubsection{Results} \label{eqns}

A final numerical assumption is an average particle mass of 1.4$m_p$ to account for the contribution of He.  The resulting numerical formulas are
\begin{eqnarray}
             & M     & = 5.6\times10^8 \sum \left(\frac{\co}{0.4}\cf\right)\left(\frac{r^2}{100~\mathrm{kpc}^2}\right)\left(\frac{N(\mathrm{H})}{10^{21}~\mathrm{cm}^{-2}}\right)\msun, \\
             & dM/dt & = 11.5 \sum \left(\frac{\co}{0.4}\cf\right)\left(\frac{r}{10~\mathrm{kpc}}\right)\left(\frac{N(\mathrm{H})}{10^{21}~\mathrm{cm}^{-2}}\right)\left(\frac{|\Delta v|}{200~\kms}\right)\smpy, \\
             & p     & = 2.2\times10^{49} \sum \left(\frac{\co}{0.4}\cf\right)\left(\frac{r^2}{100~\mathrm{kpc}^2}\right)\left(\frac{N(\mathrm{H})}{10^{21}~\mathrm{cm}^{-2}}\right)\left(\frac{|\Delta v|}{200~\kms}\right)\mathrm{dyne~s}, \\
             & dp/dt & = 1.4\times10^{34} \sum \left(\frac{\co}{0.4}\cf\right)\left(\frac{r}{10~\mathrm{kpc}}\right)\left(\frac{N(\mathrm{H})}{10^{21}~\mathrm{cm}^{-2}}\right)\left(\frac{|\Delta v|}{200~\kms}\right)^2\mathrm{dyne}, \\
             & E     & = 2.2\times10^{56} \sum \left(\frac{\co}{0.4}\cf\right)\left(\frac{r^2}{100~\mathrm{kpc}^2}\right)\left(\frac{N(\mathrm{H})}{10^{21}~\mathrm{cm}^{-2}}\right) \\
             &       & \;\;\;\;\;\;\;\;\;\;\;\times\left[\left(\frac{|\Delta v|}{200~\kms}\right)^2 + 1.5\left(\frac{b}{200~\kms}\right)^2\right]\mathrm{erg}, \nonumber \\
\mathrm{and} & dE/dt & = 1.4\times10^{41} \sum \left(\frac{\co}{0.4}\cf\right)\left(\frac{r}{10~\mathrm{kpc}}\right)\left(\frac{N(\mathrm{H})}{10^{21}~\mathrm{cm}^{-2}}\right)\left(\frac{|\Delta v|}{200~\kms}\right) \\
             &       & \;\;\;\;\;\;\;\;\;\;\;\times\left[\left(\frac{|\Delta v|}{200~\kms}\right)^2 + 1.5\left(\frac{b}{200~\kms}\right)^2\right]\mathrm{erg~s^{-1}}. \nonumber
\end{eqnarray}

The sums here are performed over the outflowing velocity components in each galaxy.  We remind the reader that we assume $r = 5$~kpc universally, but different values of \co\ for the LIRGs (0.4) and the ULIRGs (0.8).  We record these quantities for each galaxy in Table \ref{objprop}, their median values and dispersions in Table \ref{avgprop}, and their distributions in Figures \ref{histof}$c-h$.  K-S and Kuiper tests show that the distributions for each subsample are consistent with being drawn from the same parent distribution.

\subsection{Mass Entrainment Efficiency} \label{eta}

The mass entrainment efficiency compares the amount of gas entrained by the hot wind fluid to the amount of gas being turned into stars.  Star formation in these galaxies occurs at the galactic nucleus in dense concentrations of molecular gas.  However, most of the gas in a superwind is entrained from the disk and halo of the host galaxy as the wind propagates outward, either by acceleration of cold clouds or mass-loading of these clouds into the wind fluid \citep{s_ea94,ss00}.  The mass entrainment efficiency is thus a useful quantitative description of how the wind's evolution is connected to its power source (the starburst).

The star formation rates for our sample are computed from the usual Kennicutt formula relating SFR to the total infrared luminosity \citep{k98},
\begin{equation}
\mathrm{SFR}= \alpha~\frac{\lir}{5.8\times10^{9}~\lir}.
\end{equation}
The model for this scaling assumes continuous star formation, with a starburst age of $10-100$~Myr and a Salpeter initial mass function (IMF) with stellar mass limits 1 and 100~\msun.  We correct for AGN contribution to \lir; $\alpha$ equals the fraction of the infrared luminosity powered by star formation.  {\it Infrared Space Observatory} ({\it ISO}) observations show that $70\%-95\%$ of the infrared luminosity of a typical ULIRG is dust-reprocessed light from young stars \citep{g_ea98}.  These values apply to our sample, which we have selected to include objects whose spectra indicate vigorous starbursts.  We therefore assume $\alpha=0.8$ for our ULIRGs.  For the IRGs, we use $\alpha = 1.0$.

The mass entrainment efficiency is the mass outflow rate normalized to the corresponding star formation rate of a galaxy:
\begin{equation}
\eta \equiv \frac{dM/dt}{\mathrm{SFR}}.
\end{equation}
Figure \ref{histof}$i$ shows the distributions of $\eta$ for each subsample.  We measure median values of 0.33, 0.19, and 0.09 in our IRG, low-$z$ ULIRG, and high-$z$ ULIRG subsamples, respectively.  The full range of $\eta$ is three orders of magnitude, $0.01 - 10$.  The distribution of $\eta$ for the IRGs is different from that of the ULIRGs at $>$99\% confidence.  As with $dM/dt$, $\eta$ is sensitive to the assumed wind geometry and to our assumptions about the local physical state of the gas.

Note that $dM/dt$ and SFR are time-averages over the wind and starburst lifetimes, respectively.  The superwinds in these galaxies are powered initially by stellar winds, and by supernovae at times $\ga$10$^7$~yr \citep{lrd92}.  The wind lifetimes we calculate ($t_{wind} = r_2 / v$) are of order $5-50$~Myr; thus, the winds are generally supernovae-powered.  The starburst lifetimes ($\la$100~Myr) are comparable to the wind lifetimes, as the gas consumption timescales for the molecular gas in the nuclei of ULIRGs indicate.  (ULIRGs have molecular gas masses $\sim$10$^{10}$~\msun\ [\citealt{sss91,s_ea97}] and form stars at rates $\ga$150~\smpy.)  Thus, it is appropriate to directly compare our computed $dM/dt$ and SFR.  In this model, $\eta$ is also a time-average over the lifetime of the wind.

\section{OUTFLOW PROPERTIES AND HOST GALAXY PROPERTIES} \label{ofprop_v}

One of the primary purposes of this survey is to look for dependence of outflow properties on the properties of the host galaxies.  This allows us to (1) better understand the physics of outflows and (2) describe their properties using approximate analytic functions.  The latter is especially useful as input to theoretical analysis and simulations.  The galaxy properties that we use in this analysis are discussed in Paper I.  They include star formation, optical and near-infrared luminosity, circular velocity (mass), and spectral type.  Ideally, it would be useful to also look for a dependence on star formation rate surface density.  However, since most of our objects are not well-resolved, we cannot do this reliably.

In determining the presence of a correlation between two quantities, we use three tests: Pearson's correlation coefficient, Spearman's correlation coefficient, and a weighted least-squares fit.  We accept as significant those correlations for which the slope in the fit ($a \pm\delta a$, where log $X$ $=a$ log $Y+Y_0$) satisfies $a > (3\times\delta a)$ and for which the probability of no correlation using Pearson's $r$ is less than $5\%$ (in all cases for which it is $<$0.05, it is also $<$0.01).  The probability for a null result using Spearman's $r$ is also typically (though not always) low in these cases.  In Table \ref{fittab}, we list the power-law slopes  of the computed fits and correlation coefficients for each significant correlation.

In this analysis, we include a few LIRGs from \citet{hlsa00}, ULIRGs from \citet{m05}, and dwarf starbursts from \citet{sm04}.  We only use velocities from \citet{hlsa00}, since the fits to individual galaxies do not account for covering fraction, and we only include galaxies with LINER or \ion{H}{2} spectral types (4 galaxies total).  From the \citet{m05} data, we only use galaxies with measured spectral types of LINER or \ion{H}{2} and \nad\ doublet ratio ratios $R > 1.1$ which are not already in our sample (6 galaxies total; see next paragraph for justification of the doublet ratio criterion).  We use our equations to compute the properties of these galaxies starting with $\Delta v$, $\tau$, $b$, and \cf\ for each galaxy.  We do the same for the four dwarf starbursts with detectable winds in \citet{sm04}.  For the \citet{m05} ULIRGs, we assume $\co = 0.8$ and $r = 5$~kpc as we do for our ULIRGs.  For the dwarf starbursts, we assume $\co = 0.4$ (as for the LIRGs) and use the measured shell radii listed in \citet{sm04}, with the radius changing from object to object and ranging from 0.1$-$1~kpc.  To estimate errors where none are given in these references, we use median values from our own observations at similar resolution.

We note as a caveat that two of the dwarf galaxies have \nad\ doublet ratios $R\sim1.1$, and thus the lines are optically thick \citep{sm04}.  The technique used by \citet{sm04} to recover the column densities assumes Gaussians in intensity and thus does not self-consistently treat cases of high $\tau$.  The resulting column densities are thus overestimates to the actual values under the assumption of a Maxwellian velocity distribution (see Paper I for extensive discussion of this issue).

\subsection{Detection Rate} \label{detrate_v}

\begin{figure}[t]
\plotone{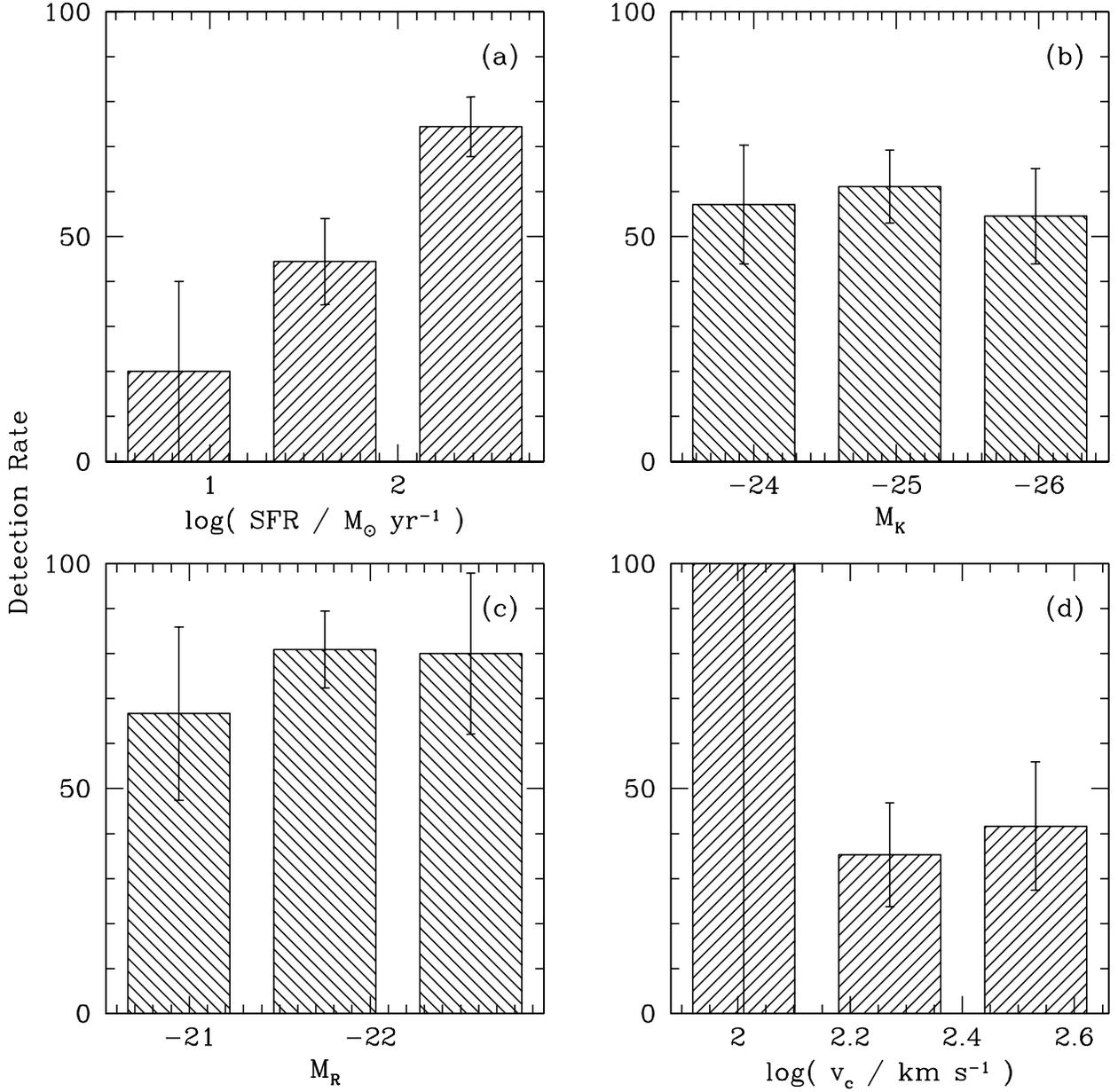}
\caption{The detection rate of winds as a function of ($a$)~star formation rate, ($b$)~$K$- or $K^\prime$-band magnitude, ($c$)~$R$-band magnitude, and ($d$)~circular velocity.  The detection rate increases as SFR increases, though at the highest values of SFR it begins to decrease.  Within the errors, it is independent of $M_{K^{(\prime)}}$, $M_R$, and $v_c$.  The error bars are 1$\sigma$, assuming a binomial distribution.  See \S\ref{detrate_v} for further discussion.}
\label{histfreq}
\end{figure}

Figure \ref{histfreq} shows the dependence of detection rate on star formation rate, $K$- or $K^\prime$-band magnitude, $R$-band magnitude, and circular velocity.  We arrive at the bins in this figure by dividing the full range of values into three bins of equal size.  The results show that detection rate increases as SFR increases.  (This conclusion is bolstered by dividing the IRGs by IR luminosity; see below.)  There is only weak or no dependence of detection rate on $M_{K^{(\prime)}}$, $M_R$, or $v_c$.  Note that, although Figure \ref{histfreq} does not show it, there is a decrease in the detection rate at the highest SFR, as seen by comparing the low-$z$ and high-$z$ ULIRGs (Table \ref{avgprop}).  We discuss this further in \S\ref{zevol}.

In \S\ref{gcf}, we argue that the detection rate likely reflects the wind opening angle rather than the frequency of occurrence of winds (which we suggest is close to unity in infrared-luminous galaxies).  The increase of detection rate with SFR is probably a result of the different wind geometry in ULIRGs, which have high SFR.  The observed difference in wind geometry could in turn reflect the nature of ULIRGs as merging galaxies.  The fact that we observe no dependence of detection rate on galactic mass implies that other physics besides the gravitational potential determine whether or not an outflow is formed.

The detection rate does {\it not} depend on spectral type.  In each subsample, the detection rate of winds in LINERs is statistically indistinguishable from that in \ion{H}{2} nuclei.  Overall, the detection rate is identical in LINERs and \ion{H}{2} nuclei: $50\%$. (We find winds in 20 of 40 LINER nuclei and 23 of 43 \ion{H}{2} nuclei.)

For the 1~Jy galaxies, {\it interaction classes} are available \citep{vks02}.  Most or all of these ULIRGs are involved in a major merger, and these classes describe what stage of the interaction that each galaxy is in.  We find that galaxies which are in the latest stages of interaction (classes IV = `merger' and V = `old merger') have the same detection rate: $(82\pm9)\%$ for class IV galaxies, and $(86\pm13)\%$ for class V galaxies.  Class III (= `pre-merger') galaxies, which are at an earlier interaction stage and have at least two identifiable nuclei, show the same detection rate within the errors: $(67\pm19)\%$.

The IRGs do not have systematic classifications attached to them.  However, from examination of DSS2 and 2MASS images, there is evidence of a major interaction or minor merger in many of them, as seen in the presence of tidal tails, very irregular stellar morphology, and/or large companion galaxies that appear to be interacting with the host.  (Many of those galaxies for which we do not observe evidence of an interaction are too distant for the images to show interesting features.)  Of the IRGs which appear to be undergoing or to have recently undergone a major interaction or minor merger, we detect a wind in 5 of 11, or $45\%$, of them.  This percentage is identical to the overall detection rate of $43\%$ in the IRGs.

Alternatively, \citet{i04} finds that all galaxies with $\lir > 10^{11.5} \lsun$ are undergoing a major interaction.  If we divide our IRG subsample by luminosity, then 8/13 galaxies (62$\pm$13\%) with $\lir > 10^{11.5} \lsun$ host winds, vs. 7/22 galaxies (32$\pm$10\%) with $\lir < 10^{11.5} \lsun$.  This could indicate some dependence of detection rate on the presence of a major interaction; however, without detailed morphological studies of the galaxies in our sample, this result is inconclusive.

In short, we conclude that there is no strong dependence of detection rate on merger stage.  The change in detection rate with SFR, which reflects ouflow geometry, could ultimately be due to the presence/absence of an interaction or strength of interaction, although we do not have firm evidence to demonstrate this.

\subsection{Outflow Velocity} \label{dv_v}

\begin{figure}[t]
\plotone{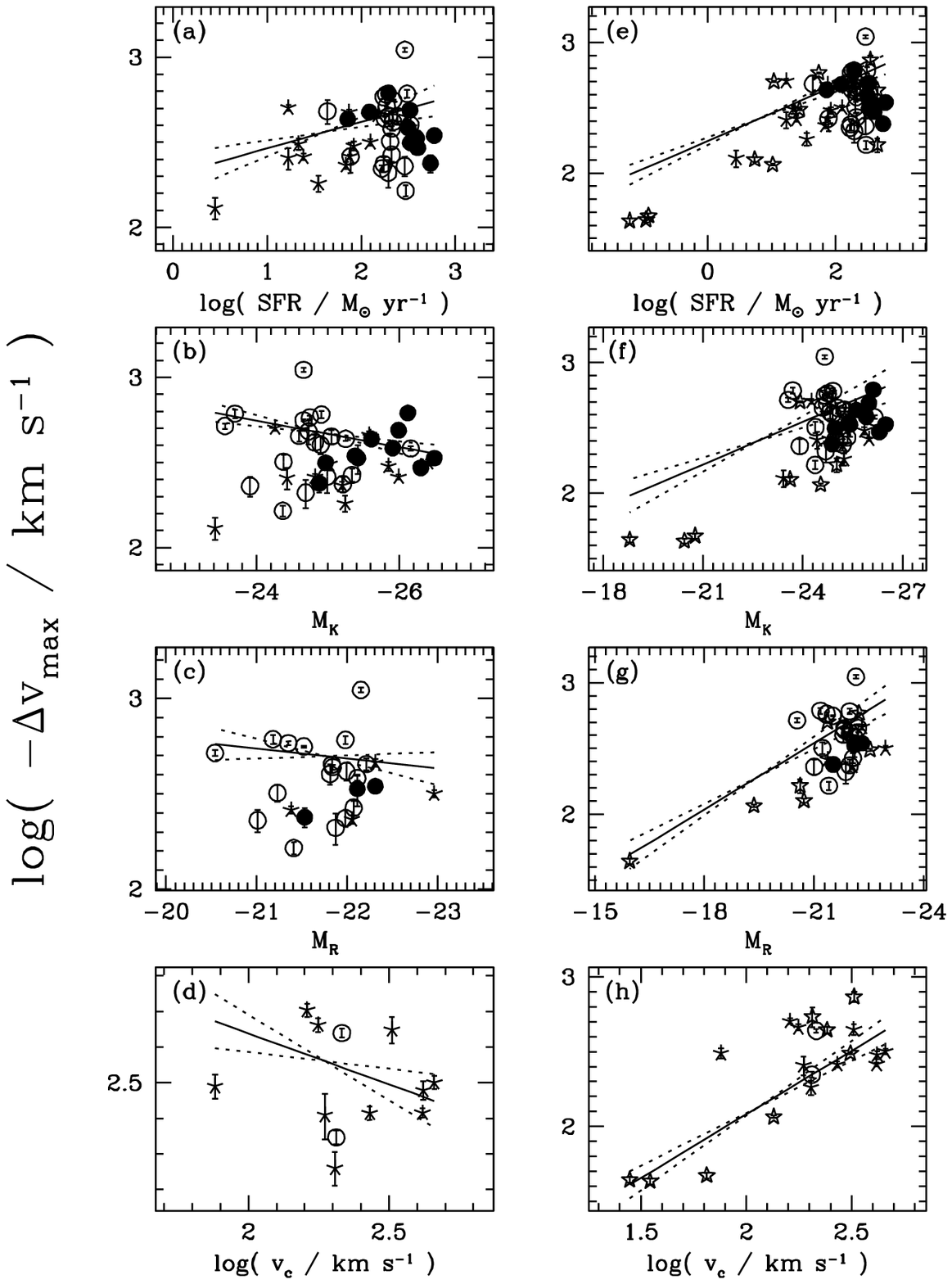}
\caption{\footnotesize{Maximum velocity ($\dvmax \equiv \Delta v - \mathrm{FWHM}/2$) vs. star formation rate, $K$- or $K^\prime$-band absolute magnitude, $R$-band magnitude, and circular velocity.  Panels $(a)-(d)$ show just our data.  Panels $(e)-(h)$ show our data plus four other LIRGs \citep{hlsa00}, four dwarf starbursts \citep{sm04}, and six other ULIRGs \citep{m05}. IRGs are shown as skeletal stars, low-$z$ ULIRGs as open circles, hi-$z$ ULIRGs as filled circles, and data from \citet{hlsa00}, \citet{sm04}, and \citet{m05} as open stars.  (In color, IRGs are red filled circles, low-$z$ ULIRGs are black, and high-$z$ ULIRGs are blue; data from other sources are black open stars.)  Error bars are 1$\sigma$.  The solid lines are weighted least-squares fits to the data; the dotted lines indicate 1$\sigma$ errors in the slope.  Fits to just our data are consistent with \dvmax\ being independent of each quantity on the horizontal axes.  However, if we add the dwarf galaxies from \citet{sm04}, correlation tests and weighted least squares fits show significant relationships between \dvmax\ and each quantity (e.g., $|\dvmax| \propto$ SFR$^{-0.2}$).  See \S\ref{dv_v} for further discussion.}}
\label{dvmax_v}
\end{figure}

In Figure \ref{dvmax_v}, we plot for each galaxy the logarithm of the maximum outflow velocity \dvmax\ as a function of SFR, $M_{K^{(\prime)}}$, $M_R$, and $v_c$.  These figures show that \dvmax\ is not significantly correlated with any of these quantities in our data.  However, if we add four dwarf galaxies \citep{sm04}, increasing the range of galaxy properties probed (e.g., the range in SFR increases from 2 to 4 orders of magnitude), we measure significant correlations of maximum velocity with each galaxy property.  The strongest correlation we find is with $v_c$: $|\dvmax| \propto v_c^{0.8\pm0.2}$.  The other dependences we compute are weak ($|a| \sim 0.1-0.2$).

\begin{figure}[t]
\plotone{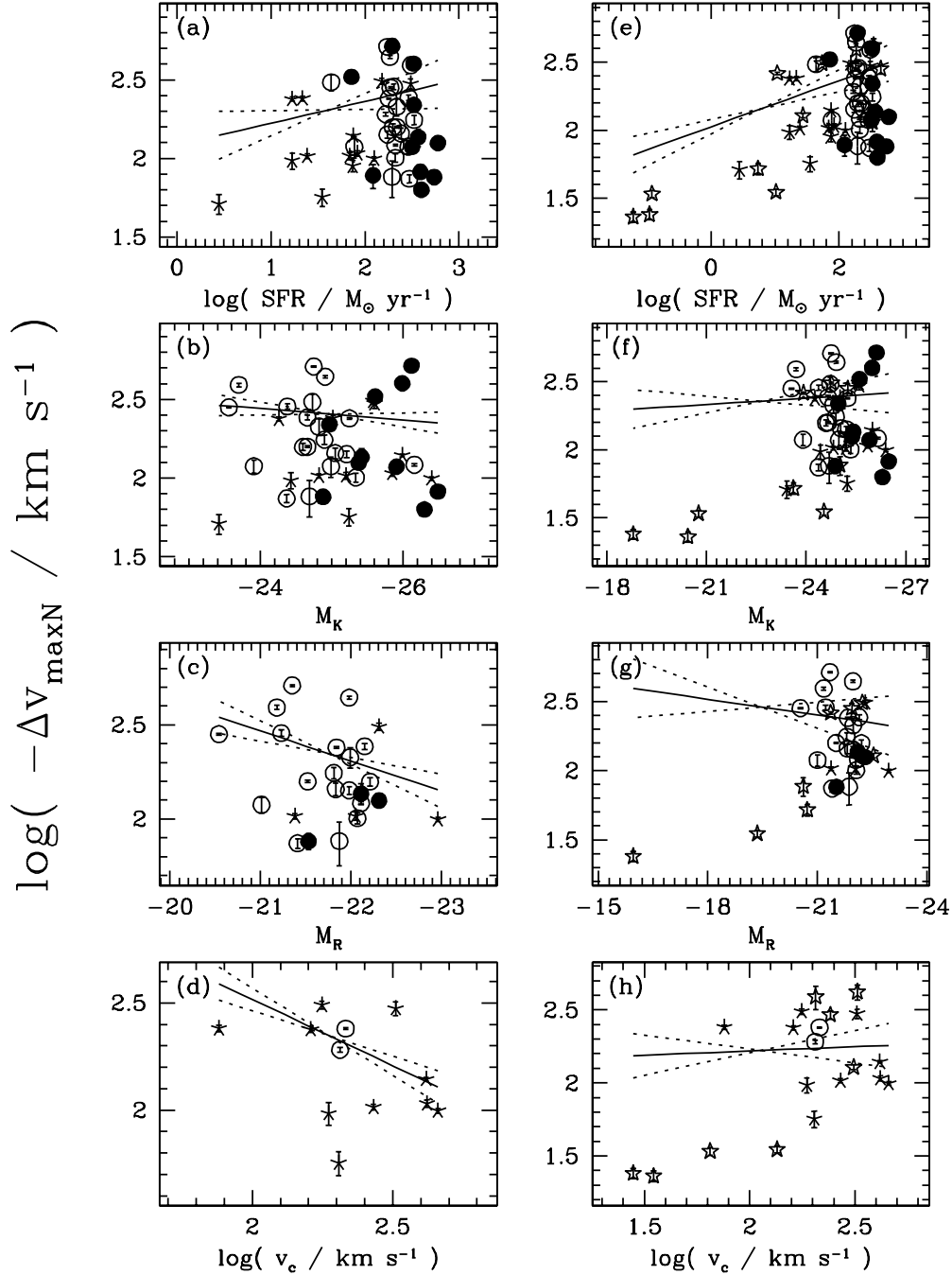}
\caption{Same as Figure \ref{dvmax_v}, but for velocity of the highest column density gas.  The symbols and lines are as in Figure \ref{dvmax_v}.  Unlike the previous plot, only our correlation tests (and not our weighted least squares fits) show a significant relationship between \dvtau\ and each quanitity when the full range of data is included.  Allowed velocities increase slowly but smoothly with SFR, luminosity, and mass, but at some characterisitic value of these galaxy properties there is a sharp increase in the allowed velocities of the optically thickest gas.  The lower right-hand corners of panels $(e)-(h)$ in this figure are not populated, but this is a selection effect; we do not include points with $\dvtau>-50$~\kms\ in our analysis because of measurement uncertainties.  See \S\ref{dv_v} for further discussion.}
\label{dvtau_v}
\end{figure}

In Figure \ref{dvtau_v}, we do the same for the velocity of the highest column density outflowing gas in each galaxy.  Again, no correlations emerge in our data.  If we add the dwarf galaxies, interesting behavior emerges.  Allowed velocities increase slowly but smoothly with SFR, luminosity, and mass, but at some characterisitic value of these galaxy properties there is a sharp increase in the allowed velocities of the optically thickest gas.  The lower right-hand corners of panels $e-h$ in this figure are not populated, but this is a selection effect; we do not include points with $\dvtau>-50$~\kms\ in our analysis because of measurement uncertainties.

In the Discussion (\S\ref{theory}), we discuss the interpretation of these correlations of velocity with galaxy properties.

Some authors have suggested that the shock-like line ratios found in the nuclei of galaxies classified as LINERs are due to the presence of shocks in outflowing gas \citep{v_ea95,lvg99,t_ea99}.  We can test this hypothesis by comparing the velocities in LINERs and \ion{H}{2} galaxies.

\begin{figure}[t]
\plotone{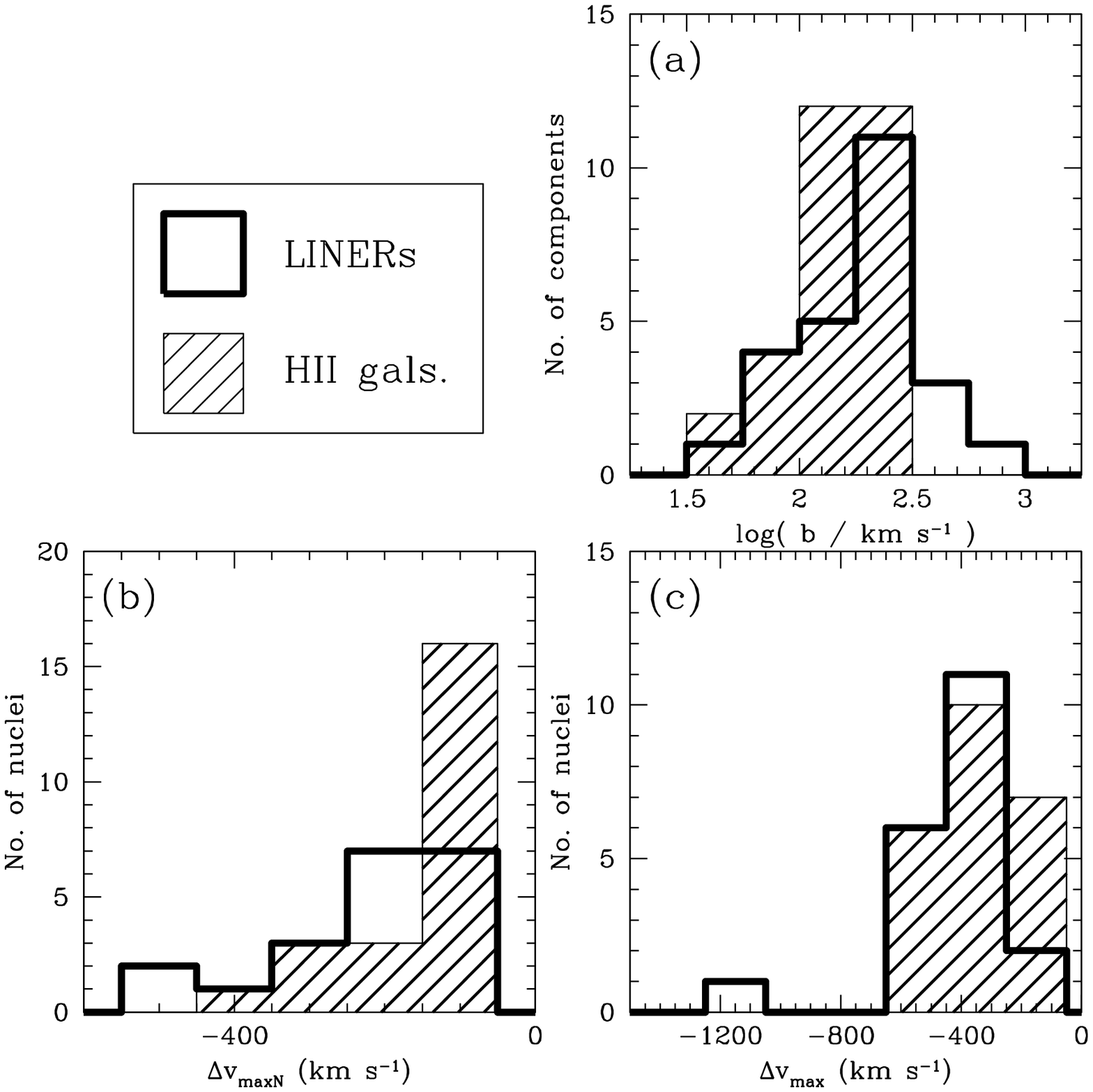}
\caption{Distributions of (a) Doppler parameter, (b) velocity of the highest column density gas, and (c) maximum velocity for LINERs and \ion{H}{2} galaxies.  The starbursting LINERs have higher median $|\dvmax|$ and $|\dvtau|$ than \ion{H}{2} galaxies by 100~\kms, while $b$ is larger by 70~\kms.  The distributions of $b$ and \dvmax\ differ at $>$95\% confidence according to both the K-S and Kuiper tests.  For \dvtau, the distributions are not significantly different.  See \S\ref{dv_v} for further discussion.}
\label{histdvlh}
\end{figure}

We find that the median maximum velocity $|\dvmax|$ in LINER nuclei is higher than in \ion{H}{2} nuclei by 125~\kms\ (393 vs. 267~\kms, respectively).  The median velocity of the highest column density gas, $|\dvtau|$, is higher in LINERs by 100~\kms\ (229 vs. 119~\kms, respectively).  Finally, the median velocity width $b$ is higher in LINERs by 70~\kms\ (224 vs. 152~\kms).  The overall distributions of maximum velocity and velocity width in all LINERs in our sample are different from those of \ion{H}{2} nuclei at $>$95\% confidence (see Figure \ref{histdvlh}).  However, the confidence level is too low for \dvtau\ to demonstrate a convincing difference.

This difference in outflow velocity and velocity dispersion could explain some or all of the physical differences between the LINER and \ion{H}{2} optical spectral classes.  As \citet{ds95} demonstrate, a modest increase in shock velocity can increase the \ntl/H$\alpha$ line ratio and push a galaxy from having an \ion{H}{2} galaxy classification to a LINER classification.  However, other effects may also contribute to the line ratios in some LINERs, including a weak AGN.

We also observe an increase in the median maximum velocity in the latest merger stages in the 1~Jy ULIRGs.  In interaction classes III, IV, and V, we observe median values of $|\dvmax|$ of 360, 350, and 450~\kms, respectively.  This conclusion is tentative, since we have $\la$10 galaxies with outflowing gas in each class.  Furthermore, the median value of \dvtau\ peaks in interaction class IV (\dvtau $=$ 130, 210, and 160~\kms\ for interaction classes III$-$V, respectively).  If there are really larger velocities later in the merger, then the acceleration of cold gas clouds is somehow more efficient in these stages.  This could be due simply to more time for the acceleration to occur, more available energy, or less pressure from surrounding gas (which has been cleared out by tidal forces or previous action of winds, or has decreased because of a decrease in density with radius).

\subsection{Mass, Momentum, Energy, and $\eta$} \label{mme_v}

\begin{figure}[t]
\plotone{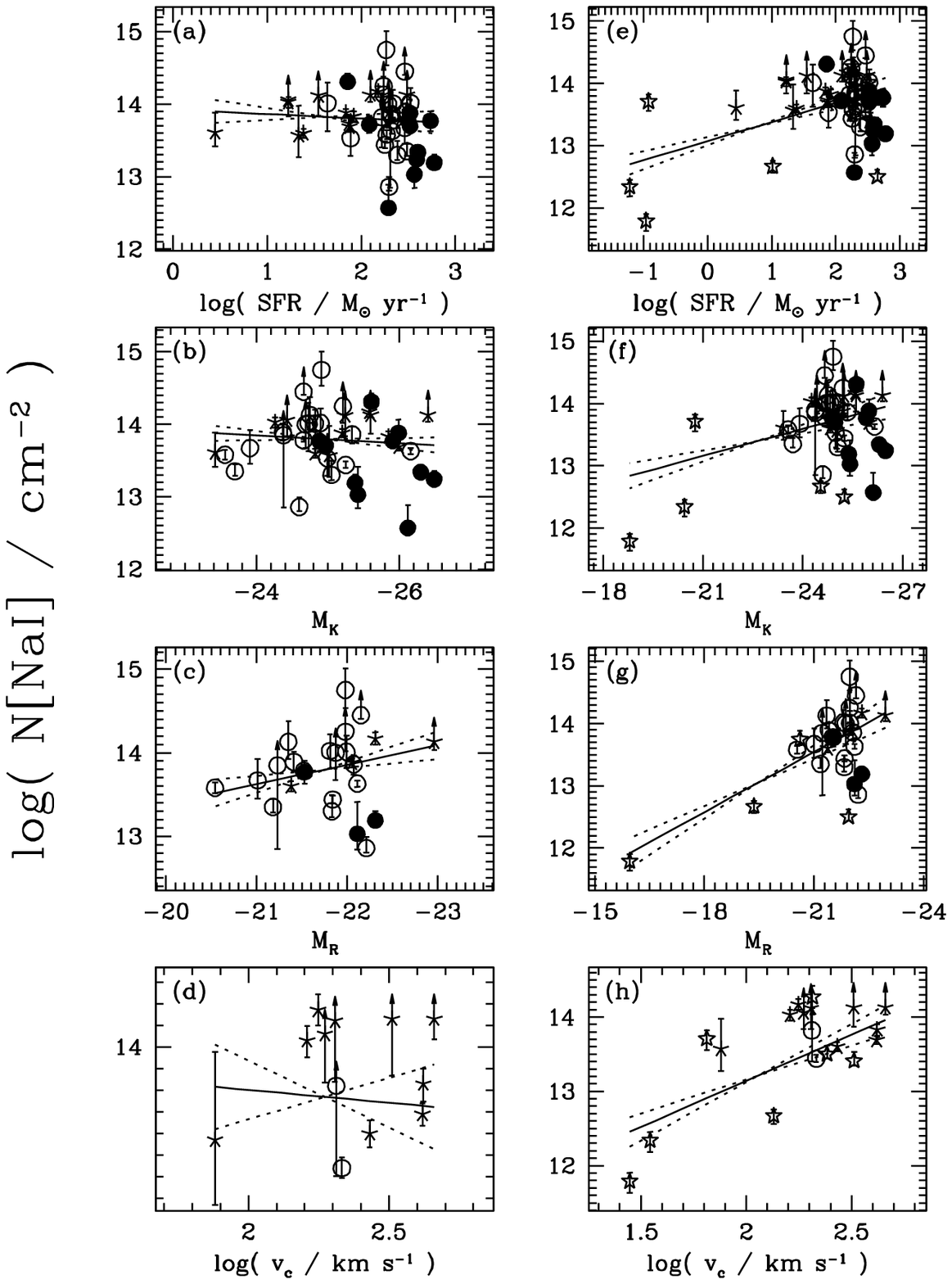}
\caption{Same as Figure \ref{dvmax_v}, but for total column density of outflowing \ion{Na}{1}.  The symbol shapes are as in Figure \ref{dvmax_v}.  The dotted arrows represent lower limits.  Fits to just our data are consistent with $N$(\ion{Na}{1}) being independent of each quantity on the horizontal axes.  If we add the dwarf galaxies from \citet{sm04}, correlation tests and weighted least squares fits show significant relationships between $N$(\ion{Na}{1}) and SFR, $M_R$, and $v_c$.  See \S\ref{mme_v} for further discussion.}
\label{nai_v}
\end{figure}

In Figure \ref{nai_v} we plot outflowing column density of \ion{Na}{1} as a function of SFR, $M_{K^{(\prime)}}$, $M_R$, and $v_c$.  This figure shows that $N$(\ion{Na}{1}) is weakly correlated with SFR, $M_R$, and galactic mass ($\propto v_c^2$): $|a| \sim 0.3-0.6$.

\begin{figure}[t]
\plotone{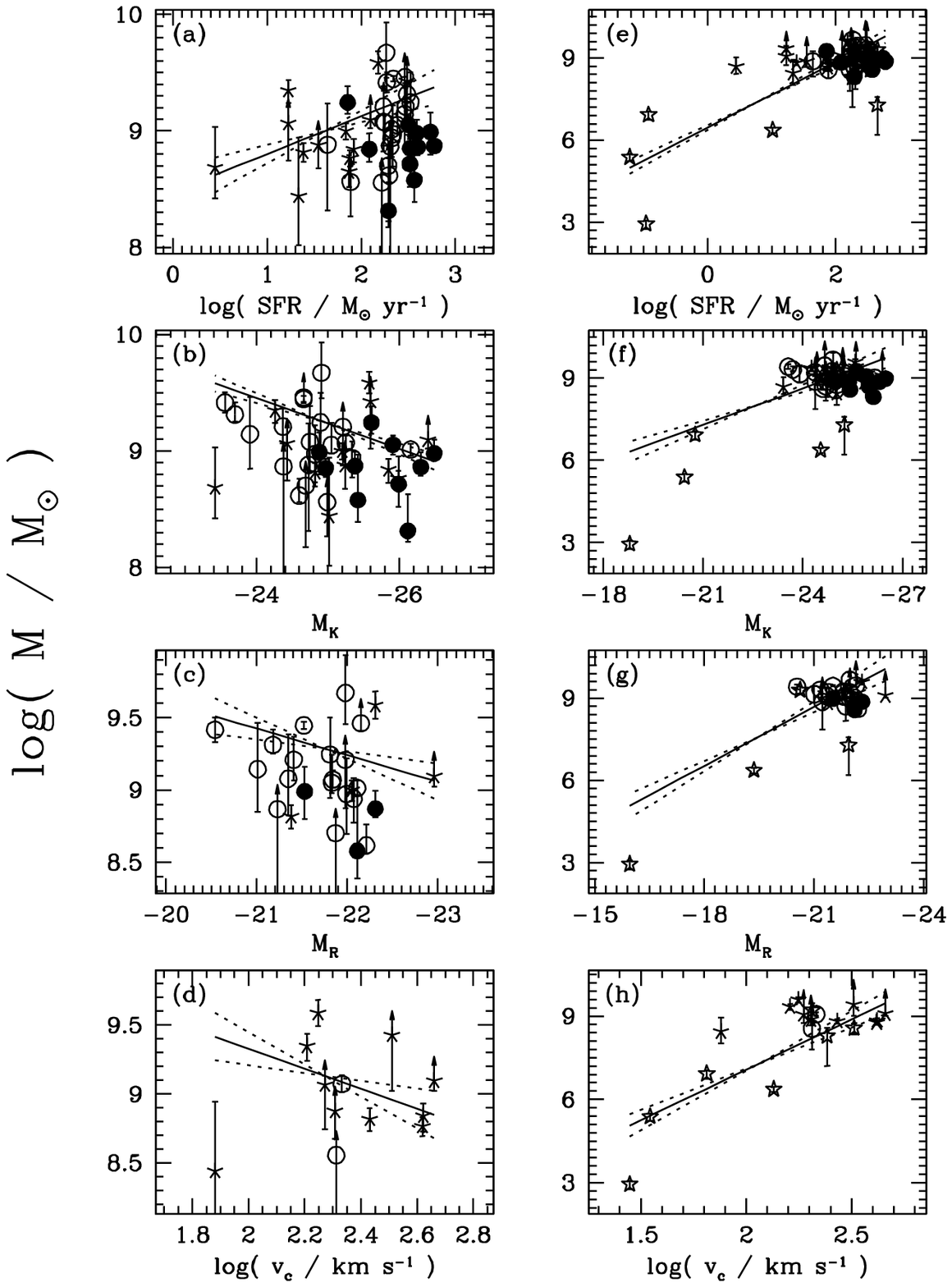}
\caption{Same as Figure \ref{dvmax_v}, but for total mass of neutral gas.  The symbol shapes are as in Figure \ref{dvmax_v}.  The dotted arrows represent lower limits.  Fits to just our data are consistent with mass being independent of each quantity on the horizontal axes.  If we add the dwarf galaxies from \citet{sm04}, correlation tests and weighted least squares fits show significant relationships between mass and each quantity.  See \S\ref{mme_v} for further discussion.}
\label{m_v}
\end{figure}

\begin{figure}[t]
\plotone{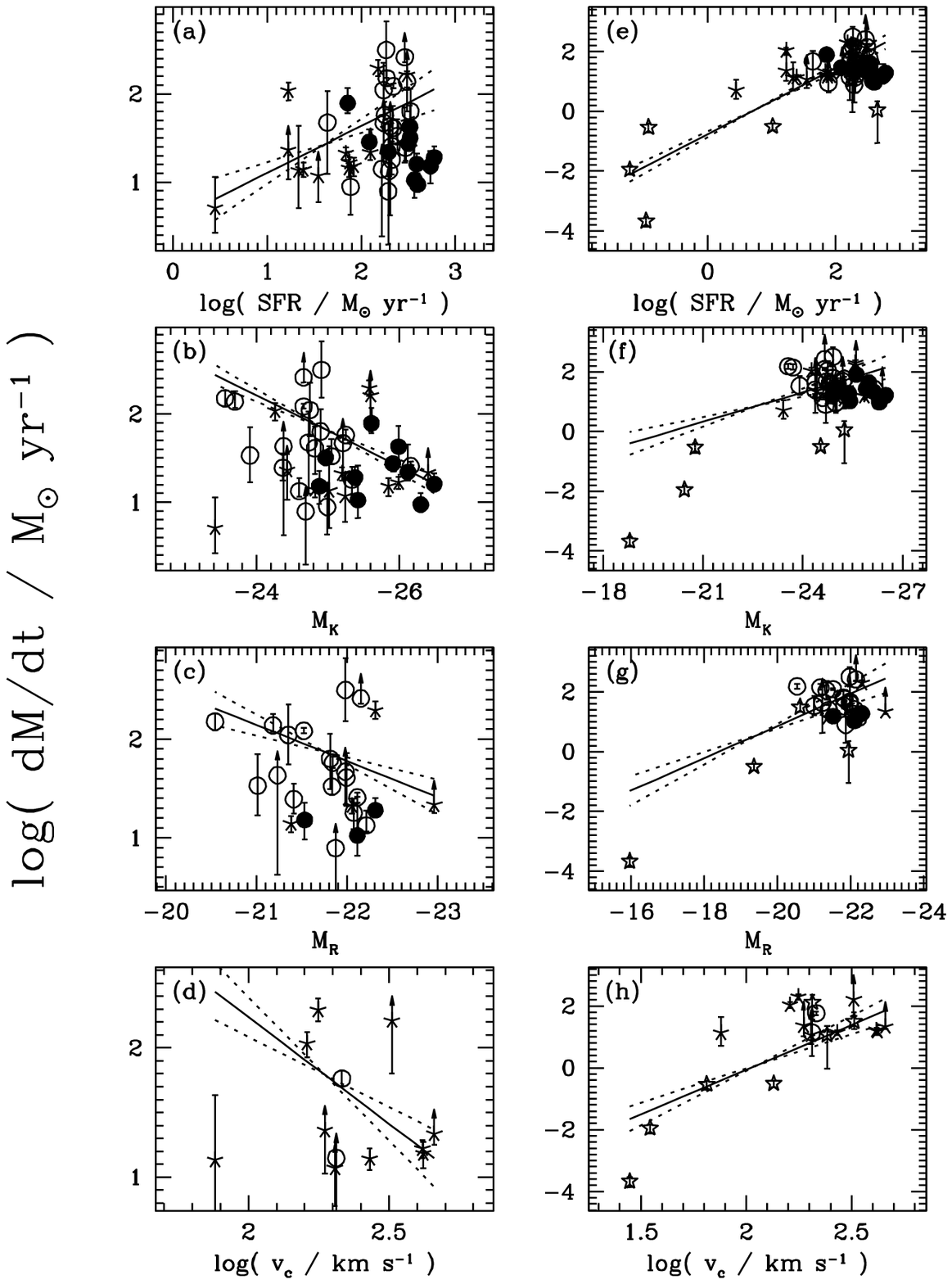}
\caption{Same as Figure \ref{dvmax_v}, but for mass outflow rate of neutral gas, averaged over the wind lifetime.  The symbol shapes are as in Figure \ref{dvmax_v}.  The dotted arrows represent lower limits.  Fits to just our data are consistent with mass outflow rate being independent of each quantity on the horizontal axes.  If we add the dwarf galaxies from \citet{sm04}, correlation tests and weighted least squares fits show significant relationships between mass outflow rate and each quantity.  See \S\ref{mme_v} for further discussion.}
\label{dmdt_v}
\end{figure}

\begin{figure}[t]
\plotone{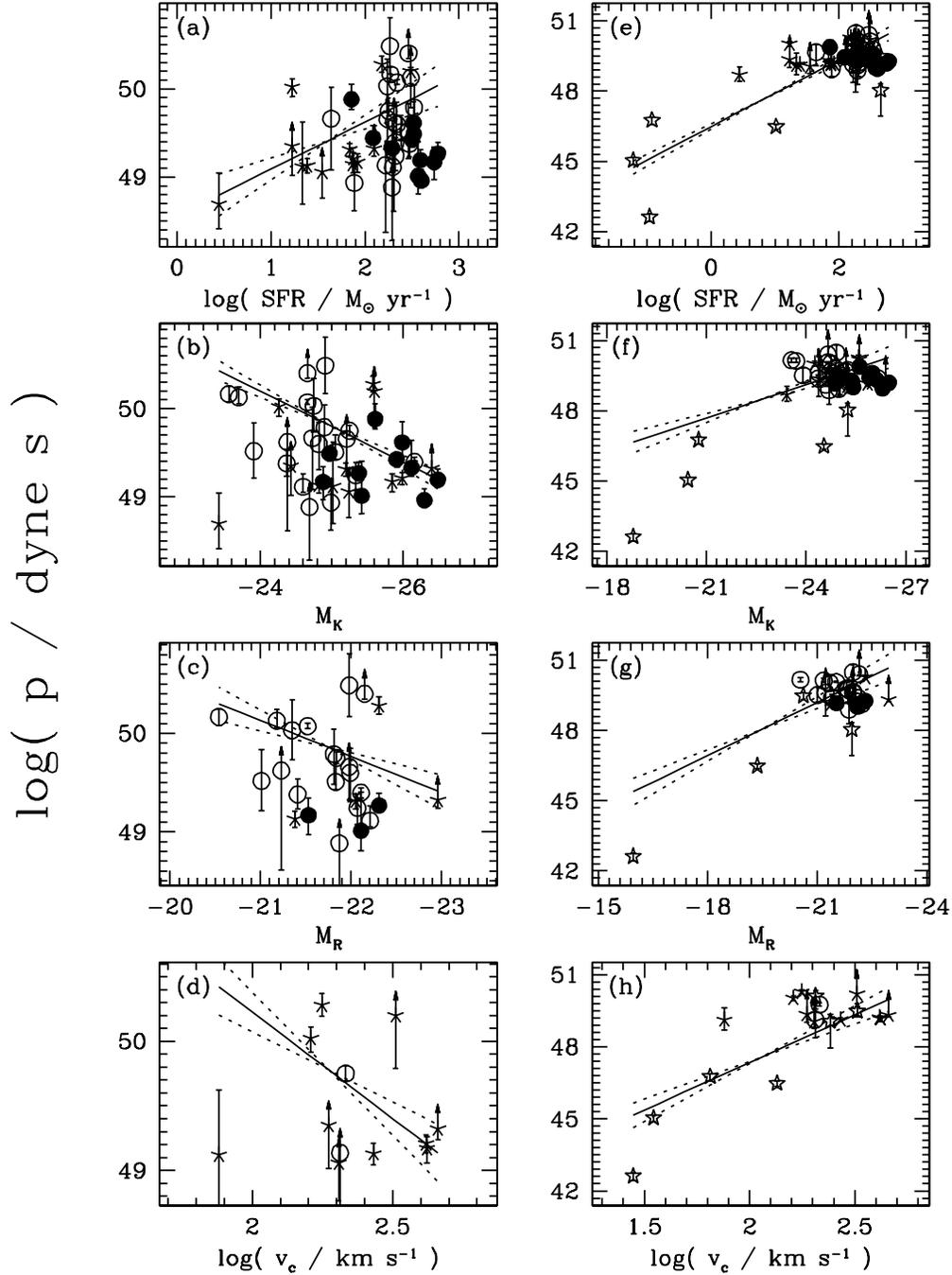}
\caption{Same as Figure \ref{dvmax_v}, but for total momentum in the neutral gas.  The symbol shapes are as in Figure \ref{dvmax_v}.  The dotted arrows represent lower limits.  Fits to just our data are consistent with momentum being independent of each quantity on the horizontal axes, but correlation tests and weighted least squares fits show significant relationships between momentum and each quantity if we include the dwarfs in \citet{sm04}.  See \S\ref{mme_v} for further discussion.}
\label{p_v}
\end{figure}

\begin{figure}[t]
\plotone{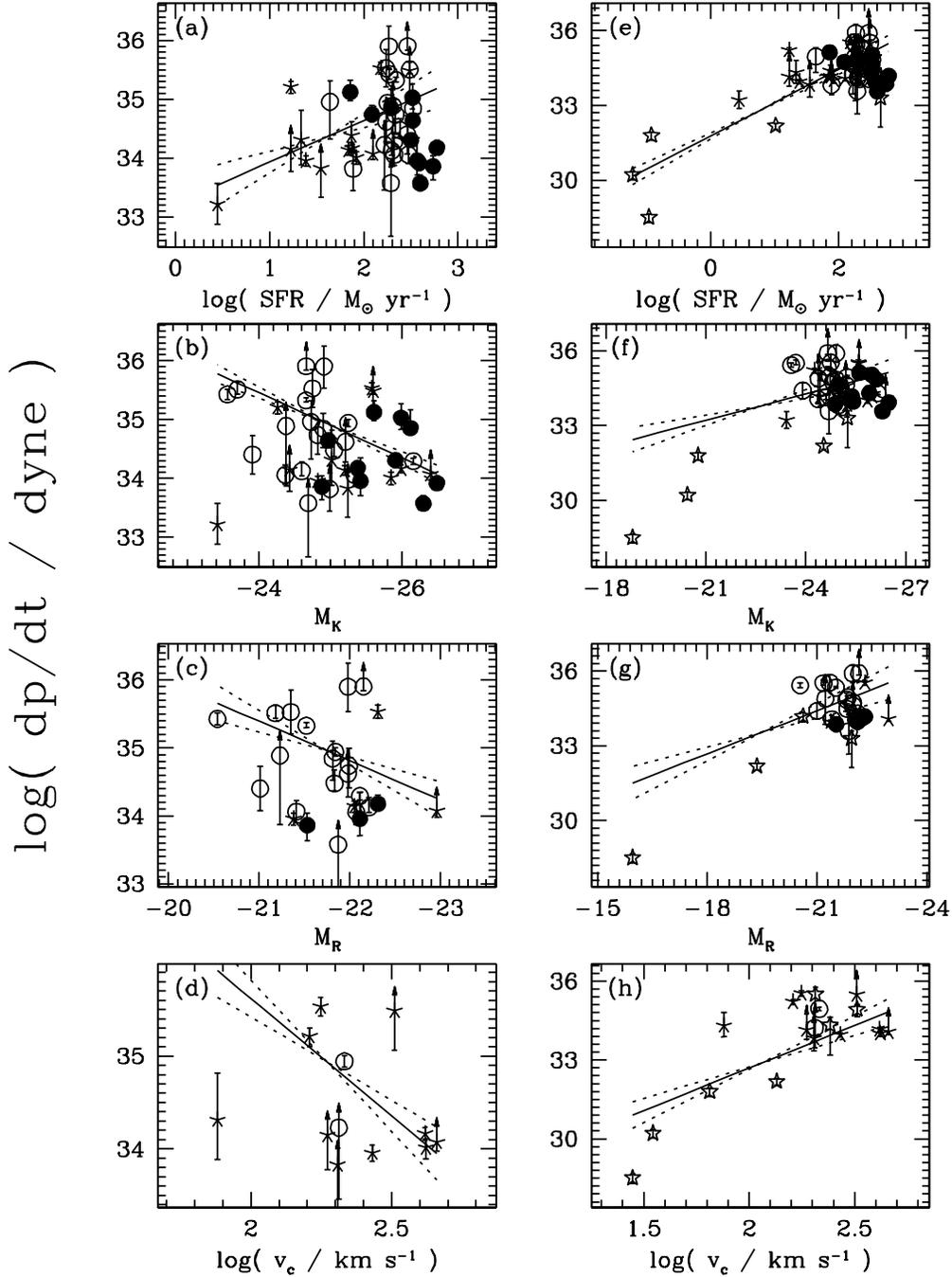}
\caption{Same as Figure \ref{dvmax_v}, but for total momentum outflow rate in the neutral gas.  The symbol shapes are as in Figure \ref{dvmax_v}.  The dotted arrows represent lower limits.  Fits to just our data are consistent with momentum outflow rate being independent of each quantity on the horizontal axes, but correlation tests and weighted least squares fits show significant relationships between momentum outflow rate and each quantity if we include the dwarfs in \citet{sm04}.  See \S\ref{mme_v} for further discussion.}
\label{dpdt_v}
\end{figure}

\begin{figure}[t]
\plotone{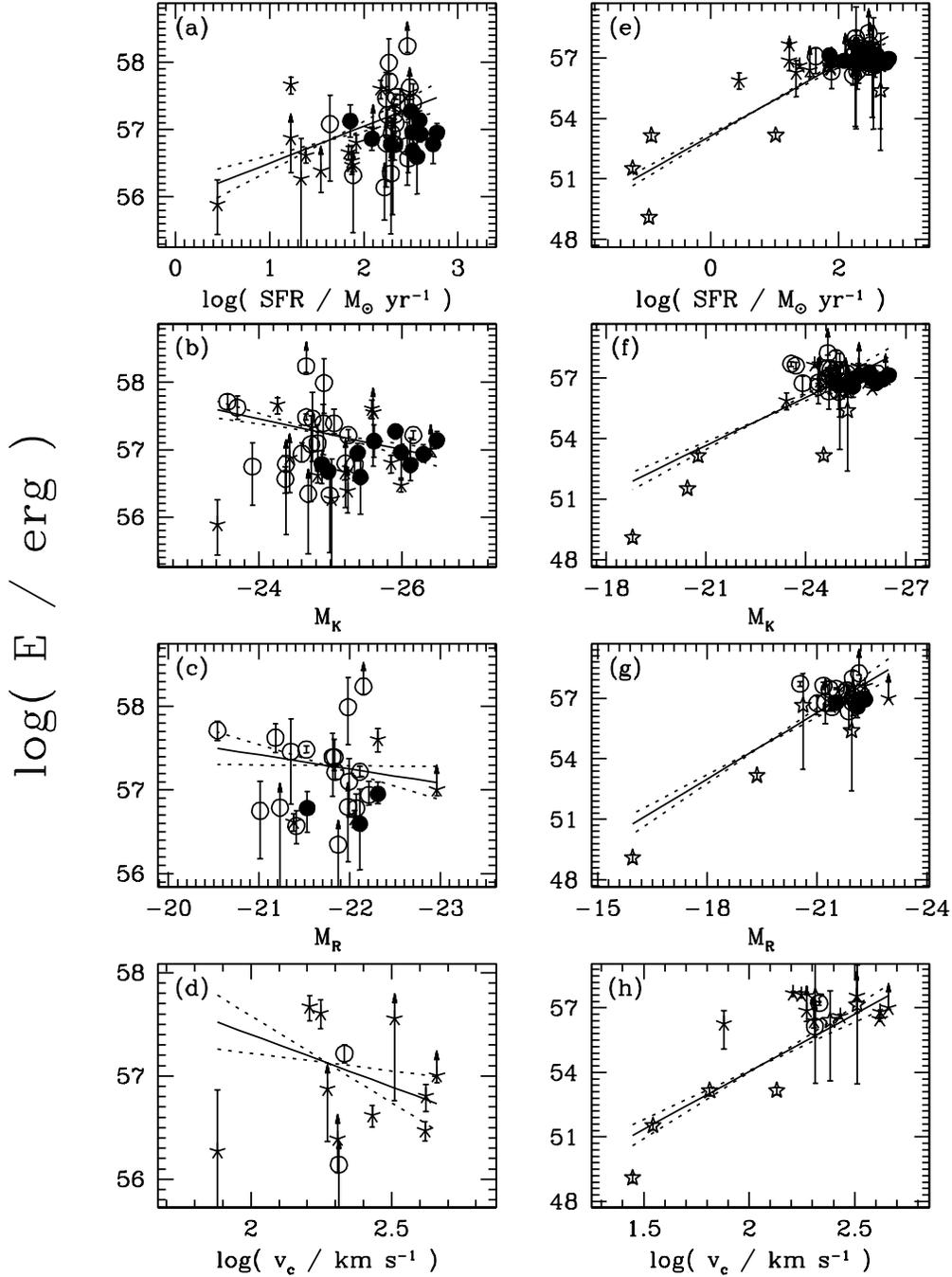}
\caption{Same as Figure \ref{dvmax_v}, but for total kinetic energy (bulk $+$ `turbulent') in the neutral gas.  The symbol shapes are as in Figure \ref{dvmax_v}.  The dotted arrows represent lower limits.  Fits to just our data are consistent with energy being independent of each quantity on the horizontal axes, but correlation tests and weighted least squares fits show significant relationships between energy and each quantity if we include the dwarfs in \citet{sm04}.  See \S\ref{mme_v} for further discussion.}
\label{e_v}
\end{figure}

\begin{figure}[t]
\plotone{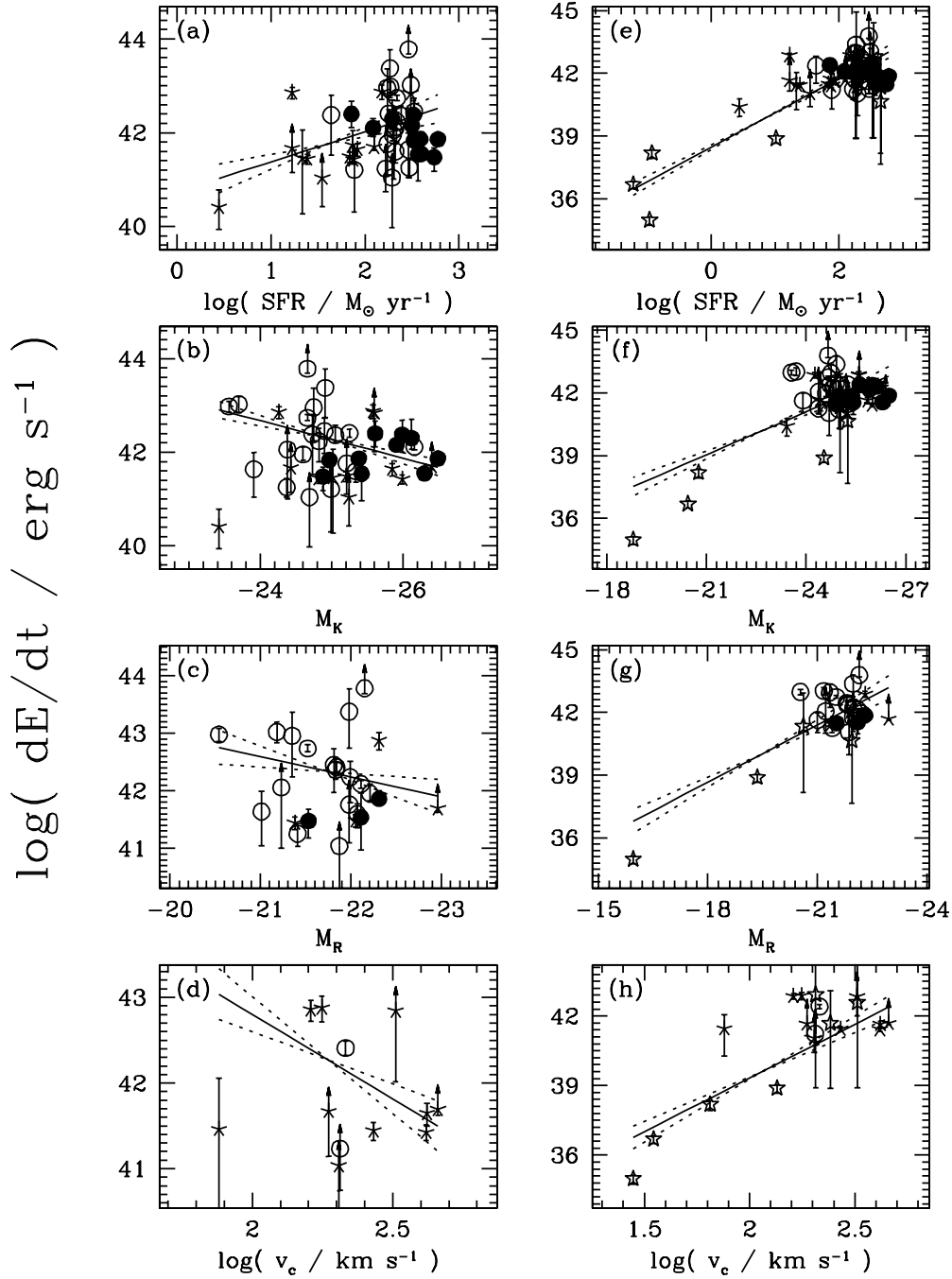}
\caption{Same as Figure \ref{dvmax_v}, but for total kinetic energy (bulk $+$ `turbulent') outflow rate in the neutral gas.  The symbol shapes are as in Figure \ref{dvmax_v}.  The dotted arrows represent lower limits.  Fits to just our data are consistent with energy outflow rate being independent of each quantity on the horizontal axes, but correlation tests and weighted least squares fits show significant relationships between energy outflow rate and each quantity if we include the dwarfs in \citet{sm04}.  See \S\ref{mme_v} for further discussion.}
\label{dedt_v}
\end{figure}

In Figures $\ref{m_v}-\ref{dedt_v}$, we plot mass, momentum, energy, and their outflow rates as a function of galaxy properties.  These figures show that for our sample considered alone, there are no significant correlations, except possibly between $E$ and SFR.  However, if we also include the four dwarf galaxies from \citet{sm04}, we find significant correlations in almost every case.  The measured correlations with $R$- and $K$-band luminosity are weak ($|a|\la 1$).  The overall slopes of mass and momentum with respect to SFR are approximately linear ($a = 1.1 - 1.4$, with $\delta a\sim0.1$).  However, the slope of energy with respect to SFR is steeper ($a = 1.6-1.8$).  The dependence of mass, momentum, and energy on galactic mass is the strongest ($a = 1.5-2.5$).

Above a characteristic star formation rate (SFR~$\ga 10-100$~\smpy), luminosity, and mass, these correlations disappear.  In other words, the masses, momenta, and energies become approximately constant as a function of galaxy properties.

\begin{figure}[t]
\plotone{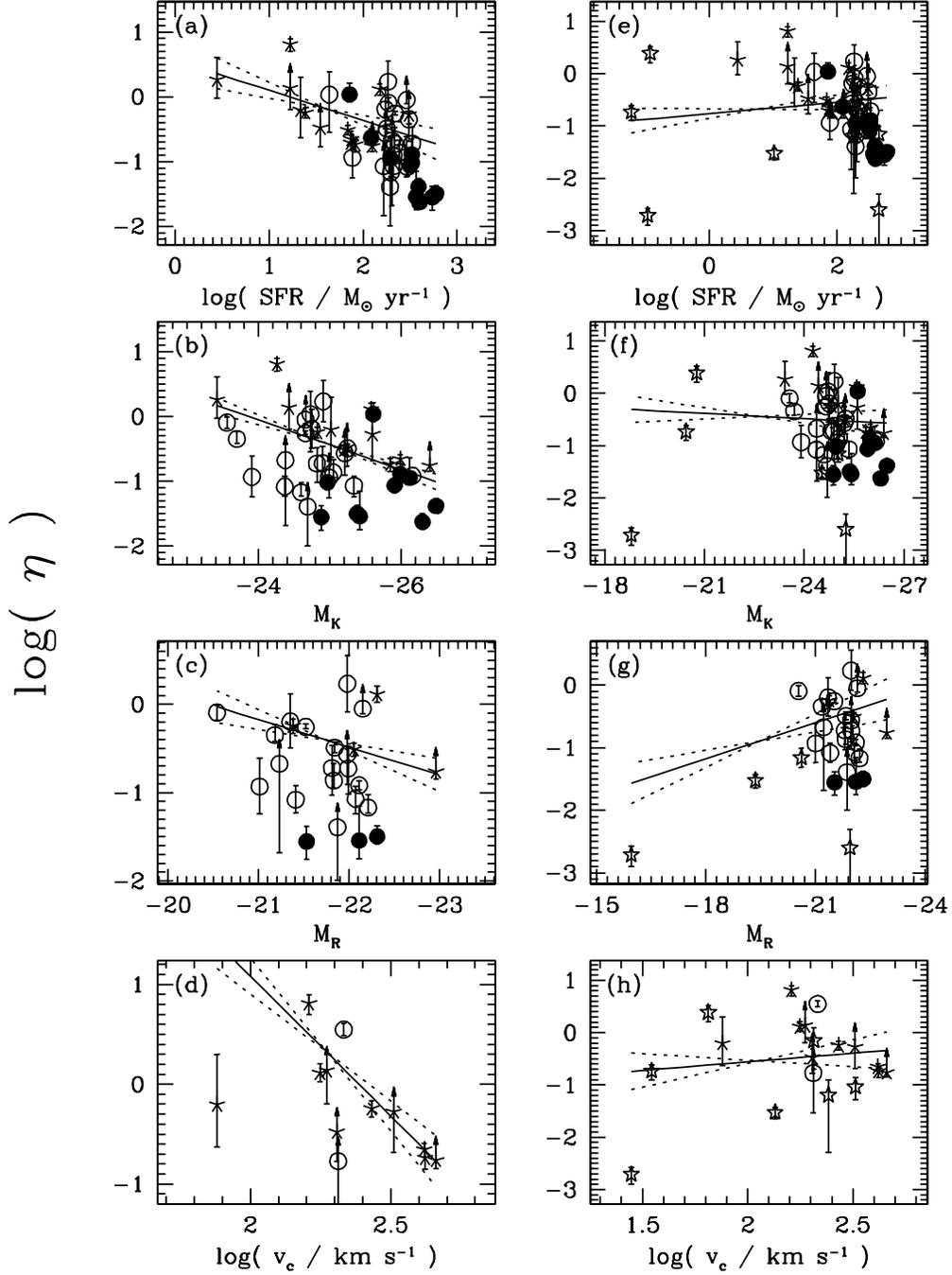}
\caption{Same as Figure \ref{dvmax_v}, but for mass entrainment efficiency, averaged over the wind lifetime.  The symbol shapes are as in Figure \ref{dvmax_v}.  The dotted arrows represent lower limits.  We find a significant relationship between $\eta$ and SFR and $K$-band magnitude {\it only in our data}; when the full data range is considered, there are no correlations between $\eta$ and the quantities on the horizontal axes.  See \S\ref{mme_v} for further discussion.}
\label{eta_v}
\end{figure}

In Figure \ref{eta_v}, we plot mass entrainment efficiency as a function of galaxy properties.  The mass entrainment efficiency is roughly constant as a function of SFR, luminosity, and mass.  However, the scatter is large (over two orders of magnitude).  Furthermore, at the highest values of SFR (i.e., in our sample alone), $\eta$ decreases as SFR and $M_{K^{(\prime)}}$ increase.

How are these correlations affected by our use of the luminosity-metallicity relationship (Paper I)?  If instead we assume solar metallicity, we find that the resulting correlations are unaffected.  The changes in slope are less than 1$\sigma$ in each case.  This convincingly confirms that non-solar metallicities do {\it not} drive the trends of outflow properties vs. galaxy properties that we observe.  However, assuming solar metallicity does move the normalization of the observed correlations upward by a factor of $\sim$2, since the galaxies in our sample have twice solar metallicity on average.  Thus, the hydrogen column density, mass, momentum, energy, and mass entrainment efficiency are higher by a factor of $\sim$2 on average under the assumption of solar metallicity.

In the Discussion (\S\ref{theory}), we interpret these correlations of mass, momentum, energy, and mass entrainment efficiency with galaxy properties.

\subsection{Correlations with Emission Lines} \label{eml}

Resolved outflows in the local universe present distinctive properties when observed in optical line emission, including limb-brightened bipolar structures and line-splitting in velocity space.  The surface brightnesses of these winds are typically much smaller than those of their background galaxies, making it difficult to see emission-line evidence of small superbubbles in many distant galaxies.  However, very luminous, extended emission-line nebulae exist in nearby LIRGs and ULIRGs such as NGC 6240 and Arp 220 \citep{ham87,ahm90,v_ea03,g_ea04} and may be powered by starburst-driven outflows.

\begin{figure}[t]
\plotone{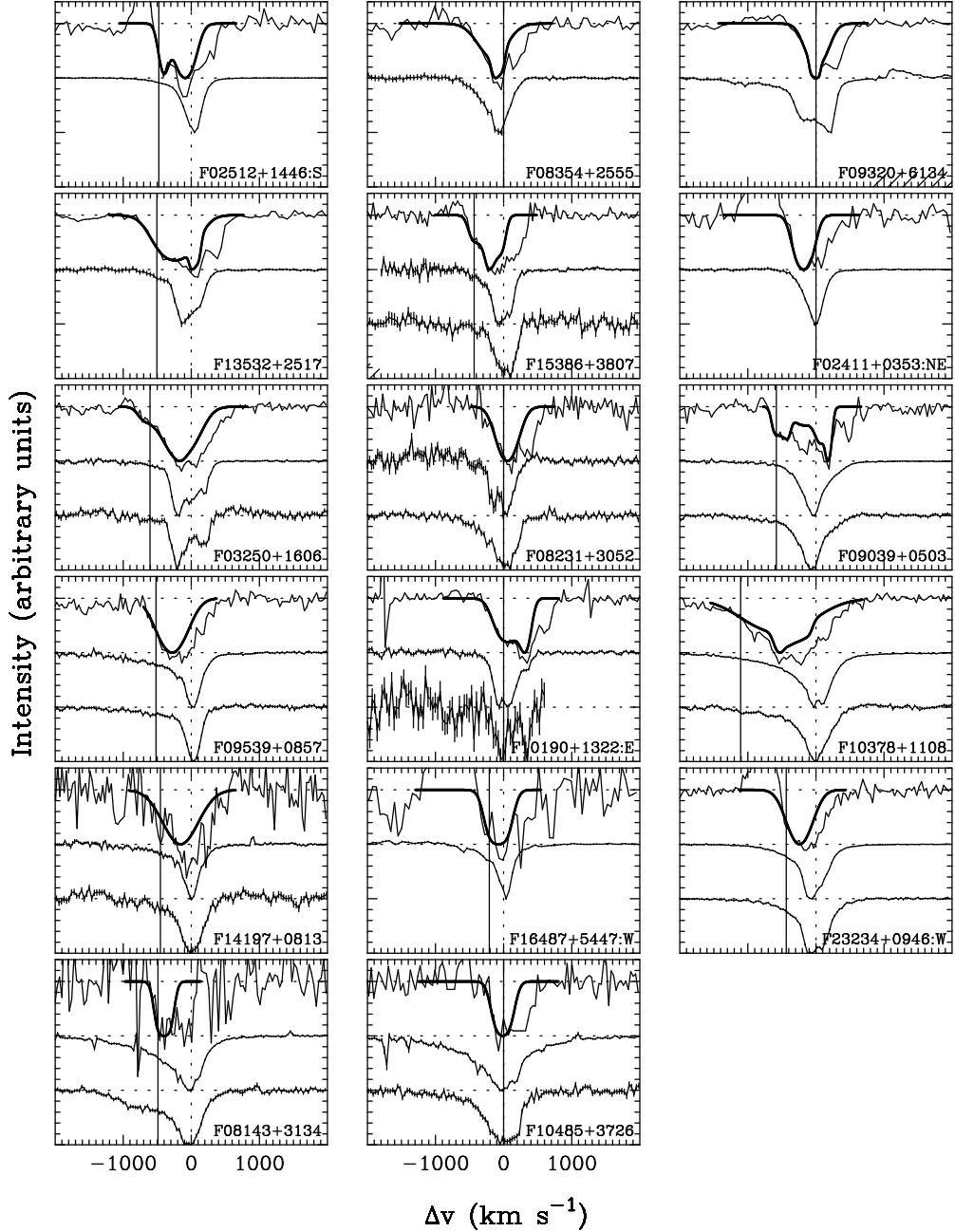}
\caption{\small{Emission-line spectra of 17 galaxies which have emission-line asymmetries in the profile wings.  The top spectrum (red) in each panel is \nad, with our fit to the \nad$_2$ $\lambda5890$ line superimposed (green).  The velocity scale is relative to the expected systemic velocity of this line.  The middle spectrum (blue) is \nt\ $\lambda6548$ for $\Delta v < 0$ and \nt\ $\lambda6583$ for $\Delta v > 0$.  The bottom spectrum (orange) is \otl.  The vertical thick line locates the neutral gas \dvmax.  The vertical error bars in the emission-line spectra are 2$\sigma$ errors.  Sixteen of these galaxies have a blue emission-line asymmetry in the profile wings, and 11 of these 16 ($70\%$) have a superwind present.  In 6 of the 11 galaxies with blue emission-line asymmetries and superwinds, the ionized gas has a higher maximum velocity than the neutral gas by several hundred \kms.  The ionized gas velocities reach 1000~\kms\ in several galaxies, as opposed to only one galaxy for the neutral gas.  See \S\ref{eml} for further discussion.}}
\label{emlspec}
\end{figure}

In Figure \ref{emlspec}, we plot inverted profiles of the \nt\ $\lambda\lambda6548,~6583$ and \otl\ lines below the \nad\ profile for 17 galaxies where the blue and red emission line wings have asymmetric profiles (i.e., one has a higher maximum velocity and/or more flux than the other) .  For each emission line, we fitted and subtracted the continuum using a low-order polynomial.  The extraction apertures for these data are the same as those for the \nad\ spectra, which cover most of the visible continuum.  The \nt\ line profile is a combination of the red half of the $\lambda6583$ line and the blue half of the $\lambda6548$ line.  The latter is not seriously contaminated by H$\alpha$, since they are 700~\kms\ apart.

In this subset of 17 galaxies, only 1 has a red emission-line asymmetry in the profile wings (F10190$+$1322:E, discussed in \S\ref{spatialinfo} and \S\ref{overlap}).  Thirteen of the other 16 galaxies, or 81\%, have \nad\ components that are blueshifted by more than 50~\kms.  Two of these are not interpreted as winds on the basis of other considerations (\S\ref{altexp}), leaving 11 of 16, or 69\%.

Thus, in galaxies with blue-asymmetric emission-line wings in \nt\ and/or \ot, 69\% possess a superwind on the basis of \nad.  The corollary: of the 45 galaxies in our sample that possess superwinds on the basis of \nad, at least 11, or $\ga$25\%, possess blue-asymmetric emission-line wings.  More galaxies may possess these wings, but they are not detected at our S/N.

These blue-asymmetric emission-line wings may very well represent the ionized phase of the outflow, which is seen prominently in the halos of nearby edge-on starbursts.  Interestingly, in half (6 out of 11) of these cases, the ionized gas has a higher maximum velocity by a few hundred \kms\ (in the other half the ionized and neutral velocities are roughly the same).  The emission line velocities reach up to $\sim$1000~\kms\ or higher in several cases.  Furthermore, in most cases the \ot\ and \nt\ profiles match well, suggesting that the low- and high-ionization states generally have the same velocities, at least in starburst outflows.

Not only are the maximum ionized gas velocities larger than the maximum neutral gas velocities, but the velocities of the bulk of the outflowing gas are, as well.  Deprojected velocities of ionized gas in edge-on starbursting galaxies are $\sim$170~\kms\ on average \citep{lh96}; for galaxies of comparable SFR (the IRGs), the velocities of the bulk of the neutral gas are 100~\kms\ on average (\S\ref{dv} and Table \ref{avgprop}).  The latter do not need to be deprojected significantly, since they are typically detected while close to face-on (\S\ref{gcf}).  Thus the ionized gas velocities are $\sim$70\% greater than the neutral-gas velocities in these galaxies.

\subsection{Comparison to \ion{H}{1} 21~cm Spectra} \label{hi}

Given the high column densities of neutral \ion{H}{1} that we infer with \nad\ ($10^{21}$~cm$^{-2}$), we might expect to also observe outflowing gas directly in emission or absorption using the \ion{H}{1} 21 cm line.  Since \ion{H}{1} data exists for many of the objects in our IRG subsample, we can search for direct evidence of neutral, outflowing \ion{H}{1} in the form of blue wings in the absorption profiles.

The kinematics of the \ion{H}{1} gas in F10565$+$2448 are the most interesting.  The systemic velocity that we measure from nebular emission lines agrees with the narrow, deep absorption trough in the \ion{H}{1} 21 cm line at $\sim$12900~\kms\ \citep{ms88}.  There appears to be a broad, blueshifted component in \ion{H}{1} that extends to roughly the same outflow velocities as seen in \nad.  There is also redshifted \ion{H}{1} in emission at $\sim$13100~\kms.  If the \nad\ and broad \ion{H}{1} absorption represent a bubble expanding in our direction along the line of sight, the redshifted \ion{H}{1} emission line could indicate a counter-bubble that is expanding away from us. 

Three other objects in our sample (F01417$+$1651, F02512$+$1446:S, and F03359$+$1523) also show \ion{H}{1} absorption in \citet{ms88}.  The S/N of the \ion{H}{1} spectra for these objects are comparable to that of F10565$+$2448, but there are no obvious broad, blueshifted components.

\section{DISCUSSION} \label{discuss}

\subsection{Alternative Explanations} \label{altexp}

The zeroth order interpretation of the blueshifted absorption lines in these galaxies is that they are produced by starburst-driven outflows.  Alternative explanations should produce velocity distributions that are symmetric about $\Delta v = 0$~\kms, for reasons we discuss below.  Thus, the maximum number of blueshifted components attributable to other phenomena is a mirror reflection of the distribution of red components seen in Figure \ref{histdv}$a$, and is therefore not significant.

\subsubsection{Gas in Rotation} \label{altexp_rot}

We observe rotation in the emission lines of most IRGs and in a sizable fraction of ULIRGs, suggesting that there are ordered gas disks in most of these systems.  In only a few cases is rotation also observed in \nad.  In a large number of these galaxies ($\sim$17 total), we see blueshifted \nad\ that is at or near the blue rotation arm of the galaxy in velocity space.  Rarely (in only $\sim$3 cases) do we observe the \nad\ near the red rotation arm.  Since there is no reason for this asymmetry in a rotating disk scenario, we conclude that this gas is in general {\it not} in simple rotation.

\subsubsection{Tidal Debris}

The red (and some of the blue) components that we observe in ULIRGs may be tidal debris, gas stirred up by vigorous interactions.  The simulations of \citet{bh91} show that much of the gas in an equal-mass prograde merger flows suddenly to a compact region at the merger center, and large tidal arms are spun out.  However, much of the extended tidal debris falls back to the merger center over Gyr time scales.  In NGC 7252, radial velocities of up to $\pm 200$~\kms\ are observed due to velocity gradients along tidal tails, the redshifted velocities representing gas falling back to the disk \citep{hm95}.

From examination of the velocity distribution of tidal material in the simulations of \citet{hm95}, we expect a narrow velocity width from a tidal tail in projection, while we observe mostly broad profiles.  However, there are two illustrative cases of possible tidal debris in ULIRGs.  F09039$+$0503 possesses a narrow (FWHM~$\sim$~30~\kms; Rupke \& Veilleux 2005, in prep.) component that is redshifted by $180-190$~\kms\ with respect to systemic.  The velocity is comparable to those measured for tidal features in NGC 7252 \citep{hm95}, and its small width is suggestive of a compact feature such as a tidal tail.  In NGC 6240, there is an emission- and absorption-line feature to the east of the disk; this feature lies atop a stellar structure that appears tidal in origin \citep[e.g.,][]{v_ea03}.  However, the evidence is only circumstantial; this galaxy also has an emission-line superbubble to the west of the galaxy \citep[e.g.,][]{v_ea03}, and it is conceivable that the \nad\ absorption and line emission to the west of the galaxy trace a counterbubble.

\subsubsection{Multiple Nuclei and Overlapping Disks} \label{overlap}

We observe evidence of overlapping disks in several double-nucleus galaxies.  In five galaxies, there is a redshifted absorbing component in the spectrum of one nucleus that is within 40~\kms\ of the redshift of the other nucleus; the nuclei with these components are F01417$+$1651:S, F02411$-$0353:NE, F16333$+$4630:W, F16474$+$3430:N, and F23234$+$0946:E.  A sixth galaxy, F08354$+$2555, contains \nad\ that is blueshifted from the \ion{H}{1} velocity but is coincident with the emission-line peak, which also corresponds to the velocity of a compact stellar object $5-10$\arcsec\ S of the nucleus; the \nad\ may be associated with either the galaxy nucleus or the southern object.  Finally, F10190$+$1322:E has redshifted absorption at $+320$~\kms\ with respect to systemic, as well as extended, redshifted emission.  The velocity of the part of the western disk that overlaps the eastern nucleus matches the redshifted absorption in velocity space \citep{m_ea01}.

This suggests that some of the components in our velocity distribution are due to the projection of gas disks along the line of sight.  Notably, these seven galaxies still contain three cases of high-velocity blueshifted absorption that are consistent with the outflow hypothesis.  In our analysis, we have removed the components that appear to be due to these overlapping disks, as we find this hypothesis more compelling than the outflow one for this small subset of components.

\subsubsection{Merger-Induced Winds}

Recently, \citet{cp_ea04} have hypothesized that the shocks produced as gas funnels to the center of the merger of two equal-mass galaxies can produce outflows of hot gas.  These shocks heat a large amount of gas to $10^{6-7}$~K, which, in analogy to a starburst-driven superwind, expands radially outward at speeds $\ga$200~\kms.  This gas could potentially entrain cold gas clouds and evolve much as a starburst-driven superwind.  However, the energy injection region may be on larger scales (several kpc vs. $\la$1~kpc for a superwind).  Further exploration of this idea is warranted, but it is currently indistinguishable in our data from the superwind hypothesis.

\subsection{Frequency of Occurrence and Global Covering Factor} \label{gcf}

The detection rate $D$ is a function of both the actual frequency of occurrence of winds $F$ and the global angular covering factor of the wind, $\Omega$.  To better understand this, we make the assumption that a wind is detected if our line-of-sight lies within the opening angle of the wind.  (For an outflow emerging perpendicular to a galactic disk, the opening angle is delineated by the biconical structure of the wind.)  Within that opening angle, there may be local clumping of the wind material.  The global angular covering factor, or solid angle subtended by the wind, is the product of the these two factors (the opening angle and local clumping).  We assume that the opening angle factor $\co$ (a fraction of 4$\pi$) is given by the detection rate (which is the case if the frequency of occurrence of winds is 100\%), and that the local clumping is given by the measured covering fraction along the line-of-sight, \cf.  The global covering factor is then given by $\Omega/4\pi = \co~\langle\cf\rangle = D~\langle\cf\rangle$ (see \S\ref{windgeom} and \citealt{ckg03}).  In the more general case, where $F$ is not necessarily unity, we can set lower limits to $F$ and $\co$ using our detection rate: $D < F < 1$ and $D < \co < 1$.

We already know that \co\ is less than unity in local galaxies ($\sim$0.4 on average; e.g., \citealt{vcb05}) and is similar to our detection rate (\S\ref{detrate}).  We also observe marginally significant correlations between the outflow detection rate and the apparent ellipticity of the galaxy in the IRG subsample.  We find that we are very likely to observe winds in galaxies that are almost face-on (8 outflows in 15 galaxies) but not at all likely in edge-on galaxies (2 out of 12 galaxies).  This is consistent with values of \co\ close to $1/2$, under the assumptions that winds occur in all starbursts and emerge perpendicular to the galaxy disk.

In ULIRGs, however, both \co\ and $F$ must be at least 0.7, since $D \ga 0.7$.  Thus, the geometry of outflows in ULIRGs is not the same as in local disk galaxies.

In either case, we conclude that winds occur in almost all starbursting, infrared-luminous galaxies (i.e., $F\sim1$).  To get the global covering factor, $\Omega$, we then fold together our detection rates and measured average covering fractions along the line of sight (see Table \ref{avgprop}).  We thus compute $\Omega/4\pi \sim 0.15$ for the IRGs and $\Omega/4\pi \sim 0.30$ for ULIRGs.

The high-redshift Lyman-break galaxies have star formation rates similar to the IRGs \citep{p_ea98}.  However, the detection rate of winds in LBGs is almost 100\% \citep{ad_ea03}.  The method of detection involves both absorption and emission lines, but it seems to imply that the opening angle and global covering factor of these winds are much higher than that in local galaxies of similar SFR.  On large scales, the opening angles of LBG winds are smaller, however, perhaps closer to 4$\pi$/3 \citep{ad_ea05}; this value is similar to that of IRGs.

\subsection{Gas Escape Fraction}

In our preliminary report \citep{rvs02}, we tentatively claimed that the fraction of gas in ULIRG superwinds that escapes the host galaxy ($f_{esc}$) is high, perhaps up to $40-50\%$.  We computed this value by estimating the circular velocity, $v_c$, in each galaxy and comparing it to the escape velocity, $v_{esc}$ (computed using a singular isothermal sphere).  However, we previously included two Seyfert 2s with high outflow velocities, whose winds are not necessarily starburst-driven \citep{rvs05b}.

In this work, we continue to use a singular isothermal sphere with $r_{max} / r = 10-100$.  If the gas absorbs at radius $r$, then the escape velocity is parameterized uniquely by $v_c$ and $r_{max} / r$.  Our procedure to calculate $f_{esc}$ is as follows: (a)~to get $v_c$, use measured values where possible; otherwise, for ULIRGs, use an average value (K. Dasyra, private communication; see Paper I for more details); (b)~use $v_c$ to compute $v_{esc}$; (c)~compute the mass or mass outflow rate of gas that has a velocity above $v_{esc}$; (d)~sum $dM/dt$ and $dM/dt_{esc}$ over all galaxies with measured $v_c$; and (e)~divide $dM/dt_{esc}^{total}$ by $dM/dt^{total}$.  (Note that $M$ and $dM/dt$ are interchangeable in this algorithm; the values of $f_{esc}$ computed using $M$ are comparable to those computed using $dM/dt$, but smaller by a factor of 2.  The discrepancy is due to the extra factor of $\Delta v$ in the definition of $dM/dt$; see \S\ref{eqns}.)

By examining Figure \ref{theory_v}$b$, we see that a significant number of galaxies have maximum velocities close to or above the predicted $v_{esc}$ for a singular isothermal sphere.  Ignoring halo drag \citep{st01} or acceleration of the wind, we compute that between 5\% and 20\% of the neutral material in these winds will escape the galaxy and enter the IGM, for $r_{max}/r = 10-100$.  These numbers are smaller than our previous estimates \citep{rvs02}, but are still significant.  The hot, freely-expanding wind that drives the entrained, neutral material (and carries the majority of the metals; \citealt{mkh02}) is even more likely to escape.  If we correct for possible projection effects in the velocities, this escape fraction is likely to increase.

Some of the maximum velocities that we measure are close to plausible values for $v_{esc}$.  Is this coincidental or suggestive of further physics?  We suggest that there is a good possibility that there is (or was) material of higher velocities, as is seen in local starburst-driven superwinds like that in NGC 3079 at velocities of up to 1500~\kms\ \citep{vc_ea94,cb_ea01}.  This high-velocity gas may be better probed by emission lines, which are more sensitive to higher-temperature, lower-density gas.  We do observe higher emission-line than absorption-line velocities in several galaxies (\S\ref{eml}).  Assuming that this gas moves radially (though it may form vortices; see \citealt{cb_ea01}), it will expand more quickly than the low-velocity gas and dissipate as it reaches large radius, giving it a low cross-section in the line-of-sight and making it difficult to detect using absorption-line probes.  It may be that the single galaxy in our sample with $\dvmax > 1000$~\kms\ is an example of this, where the gas has not completely escaped to large radii.  However, the \nad\ feature in this galaxy is quite broad and deep, and for this reason not obviously consistent with this interpretation.

\subsection{Comparison with Theory} \label{theory}

A substantial number of numerical simulations of superwinds have been performed over the years (e.g., \citealt{s_ea94,ss00}).  These models do not typically make predictions about values of mass, momentum, and energy in the cold gas.  They do show that there are large quantities of warm gas in the superwind, distributed both in filaments and clouds entrained from the disk and as a swept-up shell surrounding the wind (which is disrupted by Rayleigh-Taylor instabilities at `blowout').  Uncertainties in these models still exist because they treat the ISM as a continuous medium and do not include the microphysics of wind/cloud interactions (such as conduction and ablation leading to mass-loading of the wind fluid and cloud destruction, small-scale hydrodynamic instabilities, and radiative cooling).  New simulations are underway (Cooper, Sutherland, \& Bicknell 2005, in prep.) incorporating a fractal gas distribution, which will better treat the physics of entrainment and thus the gas mass in the wind.  This should in turn lead to accurate predictions of mass outflow rates.

In lieu of comparison to detailed wind simulations, there is a limited amount of analytic work to which we can compare our results \citep[e.g.,][]{s03,mqt05}.  Furthermore, we can also compare the properties of the entrained gas to the properties of the hot gas we expect to be produced by the starburst \citep{l_ea99}.  We will also compare to predictions based on radiation pressure.

\subsubsection{Velocities}

The gas in these winds may be driven by ram pressure from the hot free wind, by radiation pressure from the starburst on the gas and dust in the wind, or by some combination of the two.  In the case that ram pressure drives the wind, the properties of the hot gas help determine the final wind velocities.  \citet[][eq. A3]{mqt05} predict the `characteristic' velocities of clouds accelerated by the hot gas, as well as the cloud velocity as a function of radius.  The maximum velocity that these clouds can attain is close to the characteristic velocity and is largely constrained by the velocity of the hot wind that drives them.

\begin{figure}[t]
\plotone{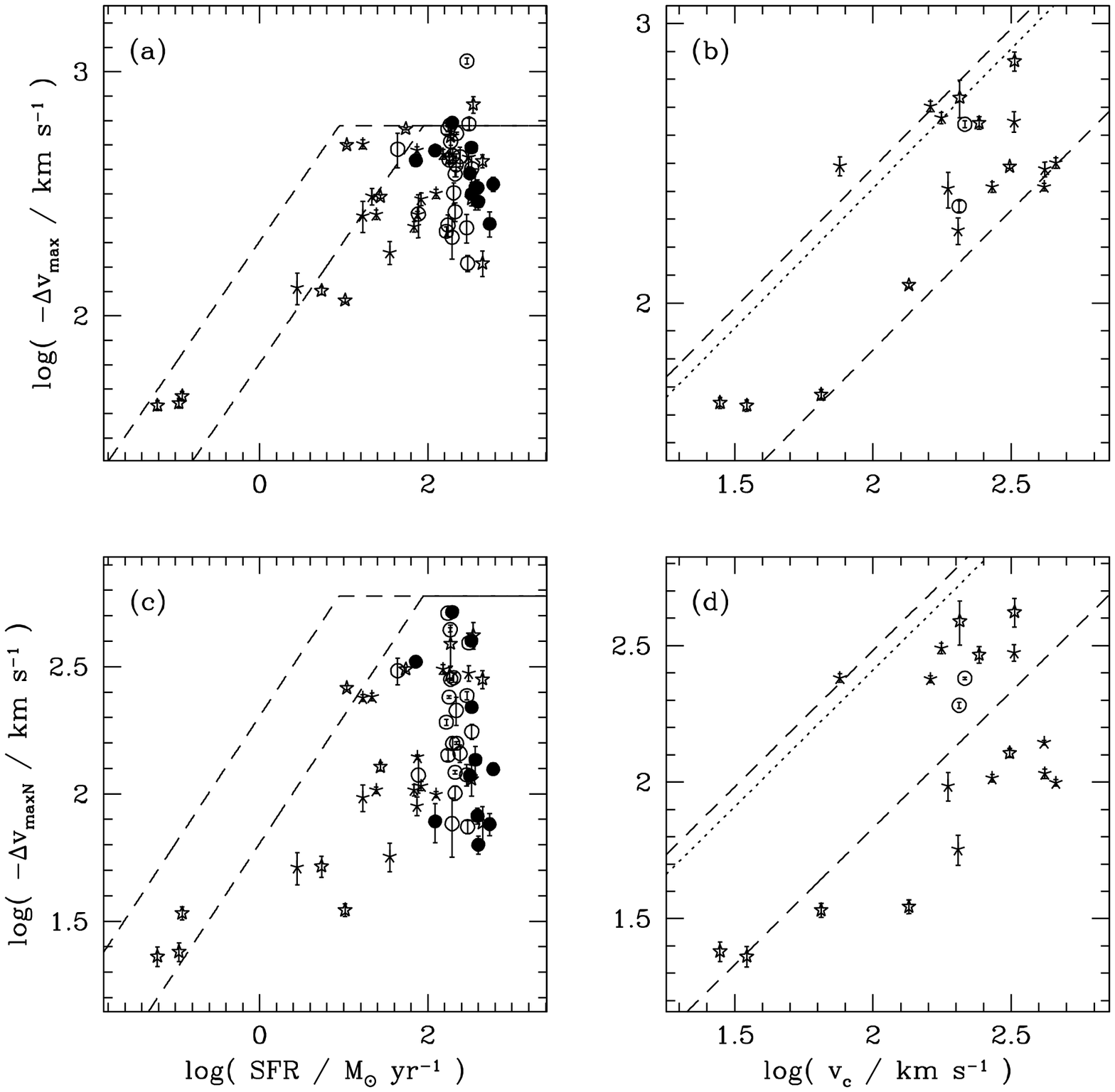}
\caption{Maximum velocity (\dvmax) and velocity of the highest column density outflowing gas (\dvtau) vs. ($a$), ($c$) star formation rate and ($b$), ($d$) circular velocity.  The dashed lines in panels ($a$) and ($c$) are characteristic velocities of ram-pressure accelerated clouds \citep{mqt05} for column densities of 10$^{20}$ (top line) and 10$^{21}$~cm$^{-2}$ (see \S\ref{theory} for other assumptions); the measured velocities are roughly consistent with these upper limits given the measured column densities.  The dashed lines in panels ($b$) and ($d$) are models of radiation pressure driving of an optically thick shell.  All the points but a few are consistent with a model where the luminosity of the starburst is between 1.05 and 2 times the critical luminosity for momentum driving of a wind \citep{mqt05}.  The dotted lines in panels ($b$) and ($d$) show the escape velocity of a singular isothermal sphere with $r_{max}/r = 10$; a significant number of points fall near or above this line.  See \S\ref{theory} for more details.}
\label{theory_v}
\end{figure}

In Figure \ref{theory_v}$a$, $c$, we plot the characteristic velocity $v_{cl}$ for clouds of column density $N$(H)~$= 10^{20}$ and 10$^{21}$~cm$^{-2}$.  We assume a constant hot wind velocity $v_{hot} = 600$~\kms\ (or $2\times10^7$~K) to be consistent with the highest velocities we measure, a cloud initial radius of 1~kpc, $\co = 0.4$, $(dM/dt_{hot})/SFR = 0.33$ \citep{l_ea99}, and a mean particle mass $\mu = 1.4$.  Note that $v_{cl}$ is set to $v_{hot}$ if it rises above it.  We see that \dvmax\ is less than or equal to $v_{cl}$ for $N$(H)~$= 10^{20}$~cm$^{-2}$, consistent with the idea that the clouds are accelerated by the hot wind until they reach a velocity near $v_{cl}$.  The range of measured $N$(H) roughly matches that predicted by the $v_{cl}$ lines.

\begin{figure}[t]
\plotone{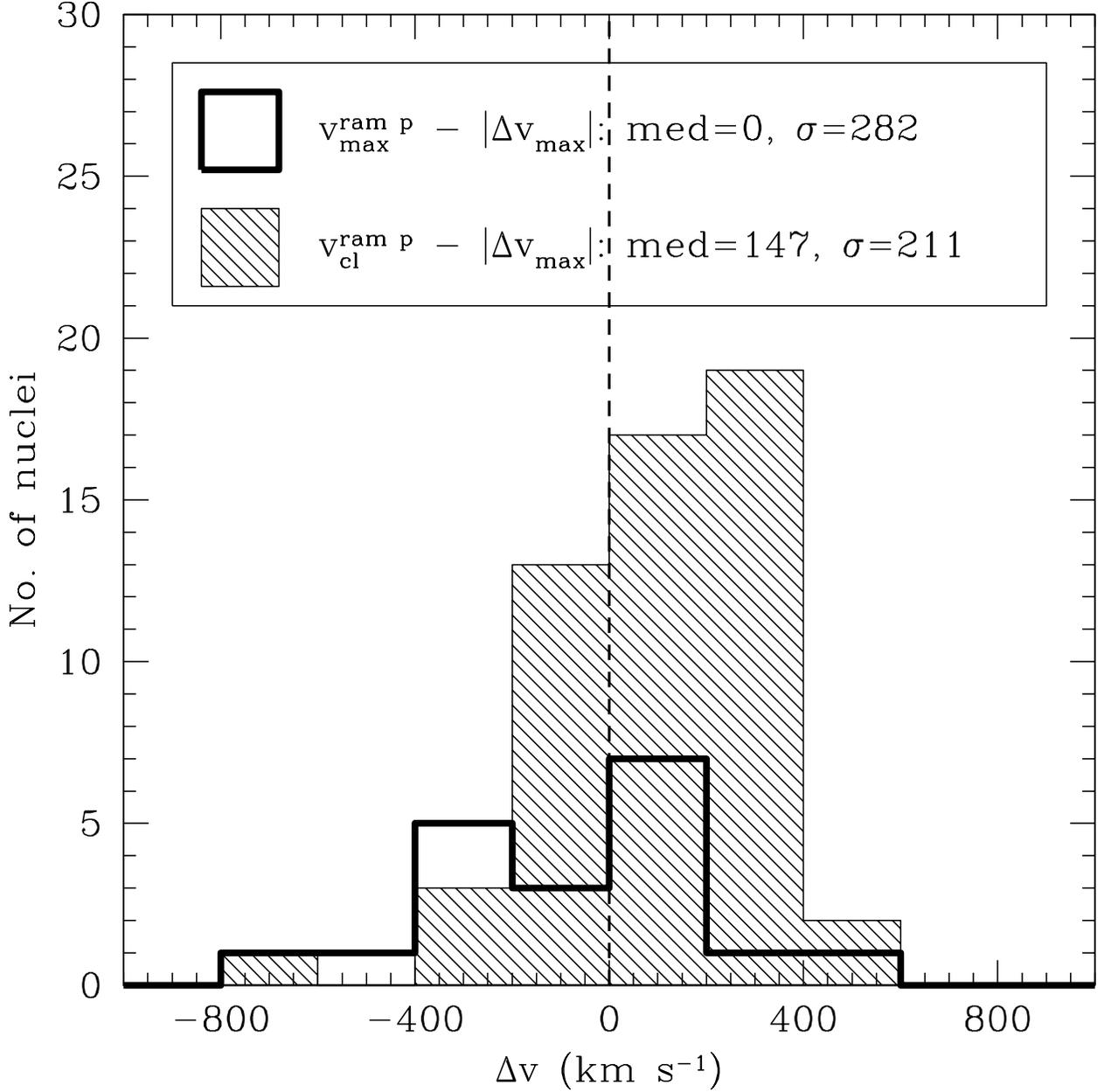}
\caption{Distribution of velocity differences between theory and observation in a model of ram-pressure driven clouds \citep{mqt05}.  The large dispersion ($200-300$~\kms) between the predicted and observed values suggests that the model is incorrect.  The smaller number of points in the $v_{max}^{ram~p}$ histogram results because $v_c$ must be available to compute this quantity.  See \S\ref{theory} for more details.}
\label{histdvth}
\end{figure}

To better incorporate column density and circular velocity information, we compute the expected characteristic and maximum velocities in this model for each galaxy.  We plot the histogram of the differences between expected and observed velocities in Figure \ref{histdvth}.  We find that the agreement between the model and observations is in general poor.  This is best indicated by the large dispersion ($200-300$~\kms) in the distributions.  Furthermore, in one-third to one-half of the cases the observed velocity is larger than the predicted velocity.  We thus conclude that we are missing important galaxy or wind properties that determine the cloud velocities, such as knowledge of the hot wind temperature in individual galaxies or the cloud initial radius.

In Figure \ref{theory_v}$b$, $d$, we plot lines for the case that an optically thick shell of gas is driven by radiation pressure.  In this case, the outflow velocity is linearly proportional to the galaxy's circular velocity, for a constant ratio of the galaxy's luminosity to the critical luminosity for driving of an outflow (eq. [17] of \citealt{mqt05}).  In most cases, the measured values of \dvmax\ fall between the lines for $L/L_{crit} = 1.05$ and 2.0, assuming that the observed clouds are at 10 times the initial radius.

The consequences of these results for distinguishing between
ram-pressure-driven and radiation-pressure-driven winds are ambiguous.
The ram pressure model clearly needs modification before it can be
matched to the data.  The radiation pressure model may fit the data, but it is underconstrained.

\subsubsection{Mass, Momenta, and Energy} \label{theory_mme_txt}

In \S\ref{mme_v}, we discuss the positive correlations we observe between outflow properties and galaxy properties.  Here, we address the interpretation of these correlations.

\begin{figure}[t]
\plotone{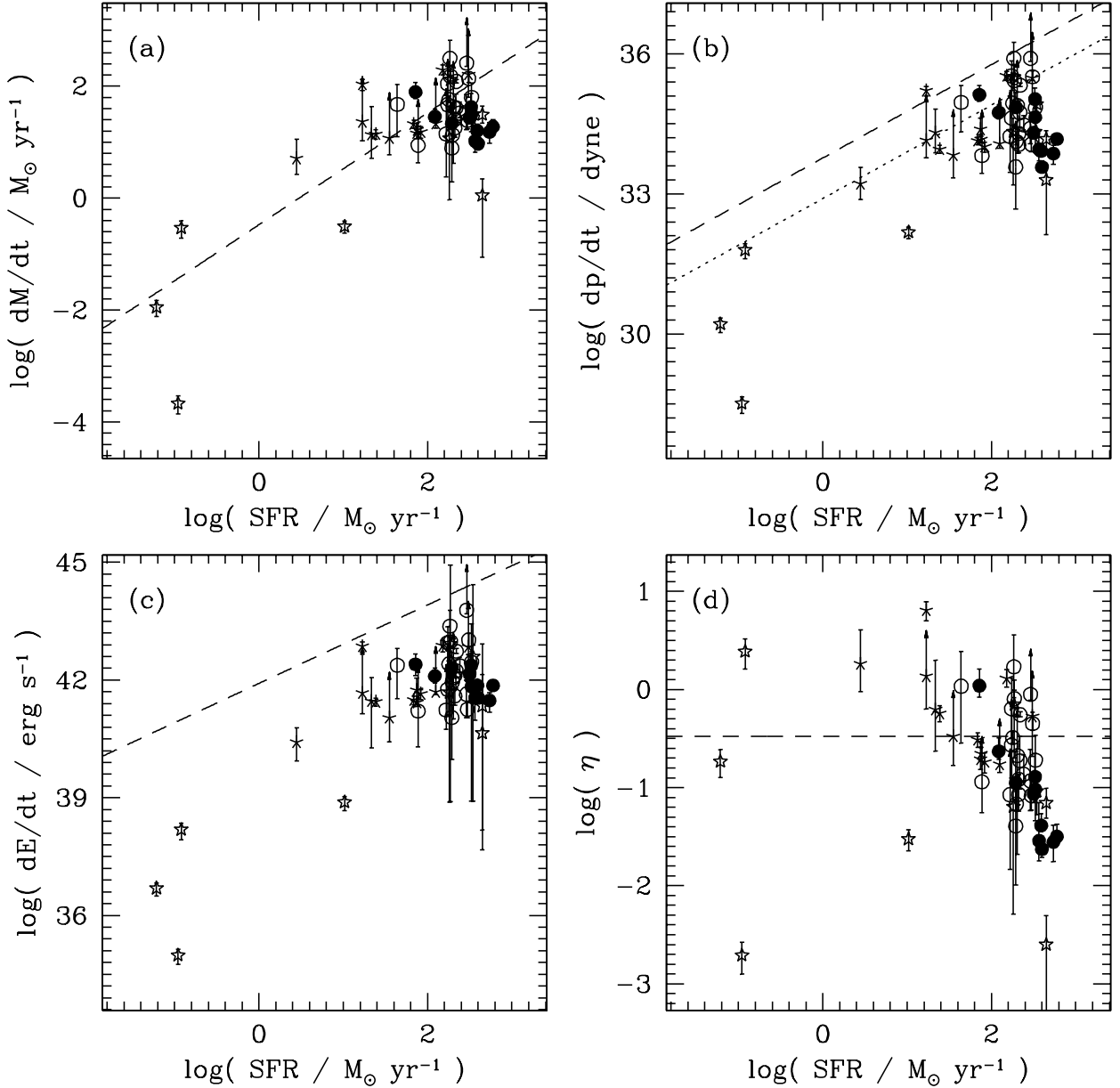}
\caption{\scriptsize{(a) Mass outflow rate, (b) momentum outflow rate, (c) kinetic energy outflow rate, and (d) mass entrainment efficiency as a function of star formation rate.   The dashed lines in $(a)-(c)$ are hot gas mass, momentum, and energy injection rates from stellar winds and supernovae for a continuous starburst of age $\ga$40~Myr, assuming twice solar metallicity and a Salpeter IMF ($1-100$~\msun; \citealt{l_ea99}).  We assume that these quantities are proportional to SFR.   Under these same assumptions, the dashed line in ($d$) is the `reheating efficiency,' or the normalized mass outflow rate in {\it hot} gas.  The dotted line in ($b$) is the momentum injection rate from radiation pressure, assuming isotropic absorption of the bolometric luminosity of the galaxy by optically thick clouds.  The mass of entrained neutral gas is on average a factor of 2 lower than the injected mass of hot gas.  The range of residuals is three orders of magnitude, however.  The observed momenta are more consistent with being driven by ram pressure than by radiation, since essentially all points fall below the predicted momenta in stellar winds and supernovae (as expected from conservation of momentum).  The range of $dp/dt$ implies that another gas phase carries most ($\ga$90\%) of the wind's momentum in most cases.  The observed energies are between 1 and 1000 times lower ($\sim$100 on average) than the energy in the hot wind.  Thus, if the wind is ram-pressure driven, it must be significantly cooled by radiation and/or another gas phase must carry most of the wind's energy.  The median observed mass entrainment efficiency in neutral gas (0.2) is comparable on average to the predicted reheating efficiency of hot gas (0.33), but the dispersion in $\eta$ is large (four orders of magnitude, or $0.001-10$).  The overall dependences of mass, momentum, and energy outflow rates on SFR are as expected (i.e., close to linear), except for $dE/dt$ ($dE/dt \sim$~SFR$^{1.6}$).  This implies an increase in the thermalization efficiency of the wind or the efficiency with which it accelerates cold gas clouds with SFR.  However, the flattening of these dependences at high SFR may be due to exhaustion of the available ambient gas, the inability of the gas clouds to accelerate beyond $\sim$600~\kms, and/or a decrease in thermalization efficiency at high SFR.  Mass entrainment efficiency is independent of SFR except for SFR~$\ga10$~\smpy, where $\eta \sim$~SFR$^{-0.5}$.  See \S\ref{theory} for more details.}}
\label{theory_mme}
\end{figure}

In Figure \ref{theory_mme}, we plot the mass, momentum, and energy injection rates ($dM_{hot}/dt$, $dp_{hot}/dt$, and $dE_{hot}/dt$) from stellar winds and supernovae as dashed lines.  We assume that these scale linearly with SFR, and we take the normalizations from the stellar synthesis models of \citet{l_ea99}.  These normalizations assume a continuous starburst of age $40$~Myr or more, a Salpeter IMF with mass limits 1 and 100~\msun, and twice solar metallicity (to match the average metallicity of our sample as computed in Paper I).  The dotted line in the plot of $dp/dt$ vs. SFR assumes optically thick radiation pressure driving over the same timescale.

The approximately linear proportionality of $M$, $dM/dt$, $p$, and $dp/dt$ on SFR (Table \ref{fittab}) is consistent with the neutral gas being driven by the hot wind fluid ejected by the starburst if the supernova rate scales with SFR.  The steeper dependence of $E$ and $dE/dt$ on SFR implies that galaxies with SFR $\sim$ 10$-$100~\smpy\ thermalize the kinetic energy of their supernovae more efficiently than dwarf starbursts and/or accelerate the cold clouds in the wind more efficiently.  The latter requires that there be fractionally more energy in other wind phases in the dwarf galaxies, since a smaller fraction of the available energy is transferred to the neutral gas clouds.  If our assumptions about the wind geometry are correct, then the increase with increasing SFR of thermalization efficiency and/or fraction of energy in the cold wind is $\sim$20 on average over two orders of magnitude in SFR.

The flattening of these relationships at the highest star formation rates (SFR~$\ga 10-100$~\smpy) implies some `saturation' effect.  The simplest explanation is that there is not enough interstellar mass in the galaxies with the highest SFR to keep up with the trend (see the plot of $M$ vs. SFR in Figure \ref{m_v}).  The outflowing gas masses in ULIRGs are $\ga$10$^9$~\msun, which is already $\ga$1\% of the dynamical mass of a ULIRG \citep{tg_ea02}.  Note that the maximum wind velocity also saturates at $\sim$600~\kms\ (Figure \ref{dvmax_v}).  There is more energy and momentum available in the hot wind at high SFR.  However, it doesn't move into the colder gas, either because there isn't enough gas to entrain or because the clouds can't be accelerated above a certain velocity \citep{mqt05,m05}.  Finally, the thermalization efficiency of the supernova kinetic energy may be smaller in the most luminous galaxies than in galaxies with moderate luminosity (by a factor of 10 on average).  This could result from the massive amounts of molecular gas in ULIRGs \citep{sss91,s_ea97}, which may absorb energy and momentum and inhibit the outflow.  The kinematics of this molecular phase may not be probed by our observations, and some of its energy will be radiated away due to its high density.

The strong dependence of these quantities on galactic mass is striking.  The increase of mass, momentum, and energy with $v_c$ implies that, contrary to expectations, the large potentials of massive galaxies {\it do not} inhibit the formation of strong, energetic winds (see also \S\ref{detrate_v}).  The fact that higher-mass galaxies have much larger and more energetic winds is related to the greater amount of star-forming and radiative energy available in massive galaxies, as well as the greater amount of ambient gas available for entrainment.  The flattening at the highest masses, as we discuss above, may imply a removal of most of the ambient gas in the way of the wind, and is also consistent with a velocity ceiling or a decrease in thermalization efficiency at the highest masses.

How do the magnitude of the predictions of the hot wind model compare with the data?  Comparing the mass in the hot and cold gas (using $dM/dt$), we see that the relative amount of entrained gas ($dM/dt_{cold}/dM/dt_{hot}$) is less than unity ($\sim$0.5) on average.  However, the range is quite large, from 0.001 to 10.  The ratio of gas mass in cold and hot gas is not typically well-known for even nearby galaxies; we show here that the range of values is probably large.  The momentum in the cold gas is always smaller or equal to the available momentum in the hot wind fluid, consistent with conservation of momentum in the ram pressure driven model.  Some galaxies appear to have most of their momentum in the cold, entrained gas.  However, on average the amount of momentum in the entrained gas is a factor of 10 less than the momentum injected into the hot wind fluid.  The range of $p_{cold}/p_{hot}$ is $\sim$0.001$-$1.  The theoretical plots of energy and $dE/dt$ vs. SFR show that the observed values are 1\% or less on average than the injected values (though again the range is large, from 0.01\% to almost 100\%).  This implies that the hot gas must carry a substantial fraction of the wind's energy and/or that the thermalization efficiency is low ($\sim$10\% or less).

\citet{s03} argues that the mass outflow rate of entrained gas may be proportional to the porosity of the ISM.  The porosity $Q$ is related to the filling factor of hot gas in the ISM by $f_{hot} \equiv 1-e^{-Q}$.  In this prescription, $dM/dt_{cold} = \beta f_{hot} \times \mathrm{SFR} $, where $\beta$ represents the amount of entrainment.  We find that $dM/dt_{cold} = 0.2~\mathrm{SFR}^{1.1\pm0.1}$ (see Figure \ref{dmdt_v} and Table \ref{fittab}), consistent with this interpretation.  Since we find that $dM/dt_{cold} \sim 0.5~dM/dt_{hot}$ on average, $\beta \sim 0.5$.  In this model, $f_{hot} \sim 0.4$, which means that the volume filled with hot gas is around 40\% of the total.  The porosity is thus $Q \sim 0.5$.  Note, however, that $\beta$ is subject to considerable uncertainty, and this result should be considered tentative.

How does our data distinguish between driving of the wind by a hot wind versus radiation pressure?  Figure \ref{theory_mme} shows the expected momentum injection rates from radiation pressure as a function of SFR.  These have the same linear dependence on SFR as the injection rates from stellar winds and supernovae, but the amount of momentum in radiation pressure is lower by a factor $\ga$10.  Thus, the winds with the highest momenta per unit SFR need some other source of momentum besides radiation pressure.  Winds with lower momenta may have some contribution to $p$ and $dp/dt$ from radiation pressure, however.

\subsubsection{Mass Entrainment Efficiency} \label{loweta} 

The `mass entrainment efficiency' is an important parameter as input to simple prescriptions in cosmological simulations, and is typically assumed to be of order unity \citep[e.g.,][]{kc98,a_ea01a,a_ea01b}.  Our results show that, on average, $\eta$ is up to an order of magnitude smaller than unity in IRGs and ULIRGs ($0.1-0.3$).  However, it is comparable to the `reheating efficiency' of the galaxy (equal to $dM/dt_{hot}$ / SFR), which is $\sim$0.33 for a starburst of age $\ga$40~Myr and twice solar metallicity with a Salpeter IMF (mass limits 1 and 100~\smpy; \citealt{l_ea99}).  The dispersion in $\eta$ is large, ranging from 0.001 to 10.

This quantity is independent of galactic mass, as shown by our data.  However, there is some evidence that it is not constant with star formation rate or $K$-band magnitude (Figure \ref{eta_v}$a$, $b$) for SFR $\ga$10~\smpy\ and $M_K < -24$, but decreases as $\eta \sim$ SFR$^{-0.5\pm0.2}$ and $\eta \sim M_K^{-0.4\pm0.1}$.  (Note that the dependence on SFR is not a 3$\sigma$ result and thus does not appear in Table \ref{fittab}.)  This decrease in $\eta$ may be due to the exhaustion of the gas reservoirs available for entrainment, the saturation of velocity at high SFR/luminosity, or a decrease in thermalization efficiency at high SFR/luminosity (\S\ref{theory_mme_txt}).

The normalization of $\eta$ is uncertain, due to possible errors in wind geometry and local physical conditions (\S\ref{mme}).  Changes in our assumptions act in different directions, however.  Assuming a thick instead of thin wind decreases $\eta$, while increasing the ionization fraction increase it.  If our values are correct, theorists typically overestimate the mass entrainment efficiency\citep[e.g.,][]{kc98,a_ea01a,a_ea01b,s03}.  Furthermore, cosmological simulations need to account for the large dispersion in $\eta$ (four orders of magnitude!) and the variation of $\eta$ with SFR that we observe, rather than assuming a single value.

\subsection{Superwinds in Mergers}

It has been postulated that superwinds play a role in the evolution of gas-rich mergers.  ULIRGs may evolve into quasars when a buried AGN turns on and breaks the obscuring screen of dust \citep[e.g.,][]{s_ea88,vks02}.  Given their high frequency of occurrence in ULIRGs, outflows could easily play a role in redistributing dust and gas, thereby increasing the escape fraction of nuclear continuum light.  If most ULIRGs do evolve into ellipticals, this gas redistribution may also destroy the central gas density spikes predicted in numerical simulations of mergers \citep{mh94}.  If left in place, these spikes would evolve into sharp upturns in the stellar surface brightness distribution in evolved ellipticals at small radii; breaks like this are not typical of elliptical surface brightness profiles \citep{hy99}.  The magnitudes of $dM/dt$ that we measure ($\la$100~\smpy\ on average) would be able to evacuate $10^{10}$~\msun\ of gas if they operate over $\ga$100~Myr (comparable to or greater than the gas consumption timescale in ULIRGs).  Roughly $10^{10}$~\msun\ of molecular gas are observed in the centers of ULIRGs \citep[e.g.,][]{sss91,s_ea97,ds98,hy99}.

Many of the best-studied and clearest examples of superwinds in the local universe are found in starbursting disk galaxies with \lir\ $< 10^{11}$~\lsun.  Given that many LIRGs, and most ULIRGs, are mergers in which the morphology and kinematics of the galaxy are highly disturbed \citep[e.g.,][]{kvs02,a_ea04,i04}, one might suppose that the physical picture of a symmetric, bipolar superwind along the galaxy's minor axis would not apply.  However, there is evidence from observations of resolved, infrared-luminous merging galaxies that these galaxies produce ordered superwind structures analogous to those in quiescent disk galaxies.  Examples include (in order of increasing \lir): NGC 520 (\citealt{hvg96}; \citealt{hvy00}), Arp 299 \citep{h_ea99,hvy00}, NGC 6240 \citep{ham87,v_ea03,g_ea04}, and Arp 220 \citep{ham87,h_ea96,hvy00,acc01,m_ea03}.  These galaxies all show spectacular signs of merging, including tidal tails and bridges and multiple nuclei.   NGC 520, Arp 299, and Arp 220 possess large-scale rotating \ion{H}{1} disks, and the \ion{H}{1} data for these galaxies show gaps along the minor axis and \ion{H}{1}-poor regions in the stellar tidal tails; this gas may have been evacuated by a superwind \citep{hvy00}.  Other evidence for superwinds comes from emission-line and X-ray data, which show minor-axis extensions, bubble-like and bow shock structures, kinematic broadening and/or line-splitting, shock-like emission line ratios, and/or tight correlations between optical and X-ray emitting filaments.  Most recently, diffuse and extended X-ray-emitting gas has been discovered in perhaps the best-studied merger, the Antennae; this gas may have a superwind origin \citep{f_ea04}.

However, the large opening angles we infer for ULIRG winds ($\co\ga0.7$) are not obviously consistent with the picture of a biconical superwind.  Alternatively, there may be multiple superwind episodes of this kind over a short period of time, which will fill out the area surrounding the galaxy.

\subsection{Redshift Evolution} \label{zevol}

One of the original purposes of this study was to explore the properties of superwinds deeper into redshift space.  Given that the number density of ULIRGs evolves strongly upward with increasing redshift \citep[e.g.,][]{ks98,c_ea03,cb_ea04,p_ea05}, we expect that winds from ULIRGs had a strong impact on the intergalactic medium and galaxy evolution at $z > 1$.  However, this assumes that the properties of winds in ULIRGs do not change with $z$.  Apart from our study, there are currently no observations of winds in ULIRGs outside of the local universe, except in a single hyperluminous infrared galaxy at high redshift \citep{s_ea03}.

We have partially accomplished this goal by observing a substantial number of galaxies at redshifts up to $z = 0.5$.  We do observe differences between our high-$z$ and low-$z$ subsamples.  Notably, winds are less frequently observed in the high-$z$ ULIRGs (with a detection rate of $46\pm14\%$ vs. $80\pm7\%$ for the low-$z$ ULIRGs), and less efficiently entrain gas (with a lower median $\eta$ than the low-$z$ ULIRGs by a factor of 2).  However, the high-$z$ galaxies have higher star formation rates (or equivalently, luminosities) than the low-$z$ ULIRGs, on average (389~\smpy\ and 225~\smpy, respectively).  This is a selection effect, since the most distant ULIRGs observed by the {\it Infrared Astronomical Satellite} ({\it IRAS}) are necessarily those with the highest intrinsic luminosities.

The differences in mass entrainment efficiency between the low- and high-$z$ subsamples are primarily a result of the variation of $\eta$ with SFR at high star formation rates (\S\ref{mme_v}), rather than redshift evolution.  In other words, the average mass entrainment efficiency in our high-$z$ ULIRGs is lower simply because they follow a trend of $\eta$ vs. SFR that is independent of redshift (at least for $z \la 1$).

The change in the rate of wind detection is less certainly attributable to SFR variations.  It is quite possible that the wind frequency of occurrence drops above a certain star formation rate.  Other factors may also play a role; for instance, more extended continuum light in a distant galaxy may leak into the slit and wash out the \nad\ absorption line.  The high-$z$ spectra generally have lower S/N than those of nearer galaxies, as well, diminishing our ability to detect \nad.  Finally, the global wind covering factor in the highest-luminosity starbursts may simply be lower (\S\ref{gcf}).

Assuming that the relationships between wind properties and galaxy properties, as well as the properties of ULIRG winds, remain constant with redshift, the cosmological impact of winds in ULIRGs depends largely on their density evolution.  Studies to date \citep[e.g.,][]{ks98,c_ea03,cb_ea04,p_ea05} show that the number density of ULIRGs rises strongly with increasing redshift and that they host a large fraction of star formation at $z > 1$ .  Thus, we expect that winds in ULIRGs, since they are common, massive, and energetic, have the potential to strongly impact the intergalactic and intracluster medium at redshifts $>$1.  Properly assessing their actual impact, however, will require a better understanding of how much of their gas escapes into the IGM/ICM.

The highest redshift winds (at $z \ga 3$) have higher velocities than those in galaxies of comparable SFR in the local universe (\S\ref{dv}).  However, their global covering factors may be similar (\S\ref{gcf}).  The higher velocities may reflect a difference in the structure of these early star forming galaxies or their surroundings from those of today.

\section{SUMMARY} \label{concl}

We have surveyed 78 starburst-dominated infrared-luminous galaxies, the largest systematic study to date of superwinds at $z \la 3$.  Our primary goal has been to study the detection rate and properties of winds as a function of the properties of the winds' host galaxies.  This provides insights into the physics and properties of superwinds that studies of individual galaxies do not allow.  Furthermore, by studying a large sample of ultra-luminous infrared galaxies (ULIRGs), we can quantify how superwinds affect galaxy evolution and the intergalactic medium at high redshift, where ULIRGs may host most of the star formation in the universe \citep{p_ea05}.

{\bf(1) Detection Rate.}  We find that superwinds are present in almost all infrared-luminous starburst galaxies.  Our detection rates are 43\% and 70\% for our IRG ($\langle \lir \rangle = 10^{11.36} \lsun$) subsample and ULIRGs, respectively.  However, these detection rates are lower limits to the actual frequency of occurrence of outflows in these galaxies and also to the opening angle of the outflows.  Because wind opening angles in local disk galaxies (as a fraction of 4$\pi$) equal the detection rate in our IRG subsample, we assume that these winds are found in almost all of our sample and that the detection rate exclusively reflects the wind geometry.  We thus compute that the global covering factor of neutral gas outflows is higher in ULIRGs ($\Omega/4\pi\sim0.3$) than in LIRGs ($\Omega/4\pi\sim0.15$).  The opening angle of high-$z$ starbursts is uncertain, but may be close to several tenths of 4$\pi$, similar to the IRGs \citealt{ad_ea03,ad_ea05}.

At the highest star formation rates, the detection rate in ULIRGs decreases (from 80\% to 46\%).  However, the overall trend is for detection rate to increase with SFR.  Detection rate does {\it not} depend on optical or near-infrared luminosity, galactic mass, spectral type, or on the merger stage in ULIRGs.

Some of the absorption we detect is related to overlapping gas disks in a multiple system or to tidal debris; these components are removed from the formal analysis.  In a handful of cases we can demonstrate that these explanations are more likely than the outflow hypothesis.  However, most alternative explanations should produce a roughly symmetric velocity distribution or one skewed to the red, and the one that we observe in our data has a strong blueward asymmetry.  The number of outflowing components that we can attribute to these alternative explanations is at most a mirror image of the distribution of red components, and thus much less than the total number of blueshifted components.

{\bf(2) Velocity.}  The maximum velocities for these winds range from 150 to 600~\kms\ (and up to 1100~\kms\ in a single case), with a median value of 350~\kms.  The distribution of velocities is not consistent with a simple constant-velocity model of projection effects.  The velocity of the gas with the highest column density in each galaxy has a median value of 140~\kms, less than half that of \dvmax.  The average projected velocities of ionized gas in local edge-on starbursts of SFR comparable to the IRGs ($\sim$170~\kms; \citealt{lh96}) are higher than the average value of \dvtau\ in IRGs (100~\kms).  High-redshift starbursts of star formation rate similar to the IRGs have higher mean velocities than both the IRGs {\it and} the ULIRGs: $\sim$300~\kms, vs. $\la$200~\kms\ \citep{ad_ea03,ss_ea03}.

The maximum velocities we measure are often close to or slightly larger than the escape velocity of the galaxy.  The escape fraction of neutral gas is thus non-zero when averaged over the entire sample: $f_{esc} \sim 5-20\%$.  Furthermore, we often observe higher velocities in the ionized gas (as probed by emission lines), which has velocities exceeding 1000~\kms\ in several galaxies.

We show that LINERs have higher median values of \dvmax\ and Doppler width $b$ (by $\sim$100~\kms) than \ion{H}{2} galaxies.  Thus, the spectral classification for these systems, based on line ratios, may simply reflect high-velocity vs. low-velocity shocks.  We also show that for ULIRGs, wind velocity may increase as the merger progresses (though the statistics are admittedly low).

The outflow velocities in starbursts rise slowly with star formation rate, luminosity, and galactic mass when we consider both dwarf starbursts \citep{sm04} and luminous galaxies.  However, in our sample alone, velocity is independent of galaxy properties.  This is consistent with a limit to the terminal velocity of neutral gas clouds above an SFR of $10-100$~\smpy\ \citep{mqt05,m05}.  However, detailed models of ram pressure driven clouds which incorporate this limit are not consistent with our data \citep{mqt05}.  Models of radiation pressure driven gas are presently underconstrained \citep{mqt05}.

{\bf(3) Mass, Momentum, Energy.}  We find that mass, momentum, energy, and their outflow rates depend on star formation rate roughly linearly, though these relationships flatten at high SFR ($\ga$10$-$100~\smpy).  This dependence is consistent with ram pressure driving of a hot gas or radiation pressure driving, since both the supernova rate and the galaxy luminosity scale with SFR.  Modulo the flattening at high SFR, the dependence of energy on star formation rate is SFR$^{1.6}$, suggesting that galaxies with SFR~$\sim$~10$-$100~\smpy\ have higher thermalization efficiency than lower-SFR galaxies, or accelerate the cold gas more effectively and thus have fractionally less energy in other wind phases.

These wind properties correlate also with on galaxy luminosity and mass.  The relationships to galactic mass are quite strong, with power law slopes of $1.5-2.5$.  Thus, the wind properties are sensitive to the galaxy mass.  However, the winds get larger and more powerful with increasing mass, rather than smaller.

At high SFR, luminosity, and galactic mass, we observe a flattening in the increase of wind properties (velocity, mass, momentum, and energy) with galaxy properties.  This flattening may be due to entrainment of all the available gas clouds, the existence of a terminal velocity in ULIRGs above which the wind cannot be accelerated, and/or a reduction in thermalization efficiency at high SFR (perhaps due to the interception of the wind by dense ambient material, like the massive quantities of molecular gas feeding the starburst).

The magnitudes of mass, momentum, energy, and their outflow rates in neutral gas are compared to those in the hot gas, as predicted by starburst models \citep{l_ea99}.  We find that the range of entrainment factors in the wind, $M_{cold} / M_{hot}$, is $\sim$0.01$-$10, with an average of a few.  The measured momenta are, on average, a factor of $\sim$10 less than, and rarely greater than, the momentum in the hot wind.  They are sometimes greater than the momentum available from radiation pressure, however, suggesting that, in many galaxies, the hot wind is likely to dominate the driving of the wind rather than radiation pressure.  The measured energies in the neutral gas are significantly less than the energy in the hot wind on average, by factors of $\sim$1$-$1000; thus, the hot wind must carry most of the wind's energy and/or the thermalization efficiency must on average be small ($\la$10\%), especially in dwarf galaxies.

The mass entrainment efficiency, $\eta\equiv (dM/dt) /$SFR, is found to range from $0.001-10$ but is roughly $0.1$ on average.  These numbers, however, are sensitive to the wind's geometry and the local physics.  This quantity is independent of galaxy properties on average but declines with star formation rate at SFR~$\ga$~10$-$100~\smpy\ and with $M_K$ at $M_K < -24$.  The large observed range of $\eta$ and the dependence on SFR/luminosity show that the prescriptions and assumptions employed in numerical simulations of galaxy formation are not completely correct.  These prescriptions typically assume a constant $\eta$ of approximately unity.

{\bf(4) Outlook}.  There is evidence for starburst-driven superwinds in local infrared-luminous mergers such as NGC 6240.  However, the complex kinematics of the ionized gas in these galaxies make it sometimes difficult to distinguish starburst-driven winds from tidal motions directly produced by the interaction.  In this sense, our data provides the most unambiguous evidence to date for frequent, massive, and energetic outflows in these systems.

To interpret our data more completely, however, we require better knowledge of local merging systems.  Hydrodynamic modeling of winds in evolving environments such as these would allow comparison of observational data to numerical models, as is possible in more quiescent systems.  More sensitive observations to probe the morphology and kinematics of faint structures in these galaxies, and three-dimensional kinematic observations of multiple gas phases will also help to disentangle the motions of different components (e.g., tidal tails vs. expanding superbubbles) as has been done for Arp 220 \citep{acc01,m_ea03,cac04}.

A better understanding of the neutral phase of these outflows is also needed, especially its geometry and small-scale structure.  We need to know how the clouds in these winds are distributed, and how they stand up under radiation, evaporation, and shock ablation.  Understanding these better will enable us to more precisely measure the properties and fate of these clouds from global measurements.

Many ULIRGs have infrared luminosities which are dominated by dust-reprocessed radiation from an AGN, rather than a starburst.  AGN also power outflows, and it would be instructive to compare the properties of outflows in starburst-dominated ULIRGs and AGN-dominated ULIRGs.  In a forthcoming paper, we present \nad\ observations of 26 ULIRGs with Seyfert nuclei and perform this comparison \citep{rvs05b}.

\acknowledgments

We thank Kalliopi Dasyra and Dong Chan Kim for supplying useful data prior to publication.  DSR is supported by NSF/CAREER grant AST-9874973.  SV is grateful for partial support of this research by a Cottrell Scholarship awarded by the Research Corporation, NASA/LTSA grant NAG 56547, and NSF/CAREER grant AST-9874973.  This research has made use of the NASA/IPAC Extragalactic Database (NED), which is operated by the JPL, Caltech, under contract with the NASA.  It also makes use of data products from the Two Micron All Sky Survey, which is a joint project of the University of Massachusetts and IPAC/Caltech, funded by the NASA and the NSF.  The authors wish to recognize and acknowledge the very significant cultural role and reverence that the summit of Mauna Kea has always had within the indigenous Hawaiian community.  We are most fortunate to have the opportunity to conduct observations from this mountain.

\clearpage

\clearpage

\epsscale{0.8}

\epsscale{1.0}

\epsscale{0.8}

\epsscale{1.0}

\clearpage

\begin{deluxetable}{cccc}
\tablecaption{Subsample Average Properties \label{avgprop}}
\tablewidth{0pt}
\tablehead{
\colhead{Quantity} & \colhead{IRGs}  & \colhead{low-$z$ ULIRGs} & \colhead{high-$z$ ULIRGs} \\
\colhead{(1)} & \colhead{(2)} & \colhead{(3)} & \colhead{(4)}
}
\startdata
Number of Galaxies & 35 & 30 & 13 \\
Detection Rate (\%) & (42$\pm$8)$\%$ & (80$\pm$7)$\%$ & (46$\pm$13)$\%$ \\
\tableline
Galaxy Properties & & & \\
\tableline
$ z $ & $ 0.031^{+0.04}_{-0.02} $ & $ 0.129^{+0.07}_{-0.04} $ & $ 0.360^{+0.07}_{-0.06} $ \\
$ \mathrm{log} [\lir/\lsun] $ & $11.36\pm0.4$ & $12.21\pm0.2$ & $12.45\pm0.2$ \\
$ \mathrm{SFR}~(\smpy) $ & $40^{+55}_{-23}$ & $225^{+95}_{-67}$ & $389^{+212}_{-137}$ \\
$ \Delta v_{max}~(\kms) $ & $301^{+145}_{-98}$ & $408^{+224}_{-144}$ & $359^{+119}_{-90}$ \\
$ \Delta v_{maxN}~(\kms) $ & $104^{+80}_{-45}$ & $167^{+122}_{-70}$ & $161^{+219}_{-93}$ \\
$ \mathrm{log} [N$(\ion{Na}{1})/cm$^{-2}$] & $13.8\pm0.2$ & $13.8\pm0.4$ & $13.5\pm0.5$ \\
$ \mathrm{log} [N$(H)/cm$^{-2}$] & $21.2\pm0.3$ & $21.2\pm0.4$ & $20.8\pm0.5$ \\
$ \mathrm{log} [M/\msun]$ & $8.8\pm0.2$ & $9.1\pm0.3$ & $8.9\pm0.3$ \\
$ dM/dt~(\smpy) $ & $17^{+20}_{-9}$ & $42^{+82}_{-28}$ & $24^{+17}_{-10}$ \\
$ \mathrm{log} [p/\mathrm{dyne~s}]$ & $49.2\pm0.3$ & $49.6\pm0.5$ & $49.4\pm0.2$ \\
$ \mathrm{log} [dp/dt/\mathrm{dyne}]$ & $34.1\pm0.5$ & $34.7\pm0.7$ & $34.5\pm0.6$ \\
$ \mathrm{log} [E/\mathrm{erg}]$ & $56.7\pm0.4$ & $57.2\pm0.5$ & $57.0\pm0.2$ \\
$ \mathrm{log} [dE/dt/\mathrm{erg~s^{-1}}]$ & $41.6\pm0.6$ & $42.2\pm0.7$ & $42.0\pm0.3$ \\
$ \eta $ & $0.33^{+0.7}_{-0.2}$ & $0.19^{+0.5}_{-0.1}$ & $0.09^{+0.1}_{-0.0}$ \\
\tableline
Component Properties & & & \\
\tableline
$ \tau $ & $1.06^{+1.4}_{-0.6}$ & $0.85^{+2.4}_{-0.6}$ & $0.34^{+0.8}_{-0.2}$ \\
$ b~(\kms) $ & $152^{+109}_{-64}$ & $196^{+170}_{-91}$ & $182^{+129}_{-75}$ \\
$ C_f $ & $0.37^{+0.2}_{-0.1}$ & $0.40^{+0.5}_{-0.2}$ & $0.45^{+0.6}_{-0.3}$ \\
\enddata
\tablecomments{For most quantities we list the median and 1$\sigma$ dispersions, under the assumption of a Gaussian distribution in the log of the quantity.  For the detection rate errors, we assume a binomial distribution.  Statistics for all quantities except $z$, \lir, and SFR are computed only for galaxies or velocity components with outflows.}
\end{deluxetable}

\begin{deluxetable}{lccrrrrrrrrrr}
\rotate
\tabletypesize{\footnotesize}
\tablecaption{Galaxy and Outflow Properties \label{objprop}}
\tablewidth{0pt}
\tablehead{
\colhead{Name} & \colhead{$z$} & \colhead{\lir} & \colhead{SFR} & \colhead{\dvtau} & \colhead{\dvmax} & \colhead{$M$} & \colhead{$dM/dt$} & \colhead{$p$} & \colhead{$dp/dt$} & \colhead{$E$} & \colhead{$dE/dt$} & \colhead{$\eta$} \\
\colhead{(1)} & \colhead{(2)} & \colhead{(3)} & \colhead{(4)} & \colhead{(5)} & \colhead{(6)} & \colhead{(7)} & \colhead{(8)} & \colhead{(9)} & \colhead{(10)} & \colhead{(11)} & \colhead{(12)} & \colhead{(13)}
}
\startdata
& & & & & {\bf IRGs} & & & & & & & \\
\tableline
       F00521$+$2858 &  0.0155 &   10.84 &      12 &       0 &       0 &    0.00 &    0.00 &    0.00 &    0.00 &    0.00 &    0.00 &   0.000 \\
       Z01092$-$0139 &  0.1533 &   11.65 &      77 &    -119 &    -261 & $>$8.56 & $>$0.95 &$>$48.93 &$>$33.82 &$>$56.32 &$>$41.21 &$>$0.115 \\
       F01250$-$0848 &  0.0488 &   11.68 &      83 &    -107 &    -301 &    8.84 &    1.18 &   49.17 &   34.01 &   56.81 &   41.65 &   0.183 \\
     F01417$+$1651:N &  0.0276 &   11.56 &      63 &       0 &       0 &    0.00 &    0.00 &    0.00 &    0.00 &    0.00 &    0.00 &   0.000 \\
     F01417$+$1651:S &  0.0271 & \nodata & \nodata & \nodata & \nodata & \nodata & \nodata & \nodata & \nodata & \nodata & \nodata & \nodata \\
       F01484$+$2220 &  0.0323 &   11.64 &      75 &    -139 &    -260 &    8.77 &    1.22 &   49.21 &   34.16 &   56.47 &   41.43 &   0.221 \\
    F02114$+$0456:SW &  0.0296 &   11.42 &      45 &       0 &       0 &    0.00 &    0.00 &    0.00 &    0.00 &    0.00 &    0.00 &   0.000 \\
       F02433$+$1544 &  0.0254 &   10.99 &      17 &     -97 &    -257 & $>$9.07 & $>$1.36 &$>$49.35 &$>$34.15 &$>$56.87 &$>$41.67 &$>$1.369 \\
       F02437$+$2122 &  0.0234 &   11.10 &      22 &    -240 &    -309 &    8.44 &    1.13 &   49.12 &   34.31 &   56.27 &   41.46 &   0.625 \\
       F02509$+$1248 &  0.0122 &   10.85 &      12 &       0 &       0 &    0.00 &    0.00 &    0.00 &    0.00 &    0.00 &    0.00 &   0.000 \\
     F02512$+$1446:S &  0.0315 &   11.63 &      74 &     -90 &    -476 &    8.65 &    1.16 &   49.15 &   34.38 &   56.54 &   41.74 &   0.194 \\
       Z03009$-$0213 &  0.1191 &   11.36 &      40 &       0 &       0 &    0.00 &    0.00 &    0.00 &    0.00 &    0.00 &    0.00 &   0.000 \\
               F1\_5 &  0.4786 &   11.85 &     122 &     -78 &    -476 &    8.84 &    1.46 &   49.44 &   34.74 &   56.86 &   42.10 &   0.234 \\
       F03359$+$1523 &  0.0357 &   11.48 &      52 &       0 &       0 &    0.00 &    0.00 &    0.00 &    0.00 &    0.00 &    0.00 &   0.000 \\
       F03514$+$1546 &  0.0222 &   11.15 &      24 &    -103 &    -260 &    8.82 &    1.14 &   49.13 &   33.96 &   56.62 &   41.45 &   0.571 \\
       F04097$+$0525 &  0.0179 &   11.13 &      23 &       0 &       0 &    0.00 &    0.00 &    0.00 &    0.00 &    0.00 &    0.00 &   0.000 \\
       F04315$-$0840 &  0.0159 &   11.60 &      69 &    -104 &    -232 &    9.00 &    1.33 &   49.31 &   34.14 &   56.66 &   41.49 &   0.308 \\
       F04326$+$1904 &  0.0247 &   11.44 &      48 &       0 &       0 &    0.00 &    0.00 &    0.00 &    0.00 &    0.00 &    0.00 &   0.000 \\
       F05187$-$1017 &  0.0285 &   11.23 &      29 &       0 &       0 &    0.00 &    0.00 &    0.00 &    0.00 &    0.00 &    0.00 &   0.000 \\
       F08354$+$2555 &  0.0182 &   11.55 &      61 &       0 &       0 &    0.00 &    0.00 &    0.00 &    0.00 &    0.00 &    0.00 &   0.000 \\
       F08498$+$3513 &  0.1895 &   11.75 &      97 &       0 &       0 &    0.00 &    0.00 &    0.00 &    0.00 &    0.00 &    0.00 &   0.000 \\
       F09120$+$4107 &  0.0085 &   10.70 &       9 &       0 &       0 &    0.00 &    0.00 &    0.00 &    0.00 &    0.00 &    0.00 &   0.000 \\
       F09320$+$6134 &  0.0395 &   11.96 &     157 &       0 &       0 &    0.00 &    0.00 &    0.00 &    0.00 &    0.00 &    0.00 &   0.000 \\
       F10015$-$0614 &  0.0166 &   11.31 &      35 &     -57 &    -182 & $>$8.88 & $>$1.07 &$>$49.06 &$>$33.83 &$>$56.39 &$>$41.04 &$>$0.334 \\
       F13532$+$2517 &  0.0293 &   10.99 &      17 &    -238 &    -506 &    9.35 &    2.04 &   50.02 &   35.21 &   57.67 &   42.86 &   6.441 \\
       F13565$+$3519 &  0.0347 &   11.33 &      37 &       0 &       0 &    0.00 &    0.00 &    0.00 &    0.00 &    0.00 &    0.00 &   0.000 \\
       F15364$+$3320 &  0.0222 &   10.21 &       3 &     -51 &    -130 &    8.69 &    0.71 &   48.70 &   33.22 &   55.89 &   40.42 &   1.827 \\
       F15386$+$3807 &  0.1828 &   11.62 &      72 &    -331 &    -433 &    9.24 &    1.90 &   49.88 &   35.12 &   57.13 &   42.40 &   1.094 \\
       F15549$+$4201 &  0.0348 &   11.10 &      22 &       0 &       0 &    0.00 &    0.00 &    0.00 &    0.00 &    0.00 &    0.00 &   0.000 \\
       F16130$+$2725 &  0.0459 &   10.68 &       8 &       0 &       0 &    0.00 &    0.00 &    0.00 &    0.00 &    0.00 &    0.00 &   0.000 \\
       F16504$+$0228 &  0.0243 &   11.86 &     125 &     -99 &    -317 & $>$9.10 & $>$1.33 &$>$49.32 &$>$34.07 &$>$57.01 &$>$41.70 &$>$0.173 \\
       F21484$-$1314 &  0.0768 &   11.50 &      55 &       0 &       0 &    0.00 &    0.00 &    0.00 &    0.00 &    0.00 &    0.00 &   0.000 \\
       F21549$-$1206 &  0.0511 &   11.19 &      27 &       0 &       0 &    0.00 &    0.00 &    0.00 &    0.00 &    0.00 &    0.00 &   0.000 \\
       F22213$-$0238 &  0.0566 &   11.33 &      37 &       0 &       0 &    0.00 &    0.00 &    0.00 &    0.00 &    0.00 &    0.00 &   0.000 \\
       F22220$-$0825 &  0.0604 &   11.40 &      44 &    -305 &    -482 &    8.88 &    1.68 &   49.66 &   34.96 &   57.08 &   42.38 &   1.081 \\
       F22338$-$1015 &  0.0623 &   11.36 &      40 &       0 &       0 &    0.00 &    0.00 &    0.00 &    0.00 &    0.00 &    0.00 &   0.000 \\
\tableline
 & & & & & {\bf low-$z$} & {\bf ULIRGs} & & & & & & \\
\tableline
       F00188$-$0856 &  0.1283 &   12.38 &     333 &    -176 &    -402 &    9.25 &    1.80 &   49.79 &   34.85 &   57.39 &   42.45 &   0.190 \\
       F01298$-$0744 &  0.1361 &   12.35 &     308 &    -392 &    -611 &    9.31 &    2.14 &   50.13 &   35.51 &   57.63 &   43.02 &   0.449 \\
    F02411$+$0353:SW &  0.1434 &   12.24 &     240 &       0 &       0 &    0.00 &    0.00 &    0.00 &    0.00 &    0.00 &    0.00 &   0.000 \\
    F02411$+$0353:NE &  0.1441 & \nodata & \nodata & \nodata & \nodata & \nodata & \nodata & \nodata & \nodata & \nodata & \nodata & \nodata \\
       F03250$+$1606 &  0.1290 &   12.13 &     185 &    -442 &    -606 &    9.67 &    2.50 &   50.49 &   35.90 &   57.99 &   43.37 &   1.708 \\
       F08231$+$3052 &  0.2478 &   12.32 &     288 &       0 &       0 &    0.00 &    0.00 &    0.00 &    0.00 &    0.00 &    0.00 &   0.000 \\
       F08474$+$1813 &  0.1454 &   12.22 &     230 &       0 &       0 &    0.00 &    0.00 &    0.00 &    0.00 &    0.00 &    0.00 &   0.000 \\
       F08591$+$5248 &  0.1574 &   12.24 &     241 &    -144 &    -449 &    9.05 &    1.52 &   49.51 &   34.48 &   57.39 &   42.36 &   0.137 \\
       F09039$+$0503 &  0.1252 &   12.10 &     173 &    -513 &    -582 &    9.08 &    2.04 &   50.03 &   35.53 &   57.46 &   42.96 &   0.638 \\
       F09116$+$0334 &  0.1454 &   12.18 &     211 &    -122 &    -382 &    9.01 &    1.41 &   49.40 &   34.29 &   57.22 &   42.12 &   0.121 \\
       F09539$+$0857 &  0.1290 &   12.13 &     187 &    -282 &    -518 &    9.42 &    2.18 &   50.17 &   35.43 &   57.71 &   42.98 &   0.805 \\
       F10091$+$4704 &  0.2451 &   12.64 &     597 &    -125 &    -347 &    8.87 &    1.28 &   49.27 &   34.18 &   56.95 &   41.86 &   0.032 \\
     F10190$+$1322:W &  0.0766 &   12.00 &     139 &       0 &       0 &    0.00 &    0.00 &    0.00 &    0.00 &    0.00 &    0.00 &   0.000 \\
     F10190$+$1322:E &  0.0759 & \nodata & \nodata & \nodata & \nodata & \nodata & \nodata & \nodata & \nodata & \nodata & \nodata & \nodata \\
       F10378$+$1108 &  0.1363 &   12.32 &     291 &    -243 &   -1106 & $>$9.46 & $>$2.41 &$>$50.40 &$>$35.90 &$>$58.24 &$>$43.78 &$>$0.893 \\
       F10494$+$4424 &  0.0919 &   12.17 &     203 &    -286 &    -319 & $>$8.87 & $>$1.63 &$>$49.62 &$>$34.89 &$>$56.79 &$>$42.06 &$>$0.212 \\
       F10565$+$2448 &  0.0430 &   12.04 &     151 &    -309 &    -460 &    9.59 &    2.29 &   50.28 &   35.53 &   57.61 &   42.88 &   1.296 \\
       F11028$+$3130 &  0.1986 &   12.43 &     368 &       0 &       0 &    0.00 &    0.00 &    0.00 &    0.00 &    0.00 &    0.00 &   0.000 \\
       F11387$+$4116 &  0.1489 &   12.20 &     219 &    -158 &    -559 &    9.45 &    2.08 &   50.07 &   35.33 &   57.48 &   42.73 &   0.553 \\
       F11506$+$1331 &  0.1274 &   12.33 &     297 &     -74 &    -164 &    9.21 &    1.39 &   49.38 &   34.06 &   56.57 &   41.25 &   0.083 \\
       F11582$+$3020 &  0.2234 &   12.60 &     544 &     -76 &    -238 &    8.99 &    1.18 &   49.17 &   33.86 &   56.79 &   41.48 &   0.028 \\
       F14060$+$2919 &  0.1169 &   12.10 &     173 &    -142 &    -236 & $>$9.21 & $>$1.67 &$>$49.66 &$>$34.62 &$>$56.80 &$>$41.76 &$>$0.270 \\
       F14197$+$0813 &  0.1305 &   12.16 &     197 &    -158 &    -452 &    8.62 &    1.13 &   49.11 &   34.12 &   56.94 &   41.95 &   0.068 \\
       F15206$+$3342 &  0.1254 &   12.20 &     218 &    -213 &    -416 &    8.97 &    1.61 &   49.60 &   34.74 &   57.10 &   42.24 &   0.188 \\
     F16333$+$4630:W &  0.1908 &   12.43 &     367 &    -136 &    -336 &    8.58 &    1.02 &   49.01 &   33.96 &   56.60 &   41.54 &   0.029 \\
     F16333$+$4630:E &  0.1908 & \nodata & \nodata & \nodata & \nodata & \nodata & \nodata & \nodata & \nodata & \nodata & \nodata & \nodata \\
     F16474$+$3430:S &  0.1121 &   12.18 &     211 &    -101 &    -267 &    8.94 &    1.25 &   49.24 &   34.06 &   56.78 &   41.60 &   0.085 \\
     F16474$+$3430:N &  0.1126 & \nodata & \nodata & \nodata & \nodata & \nodata & \nodata & \nodata & \nodata & \nodata & \nodata & \nodata \\
     F16487$+$5447:W &  0.1038 &   12.15 &     194 &     -76 &    -210 & $>$8.70 & $>$0.90 &$>$48.88 &$>$33.58 &$>$56.35 &$>$41.04 &$>$0.041 \\
     F16487$+$5447:E &  0.1038 & \nodata & \nodata & \nodata & \nodata & \nodata & \nodata & \nodata & \nodata & \nodata & \nodata & \nodata \\
       F17068$+$4027 &  0.1794 &   12.38 &     334 &       0 &       0 &    0.00 &    0.00 &    0.00 &    0.00 &    0.00 &    0.00 &   0.000 \\
       F17207$-$0014 &  0.0428 &   12.35 &     306 &    -298 &    -446 & $>$9.43 & $>$2.21 &$>$50.20 &$>$35.49 &$>$57.56 &$>$42.84 &$>$0.532 \\
     I20046$-$0623:W &  0.0840 &   12.08 &     166 &    -191 &    -222 & $>$8.56 & $>$1.15 &$>$49.14 &$>$34.23 &$>$56.14 &$>$41.23 &$>$0.085 \\
     I20046$-$0623:E &  0.0847 & \nodata & \nodata & \nodata & \nodata & \nodata & \nodata & \nodata & \nodata & \nodata & \nodata & \nodata \\
       F20414$-$1651 &  0.0872 &   12.32 &     290 &    -119 &    -230 &    9.14 &    1.53 &   49.52 &   34.40 &   56.75 &   41.64 &   0.117 \\
     F23234$+$0946:W &  0.1279 &   12.11 &     179 &    -240 &    -436 &    9.07 &    1.76 &   49.75 &   34.94 &   57.22 &   42.41 &   0.324 \\
     F23234$+$0946:E &  0.1277 & \nodata & \nodata & \nodata & \nodata & \nodata & \nodata & \nodata & \nodata & \nodata & \nodata & \nodata \\
\tableline
 & & & & & {\bf high-$z$} & {\bf ULIRGs} & & & & & & \\
\tableline
       F01462$+$0014 &  0.2797 &   12.34 &     302 &       0 &       0 &    0.00 &    0.00 &    0.00 &    0.00 &    0.00 &    0.00 &   0.000 \\
       Z02376$-$0054 &  0.4104 &   12.64 &     595 &       0 &       0 &    0.00 &    0.00 &    0.00 &    0.00 &    0.00 &    0.00 &   0.000 \\
       Z03151$-$0140 &  0.2653 &   12.15 &     195 &    -518 &    -618 &    8.31 &    1.34 &   49.33 &   34.85 &   56.78 &   42.30 &   0.112 \\
       F04313$-$1649 &  0.2672 &   12.66 &     626 &       0 &       0 &    0.00 &    0.00 &    0.00 &    0.00 &    0.00 &    0.00 &   0.000 \\
       F07353$+$2903 &  0.3348 &   12.36 &     316 &    -118 &    -384 &    9.05 &    1.44 &   49.42 &   34.31 &   57.27 &   42.16 &   0.087 \\
     F07449$+$3350:W &  0.3571 &   12.75 &     776 &       0 &       0 &    0.00 &    0.00 &    0.00 &    0.00 &    0.00 &    0.00 &   0.000 \\
     F07449$+$3350:E &  0.3571 & \nodata & \nodata & \nodata & \nodata & \nodata & \nodata & \nodata & \nodata & \nodata & \nodata & \nodata \\
       F08136$+$3110 &  0.4070 &   12.46 &     398 &     -63 &    -294 &    8.86 &    0.97 &   48.96 &   33.57 &   56.94 &   41.55 &   0.024 \\
       F08143$+$3134 &  0.3606 &   12.38 &     329 &    -400 &    -489 &    8.72 &    1.63 &   49.62 &   35.03 &   56.96 &   42.38 &   0.129 \\
       F08208$+$3211 &  0.3955 &   12.49 &     426 &       0 &       0 &    0.00 &    0.00 &    0.00 &    0.00 &    0.00 &    0.00 &   0.000 \\
       F09567$+$4119 &  0.3605 &   12.45 &     389 &     -82 &    -335 &    8.98 &    1.20 &   49.19 &   33.92 &   57.14 &   41.87 &   0.041 \\
       F10156$+$3705 &  0.4895 &   12.82 &     908 &       0 &       0 &    0.00 &    0.00 &    0.00 &    0.00 &    0.00 &    0.00 &   0.000 \\
       F10485$+$3726 &  0.3559 &   12.34 &     303 &       0 &       0 &    0.00 &    0.00 &    0.00 &    0.00 &    0.00 &    0.00 &   0.000 \\
       F16576$+$3553 &  0.3710 &   12.38 &     331 &    -219 &    -315 &    8.85 &    1.50 &   49.49 &   34.64 &   56.68 &   41.83 &   0.096 \\
\enddata
\tablecomments{ Col.(1): {\it IRAS} Faint Source Catalog label, plus nuclear ID (e.g., N $=$ north).  Only 1 object is not found in the FSC.  Col.(2): Heliocentric redshift (see Paper I).  Col.(3): Infrared luminosity, in logarithmic units of \lsun.  Col.(4): Star formation rate, computed from the infrared luminosity using a correction for AGN contribution to \lir\ (\S\ref{eta}).  Col.(5): Velocity of the highest column density gas in the outflow, \dvtau, in \kms\ (\S\ref{dv}).  Col.(6): Maximum velocity in the outflow, $\dvmax \equiv \Delta v - \mathrm{FWHM}/2$, in \kms\ (\S\ref{dv}).  Col.(7): Log of mass, in \msun\ (\S\ref{eqns}).  Col.(8): Log of mass outflow rate, in \smpy\ (\S\ref{eqns}).  Col.(9): Log of momentum, in dyne~s (\S\ref{eqns}).  Col.(10): Log of momentum outflow rate, in dyne (\S\ref{eqns}).  Col.(11): Log of total kinetic energy, in erg (\S\ref{eqns}).  Col.(12): Log of energy outflow rate, in erg~s$^{-1}$ (\S\ref{eqns}).  Col.(13): Mass entrainment efficiency; $\eta \equiv (dM/dt) / \mathrm{SFR}$ (\S\ref{eta}). }
\end{deluxetable}

\begin{deluxetable}{rlrrr}
\tablecaption{Results of Fitting \label{fittab}}
\tabletypesize{\scriptsize}
\tablewidth{0pt}
\tablehead{
\colhead{Axes} & \colhead{$N$} & \colhead{$a\pm\delta a$} & \colhead{$r_p$} & \colhead{$r_s$} \\
\colhead{(1)} & \colhead{(2)} & \colhead{(3)} & \colhead{(4)} & \colhead{(5)}
}
\startdata
      $N$(\ion{Na}{1}) vs. SFR &  55     &  0.31$\pm$0.08 &  0.37 & -0.10 \\
                   $M$ vs. SFR &  55     &  1.20$\pm$0.12 &  0.75 &  0.28 \\
               $dM/dt$ vs. SFR &  55     &  1.11$\pm$0.12 &  0.75 &  0.26 \\
                   $p$ vs. SFR &  55     &  1.43$\pm$0.14 &  0.78 &  0.26 \\
               $dp/dt$ vs. SFR &  55     &  1.34$\pm$0.16 &  0.75 &  0.18 \\
                   $E$ vs. SFR &  45$^a$ &  0.55$\pm$0.18 &  0.37 &  0.29 \\
                   $E$ vs. SFR &  55     &  1.79$\pm$0.14 &  0.81 &  0.35 \\
               $dE/dt$ vs. SFR &  55     &  1.64$\pm$0.15 &  0.80 &  0.29 \\
      $\Delta v_{max}$ vs. SFR &  59$^b$ &  0.21$\pm$0.04 &  0.73 &  0.28 \\
                 $M$ vs. $M_K$ &  48     & -0.44$\pm$0.10 & -0.74 & -0.03 \\
             $dM/dt$ vs. $M_K$ &  48     & -0.33$\pm$0.10 & -0.70 & -0.04 \\
                 $p$ vs. $M_K$ &  48     & -0.47$\pm$0.12 & -0.74 & -0.04 \\
                 $E$ vs. $M_K$ &  48     & -0.79$\pm$0.11 & -0.78 & -0.18 \\
             $dE/dt$ vs. $M_K$ &  48     & -0.69$\pm$0.12 & -0.74 & -0.14 \\
              $\eta$ vs. $M_K$ &  42$^a$ &  0.38$\pm$0.08 &  0.42 &  0.37 \\
    $\Delta v_{max}$ vs. $M_K$ &  52$^b$ & -0.11$\pm$0.03 & -0.66 & -0.20 \\
    $N$(\ion{Na}{1}) vs. $M_R$ &  29     & -0.32$\pm$0.07 & -0.61 & -0.31 \\
                 $M$ vs. $M_R$ &  29     & -0.72$\pm$0.13 & -0.83 & -0.03 \\
             $dM/dt$ vs. $M_R$ &  29     & -0.54$\pm$0.14 & -0.78 & -0.02 \\
                 $p$ vs. $M_R$ &  29     & -0.76$\pm$0.17 & -0.83 & -0.02 \\
                 $E$ vs. $M_R$ &  29     & -1.10$\pm$0.15 & -0.86 & -0.24 \\
             $dE/dt$ vs. $M_R$ &  29     & -0.91$\pm$0.16 & -0.80 & -0.21 \\
    $\Delta v_{max}$ vs. $M_R$ &  33$^b$ & -0.17$\pm$0.03 & -0.68 & -0.29 \\
    $N$(\ion{Na}{1}) vs. $v_c$ &  19     &  1.24$\pm$0.33 &  0.70 &  0.33 \\
                 $M$ vs. $v_c$ &  19     &  3.64$\pm$0.67 &  0.81 &  0.45 \\
             $dM/dt$ vs. $v_c$ &  19     &  2.88$\pm$0.67 &  0.80 &  0.48 \\
                 $p$ vs. $v_c$ &  19     &  3.97$\pm$0.84 &  0.81 &  0.47 \\
             $dp/dt$ vs. $v_c$ &  19     &  3.23$\pm$0.83 &  0.78 &  0.34 \\
                 $E$ vs. $v_c$ &  19     &  5.34$\pm$0.82 &  0.83 &  0.50 \\
             $dE/dt$ vs. $v_c$ &  19     &  4.64$\pm$0.79 &  0.81 &  0.46 \\
    $\Delta v_{max}$ vs. $v_c$ &  20$^b$ &  0.85$\pm$0.15 &  0.77 &  0.48 \\
\enddata
\tablecomments{Col.(1): Variables involved in fit: outflow property vs. host galaxy property. Col.(2): Number of points (galaxies or nuclei).  Unless otherwise noted, points come from our sample, \citet[][4 dwarf starbursts]{sm04}, and \citet[][6 ULIRGs]{m05}. Col.(3): Slope and 1$\sigma$ error ($X \sim Y^{a\pm\delta a}$). Col.(4): Pearson's (parametric) correlation coefficient.  Col.(5): Spearman's (non-parametric) correlation coefficient.}
\tablenotetext{a}{Our data only (IRGs and ULIRGs).  The correlation of $\eta$ with SFR is not really a correlation {\it per se}, but only a consequence of the lack of dependence of $dM/dt$ on SFR in our data.}
\tablenotetext{b}{Includes data from \citet{hlsa00}.}
\end{deluxetable}

\end{document}